%% file: main.tex
\pdfoutput=1
\documentclass[11pt,a4paper]{article}
\usepackage[utf8]{inputenc}
\usepackage[T1]{fontenc}
\usepackage{amsmath,amssymb,amsfonts,bm,slashed}
\usepackage{graphicx,booktabs,xcolor}
\usepackage[margin=2.5cm]{geometry}
\usepackage[colorlinks=true,linkcolor=blue!60!black,citecolor=green!40!black,urlcolor=blue!60!black]{hyperref}
\usepackage{cite}
\numberwithin{equation}{section}

\newcommand{\Mttj}{M_{t\bar t j}}
\newcommand{\as}{\alpha_s}
\newcommand{\MSbar}{\overline{\rm MS}}
\newcommand{\Nbar}{\bar N}
\newcommand{\order}[1]{\mathcal{O}(#1)}

\title{\mbox{}\hfill{\normalsize KA-TP-31-2026}\\[1.5em]
Threshold resummation for top-quark pair production\\ in association with a jet at the LHC}
\author{Bakar Chargeishvili\thanks{\href{mailto:bakar.chargeishvili@kit.edu}{bakar.chargeishvili@kit.edu}}\\[2mm]
{\small\itshape Institute for Theoretical Physics, Karlsruhe Institute of Technology (KIT),}\\
{\small\itshape Wolfgang-Gaede-Stra\ss e 1, 76131 Karlsruhe, Germany}}
\date{November 8, 2026}

\usepackage{xcolor}

\begin{document}
\maketitle

\begin{abstract}
We study threshold resummation for the invariant-mass distribution of a top-quark pair and a jet at the LHC. We perform the
resummation at next-to-next-to-leading logarithmic accuracy in soft-collinear effective theory, with the logarithms of the jet radius treated at fixed
order, and match it to the exact next-to-leading-order (NLO) cross section. We correct the one-loop soft anomalous-dimension matrices in the literature and validate the hard functions. We compare the leading-power kernel with exact real emission, and validate this comparison on Drell--Yan production. At invariant masses up to about 1~TeV with a fixed jet threshold the method does
not apply. At 1--4~TeV with a jet threshold of 15\% of the invariant mass, the kernel describes the exact real-emission density close to threshold but
fails once the emitted energy exceeds about 100--200~GeV. We trace part of this failure to initial states absent from the kernel, which depend on $\tau=M^2/S$, with $M$ the invariant mass and $\sqrt{S}$ the hadronic centre-of-mass energy. In addition, at fixed $\tau$ the difference between the kernel and the exact result grows with the ratio $M/m_t$ of the invariant mass to the top-quark mass. As a consequence, the resummed correction beyond NLO, for a scale equal to the invariant mass, ranges from 1--2\% to 17--23\% for different treatments of this region. The upper end becomes 13--16\% for a jet threshold of 20\% of the invariant mass and 23--33\% for 10\%. The range is not a bound, and the correction is
not under perturbative control. We therefore keep NLO as the central prediction. We compare the possible effect of resummation with the NLO scale uncertainty, with the known next-to-next-to-leading-order corrections, and
with NLO matched to a parton shower.
\end{abstract}

\clearpage
\tableofcontents
\clearpage

\input{sections/introduction}
\input{sections/framework}
\input{sections/ingredients}
\input{sections/resummed_formula}
\input{sections/implementation}
\input{sections/threshold_dominance}
\input{sections/leading_power_validity}
\input{sections/damping}
\input{sections/results}
\input{sections/parton_shower}
\input{sections/conclusions}

\section*{Acknowledgements}
The author thanks Sven-Olaf Moch for discussions and comments on the manuscript.

The work of B.C. was supported by the Deutsche Forschungsgemeinschaft (DFG) under grant 396021762 --- TRR 257 ``Particle Physics Phenomenology after the Higgs Discovery''.

\appendix
\input{sections/appendix_anomalous}
\input{sections/appendix_mellin}
\input{sections/appendix_mg5}

\bibliographystyle{JHEP}
\bibliography{refs}
\end{document}

%% file: sections/introduction.tex
\section{Introduction}
\label{sec:intro}

Top-quark pair production in association with a hard jet, $pp\to t\bar t+\text{jet}$, plays a central role in the
top-quark programme at the Large Hadron Collider (LHC). The cross section is large. At $\sqrt S=13$~TeV the
production of a $t\bar t$ pair together with a jet of transverse momentum $p_T>30$~GeV and rapidity $|y|<2.5$
amounts to about $360$~pb~\cite{Badger:2025ilt}, a substantial fraction of the total $t\bar t$ cross
section~\cite{Czakon:2013goa}. The process is therefore an important signal and a large background to many
searches. It is also a precision tool. The normalised distribution in the invariant mass $\Mttj$ of the
$t\bar t+\text{jet}$ system, usually expressed through $\rho_s=2m_0/\Mttj$ with a fixed reference mass $m_0$, is
sensitive to the top-quark mass~\cite{Alioli:2013mxa}. It has been used to determine the top-quark mass in the
pole and $\MSbar$ schemes~\cite{ATLAS:2015pfy,ATLAS:2019guf,CMS:2022emx}.

The theoretical description of this process has advanced steadily. The next-to-leading-order (NLO) QCD
corrections were computed in refs.~\cite{Dittmaier:2007wz,Melnikov:2010iu}, the latter including the decays of the
top quarks. Off-shell effects were included in ref.~\cite{Bevilacqua:2015qha}, and NLO predictions were matched to
parton showers in refs.~\cite{Kardos:2011qa,Alioli:2011as}. NLO predictions are today available in
general-purpose frameworks such as \textsc{MadGraph5\_aMC@NLO}~\cite{Alwall:2014hca}. The NLO corrections are
sizeable. For the cross section of ref.~\cite{Badger:2025ilt}, the variation of the unphysical renormalisation and factorisation scales
spans a range of 10--32\% of the NLO result, depending on the central scale. The next-to-next-to-leading-order (NNLO)
corrections have recently been computed~\cite{Badger:2025ilt}, with the two-loop amplitudes evaluated in the
leading-colour approximation and the missing subleading-colour terms estimated to affect the differential distributions by
1--4\%. These corrections reduce the corresponding range of the scale variation to 5--12\%.

A complementary way of improving fixed-order predictions is the resummation of logarithmically enhanced
contributions to all orders in the strong coupling $\as$. When the invariant mass of the observed final state,
$\Mttj^2$, approaches the partonic centre-of-mass energy squared $\hat s$ (the \emph{partonic threshold}),
additional real radiation is restricted to be soft. The incomplete cancellation between real and virtual soft
corrections then leaves logarithms of $1-z$, with $z=\Mttj^2/\hat s$, which become large as $z\to1$. At each order
in $\as$ the leading logarithms (LL) are those with the highest power of $\ln(1-z)$, followed by the
next-to-leading (NLL), next-to-next-to-leading (NNLL) logarithms and so on. Neglecting all terms suppressed by
powers of $1-z$ is referred to as the \emph{leading-power} approximation.
The resummation of these logarithms was developed in refs.~\cite{Sterman:1986aj,Catani:1989ne} and extended to
processes with several coloured particles, where the soft radiation is sensitive to how the colour charges are
connected in the hard process, in ref.~\cite{Kidonakis:1997gm}. It was later reformulated in soft-collinear
effective theory (SCET)~\cite{Bauer:2000yr,Bauer:2001yt,Bauer:2002nz,Becher:2006mr,Becher:2007ty}, for which
ref.~\cite{Becher:2014oda} provides a pedagogical introduction.
For top-quark pair production, threshold resummation was carried out at NLL and NNLL accuracy for the total
cross section~\cite{Bonciani:1998vc,Kidonakis:2001nj,Kidonakis:2010dk,Beneke:2011mq}. For the pair invariant-mass distribution,
the factorisation was derived and the NNLO threshold terms were obtained in ref.~\cite{Ahrens:2009uz}, and NNLL
resummation was performed in ref.~\cite{Ahrens:2010zv}. This was later combined with the resummation of
logarithms of $m_t/M_{t\bar t}$, which are important for boosted top quarks~\cite{Ferroglia:2012ku,Pecjak:2016nee},
and matched to NNLO~\cite{Czakon:2018nun}.

Threshold resummation has also been extended to $2\to3$ processes with top quarks in the final state, mainly
the associated production of a top-quark pair with a heavy colourless boson. For $t\bar tW^\pm$ it was performed at NNLL accuracy in SCET
in ref.~\cite{Li:2014ula}. For $t\bar tH$, the soft anomalous dimension for two massive coloured particles produced
with a colour singlet was given in ref.~\cite{Kulesza:2015vda}, together with the NLL resummation of the total cross
section. NNLL resummation has since been carried out for $t\bar tH$, $t\bar tW^\pm$ and $t\bar tZ$, both in
SCET~\cite{Broggio:2015lya,Broggio:2016zgg,Broggio:2016lfj,Broggio:2017kzi} and in the direct QCD Mellin-space
approach~\cite{Kulesza:2017ukk,Kulesza:2018tqz}, and has been combined with the complete NLO QCD and electroweak
corrections~\cite{Broggio:2019ewu,Kulesza:2020nfh}. A combined resummation of soft and Coulomb corrections to the total
$t\bar tH$ cross section was performed in ref.~\cite{Ju:2019lwp}. In single-particle-inclusive (1PI) kinematics,
soft-gluon resummation was formulated for general $2\to n$ processes in ref.~\cite{Forslund:2020lnu} and applied to
approximate higher-order corrections for $t\bar t\gamma$, $t\bar tW^\pm$, $t\bar tZ$ and
$t\bar tH$~\cite{Kidonakis:2022qvz,Kidonakis:2023jpj,Kidonakis:2024lht,Kidonakis:2025zha}, and for single-top
production with a light quark and a Higgs boson, a photon or a $Z$
boson~\cite{Forslund:2021evo,Kidonakis:2022ocq,Kidonakis:2022ljg}. In all of these processes at most four partons
carry colour, so the colour space is that of a $2\to2$ process. For $t\bar t$ associated production, the final state
in addition contains no massless coloured parton. Neither simplification holds for $t\bar t+\text{jet}$, which has
five coloured partons, including gluons.

For $t\bar t+\text{jet}$, the first process-specific ingredient of threshold resummation is the one-loop soft
anomalous-dimension matrix. This matrix governs how the colour structure of the hard process evolves under soft
radiation from the five coloured external partons, two of them massive. It was computed in
ref.~\cite{Chargeishvili:2022ngl}, building on a master's thesis~\cite{Chargeishvili:2021msc} in which the
kinematics near threshold, the factorisation into hard, soft and jet functions, and the colour bases for the
$q\bar q\to t\bar tg$ and $gg\to t\bar tg$ channels were worked out. In the course of the present work we found
three errors in the one-loop matrices of ref.~\cite{Chargeishvili:2022ngl}, which we correct in
section~\ref{sec:soft}. The colour algebra of this process, with colour spaces of dimension up to eleven for
general $N_c$, was one of the motivations for the tool \textsc{OrthoBase}~\cite{Chargeishvili:2024pnq}, which
constructs orthogonal multiplet bases in $\mathrm{SU}(N_c)$ colour space. The present paper continues this
programme to a first phenomenological study. We derive and implement the threshold resummation of
$d\sigma/d\Mttj$ at NNLL accuracy in the threshold logarithms, with the logarithms of the jet radius treated at
fixed order. We match the result to the exact NLO cross section and investigate its validity at the LHC.

We address whether threshold resummation improves the description of
$d\sigma/d\Mttj$ at the LHC, and in which region. The answer is not obvious, for two reasons. First, the partonic
threshold is not enforced kinematically unless $\tau=\Mttj^2/S$ approaches one. Threshold logarithms dominate
only dynamically, when the parton luminosity falls steeply enough with increasing partonic energy to suppress
contributions far from threshold~\cite{Becher:2007ty,Bonvini:2010tp}. Whether this happens depends on $\tau$ and on
the partonic channel. Second, the final state contains a jet, a collimated spray of radiation defined by a jet
algorithm with radius $R$. In the threshold limit the soft radiation must be split further into a part at wide
angles and a part collinear to the jet~\cite{Becher:2015hka,Chien:2015cka}. This separation is consistent only if
the jet is energetic compared with the soft radiation.

Our main findings, obtained for $\sqrt S=13.6$~TeV, anti-$k_T$ jets with $R=0.4$, and a central scale
$\mu=\Mttj$, are the following.
\begin{enumerate}
\item For invariant masses up to about 1~TeV and a fixed jet threshold of a few tens of GeV, threshold resummation
  is not applicable, for two independent reasons. The parton luminosity is too flat at small $\tau$ to make the
  threshold region dominant: the threshold approximation overestimates the exact NLO correction by about 40\% at
  $\Mttj=600$~GeV and $\sqrt S=13$~TeV, and this mismatch decreases only at much lower collider energies. In addition, the jet typically carries
  only about 20\% of the available energy. The scale associated with radiation at the jet boundary then exceeds
  the scale of the jet itself, and the evolution between these scales produces a large but unphysical enhancement.
\item For a jet threshold that grows with the invariant mass, $p_T^j>0.15\,\Mttj$, and for $\Mttj=1$--4~TeV, both
  problems are reduced. A comparison with the exact NLO real-emission spectrum, whose method we validate on the
  Drell--Yan process, shows that the leading-power approximation is accurate close to threshold, to within a few per cent at $z>0.95$ (3--4\% at 4~TeV). It fails, however, once the
  energy radiated away from the observed system exceeds about 100--200~GeV. The difference grows with $\Mttj/m_t$ at fixed $\tau$, so that a steeper luminosity alone does not remove it. Part of it comes from initial states absent from the leading-power kernel, which a simple collinear expression describes.
\item The resummed correction beyond NLO depends strongly on how the region beyond this onset is treated. Between a
  prescription that suppresses the resummation there and the unsuppressed result, the correction ranges from
  1--2\% to 17--23\% of the NLO cross section; this range is not a bound, since other prescriptions give values outside it in both directions. Its upper end also depends on the jet threshold, being 13--16\% for $p_T^j>0.2\,\Mttj$ and 23--33\% for $p_T^j>0.1\,\Mttj$. The correction is thus
  not under perturbative control. Its upper end is comparable with the upward NLO scale uncertainty, whereas the damped result lies far inside this band. These values hold for $\mu=\Mttj$: lowering the NLO scale to $\Mttj/2$ raises NLO by 20--22\%, about the size of the upper end, and the correction matched at that scale is only $+1.3$ to $-3.5$\%.
\item Logarithms of the top-quark mass contained in the leading-power kernel, which are resummed for boosted top-quark pairs in
  refs.~\cite{Ferroglia:2012ku,Pecjak:2016nee,Czakon:2018nun}, do not appear to have a large effect here. In our estimate, which uses one global factor per mass, they cancel to good accuracy
  between the hard-collinear matching coefficient of the top quark and the corresponding soft-collinear function, and the top
  quarks are in any case only moderately boosted for a large part of the cross section.
\item Matching NLO to a parton shower reduces the cross section for $R=0.4$ relative to NLO, by 10--24\% (parton level, stable top quarks), with a shower uncertainty of about $-6\%/{+14\%}$. This is comparable to or larger than the
  resummed correction. The reduction largely disappears for $R=1.0$, which suggests logarithms of the jet radius
  of the kind discussed in refs.~\cite{Dasgupta:2014yra,Dasgupta:2016bnd,Kang:2016mcy}.
\end{enumerate}

The paper is organised as follows. Section~\ref{sec:framework} describes the
kinematics and the factorisation theorem. Section~\ref{sec:ingredients} discusses its ingredients: the hard
function computed from one-loop amplitudes, the soft function and its colour structure (including the corrections
to ref.~\cite{Chargeishvili:2022ngl}), the jet and soft-collinear functions, and non-global
logarithms~\cite{Dasgupta:2001sh}. Section~\ref{sec:resummation} explains how the renormalisation-group equations
of these functions are solved, how the resummed cross section is assembled and matched to NLO, and how logarithms
of the top-quark mass are treated. Section~\ref{sec:implementation} describes the numerical implementation and its
validation. Sections~\ref{sec:dominance} and~\ref{sec:lp-validity} contain the core of the analysis. They examine
when the threshold region dominates, and how well the leading-power approximation reproduces the exact NLO
corrections, including the validation of our method on Drell--Yan production. Section~\ref{sec:damping} discusses
how the resummation can be restricted to the region where it is valid. Section~\ref{sec:results} presents our
predictions for $d\sigma/d\Mttj$, and section~\ref{sec:shower} compares them with NLO matched to a parton shower.
We conclude in section~\ref{sec:conclusions}. Technical details are collected in the appendices.

%% file: sections/framework.tex
\section{Kinematics and factorisation}
\label{sec:framework}

In this section we define the observable and the kinematic variables, and explain in which sense the partonic
cross section is dominated by configurations close to threshold. We then state the factorisation theorem on which
the resummation is based. The individual functions entering it are discussed in section~\ref{sec:ingredients}.

\subsection{Process, observable and threshold variable}
\label{sec:kinematics}

We consider the hadronic process
\begin{equation}
  h_1(P_1) + h_2(P_2) \;\to\; t(p_3) + \bar t(p_4) + \text{jet}(p_5) + X\,,
  \label{eq:process}
\end{equation}
with $S=(P_1+P_2)^2$. At lowest order in $\as$ it proceeds through the partonic channels
\begin{equation}
  q\bar q \to t\bar t g\,,\qquad gg\to t\bar t g\,,\qquad qg\to t\bar t q\,,
  \label{eq:channels}
\end{equation}
together with the channels obtained from them by charge conjugation and by exchanging the incoming partons.
Here $q$ denotes any of the five light quark flavours, all treated as massless. The jet is defined by the
anti-$k_T$ algorithm~\cite{Cacciari:2008gp} with radius parameter $R$ in the rapidity--azimuth plane, and we require
the leading jet to have $p_T^{j}>p_{T,\min}$. The top quarks are treated as stable on-shell particles of mass $m_t$.
Our observable is the distribution in the invariant mass
\begin{equation}
  \Mttj^2 \equiv M^2 = (p_3+p_4+p_5)^2
  \label{eq:M}
\end{equation}
of the two top quarks and the leading jet, where $p_5$ denotes the jet momentum. This choice is closely related to
the variable $\rho_s=2m_0/\Mttj$ used for top-quark mass determinations~\cite{Alioli:2013mxa}. Because we do not
impose rapidity cuts, all selection criteria are invariant under boosts along the beam axis. We exploit this
property in eq.~\eqref{eq:hadronic} below.

Threshold resummation requires a choice of the variable that measures the distance from threshold. In
ref.~\cite{Chargeishvili:2021msc} two classes of kinematics were contrasted for this process. In
\emph{one-particle-inclusive} (1PI) kinematics a single final-state object is measured, for instance the jet
transverse momentum, and the threshold variable measures the invariant mass of everything else. In
\emph{pair-invariant-mass} (PIM) kinematics the invariant mass of a system of final-state objects is measured, as in
eq.~\eqref{eq:M}. Ref.~\cite{Chargeishvili:2021msc} derived the factorisation for 1PI kinematics (the general $2\to n$ formalism in 1PI kinematics is given in
ref.~\cite{Forslund:2020lnu}) and left the PIM
case, which is the natural one for $\Mttj$, for future work. Here we adopt PIM kinematics, as was done for
$t\bar t$ production in refs.~\cite{Ahrens:2009uz,Ahrens:2010zv}. The hard function and its anomalous dimension do
not depend on the choice of threshold variable, and neither does the colour mixing of the soft function, so the
relevant colour structures carry over (see section~\ref{sec:soft} for the relation to
ref.~\cite{Chargeishvili:2022ngl}). The soft and jet functions do depend
on it, and are discussed for PIM kinematics in section~\ref{sec:ingredients}.

Let $p_1=x_1P_1$ and $p_2=x_2P_2$ be the momenta of the incoming partons and $\hat s=(p_1+p_2)^2=x_1x_2S$ the
partonic centre-of-mass energy squared. We define the threshold variable
\begin{equation}
  z \equiv \frac{M^2}{\hat s}\,,\qquad \tau\equiv \frac{M^2}{S}\le z\le 1\,.
  \label{eq:z}
\end{equation}
At lowest order the final state consists of exactly the three particles in eq.~\eqref{eq:M}, so $z=1$. Beyond
lowest order additional partons, collectively denoted by $X$, are radiated. Energy--momentum conservation gives
$\hat s-M^2=2\,p_{t\bar tj}\cdot p_X+p_X^2$ with $p_{t\bar tj}=p_3+p_4+p_5$. In the partonic centre-of-mass frame,
where $\vec p_{t\bar tj}=-\vec p_X$, this becomes the exact relation
\begin{equation}
  1-z = \frac{2E_X}{\sqrt{\hat s}}-\frac{p_X^2}{\hat s}\,.
  \label{eq:sminusM}
\end{equation}
Near threshold the second term is suppressed relative to the first by a further power of the small energy of the
radiation. The limit $z\to 1$ therefore forces all radiation in $X$ to be \emph{soft}: its total energy
$E_X\simeq \sqrt{\hat s}\,(1-z)/2$ is small compared with $M$. This constraint applies only to radiation outside the
jet. Partons clustered into the leading jet are part of $p_5$ and hence of $M$ itself, so they are not restricted
by $z\to1$. This is why the jet algorithm and its radius $R$ enter the threshold factorisation, as discussed in
section~\ref{sec:factorisation}.

\subsection{Hadronic cross section and parton luminosities}
\label{sec:hadronic}

Collinear factorisation expresses the hadronic cross section as a convolution of parton distribution functions
(PDFs) $f_{i/h}(x,\mu_f)$ with partonic cross sections, where $\mu_f$ is the factorisation scale. For the
distribution in $M$ we write it as
\begin{equation}
  \frac{d\sigma}{dM}
  = \frac{1}{M^3}\sum_{\rm ch}\frac{1}{\mathcal N_{\rm ch}}\int_\tau^1\frac{dz}{z}\;
    \mathcal L_{\rm ch}\!\left(\frac{\tau}{z},\mu_f\right)\, C_{\rm ch}(z,\mu_f)\,,
  \label{eq:hadronic}
\end{equation}
where $\mathcal N_{\rm ch}$ is the number of spin and colour states of the incoming partons
($\mathcal N=36$, 256 and 96 for $q\bar q$, $gg$ and $qg$), and the weighted parton luminosity is
\begin{equation}
  \mathcal L_{ij}(y,\mu_f) = y\int_y^1\frac{dx}{x}\,f_{i/h_1}(x,\mu_f)\,f_{j/h_2}\!\left(\frac{y}{x},\mu_f\right).
  \label{eq:lumi}
\end{equation}
The luminosity of each channel is summed over quark flavours and, for distinct incoming partons ($i\neq j$), over
the two assignments of the partons to the hadrons. The two assignments give the same partonic contribution
because the selection criteria are invariant under boosts along, and reflections of, the beam axis. The function
$C_{\rm ch}(z,\mu_f)$ is the hard-scattering kernel, integrated over the final-state phase space at fixed $M$. At
lowest order,
\begin{equation}
  C^{(0)}_{\rm ch}(z) = \delta(1-z)\int d\Phi_3(M)\sum_{\rm col,\,spin}|\mathcal M_{\rm ch}(\Phi_3)|^2\,,
  \label{eq:CLO}
\end{equation}
where $d\Phi_3(M)$ is the Lorentz-invariant phase space of $t\bar tj$ at partonic energy $\sqrt{\hat s}=M$
including the jet cuts, and the sum runs over the colours and spins of all external particles. Equation~\eqref{eq:hadronic}
then reduces to the familiar product of a luminosity and a Born phase-space integral,
$d\sigma/dM=M^{-3}\sum_{\rm ch}\mathcal N_{\rm ch}^{-1}\mathcal L_{\rm ch}(\tau)\int d\Phi_3\sum|\mathcal M_{\rm ch}|^2$.
Beyond lowest order it is often convenient to keep the dependence on the Born phase-space point. We do this by
defining $C(z,\Phi_3)$ through the factorisation theorem below, in which all functions are evaluated with the
Born kinematics $\Phi_3$ of the $t\bar tj$ system in its rest frame. Only the $\Phi_3$-integrated kernel is
a physical quantity; the unintegrated one is defined up to terms suppressed by powers of $1-z$.

Equation~\eqref{eq:hadronic} makes clear in what sense threshold resummation can be relevant at a hadron collider.
The partonic kernel $C(z)$ is convolved with the luminosity evaluated at $\tau/z$. Contributions far from threshold
($z\ll 1$) therefore enter with $\mathcal L(\tau/z)$ probed at a larger momentum fraction. If $\mathcal L(y)$
falls steeply with $y$, these contributions are suppressed and the integral is dominated by $z\to 1$, even though
nothing in the kinematics forces the partonic process to be close to threshold. This mechanism is known as
\emph{dynamical threshold enhancement}~\cite{Becher:2007ty}. It is most transparent in Mellin space. Taking moments
$\widetilde F(N)\equiv\int_0^1 dx\,x^{N-1}F(x)$ with respect to $\tau$ and writing
$\tau^{N-1}=(\tau/z)^{N-1}z^{N-1}$, the convolution in eq.~\eqref{eq:hadronic} turns into a product,
\begin{equation}
  \int_0^1 d\tau\,\tau^{N-1}\int_\tau^1\frac{dz}{z}\,\mathcal L\!\left(\frac{\tau}{z}\right)C(z)
  = \widetilde{\mathcal L}(N)\,\widetilde C(N)\,.
  \label{eq:mellin}
\end{equation}
The hadronic cross section is recovered by the inverse transform,
\begin{equation}
  \frac{d\sigma}{dM}\propto\frac{1}{2\pi i}\int dN\,\tau^{-N}\widetilde{\mathcal L}(N)\widetilde C(N)\,.
\end{equation}
The integrand
$\tau^{-N}\widetilde{\mathcal L}(N)$ has a saddle point on the real axis at some $N_0$, which is larger the faster the
luminosity falls. The partonic kernel is therefore probed mostly at Mellin moments around $N_0$. Large moments
correspond to the threshold region: the threshold logarithms
$\big[\ln^k(1-z)/(1-z)\big]_+$ become $(-1)^{k+1}\ln^{k+1}\Nbar/(k+1)$ plus terms with fewer logarithms, where
$\Nbar=Ne^{\gamma_E}$~\cite{Catani:1996yz}. The size of $N_0$, and hence the importance of the threshold region,
depends on $\tau$ and on the partonic channel, and has to be assessed case by case~\cite{Becher:2007ty,Bonvini:2010tp}.
For $t\bar t+$jet at the LHC we do so quantitatively in section~\ref{sec:dominance}.

\subsection{Modes and the factorisation theorem}
\label{sec:factorisation}

Close to threshold the partonic kernel factorises into functions that each depend on a single characteristic
scale. The functions are matrix elements of fields in soft-collinear effective theory~\cite{Becher:2014oda}, and they
are defined in dimensional regularisation and renormalised in the modified minimal-subtraction ($\MSbar$) scheme.
To identify the scales, consider which radiation is allowed for $z\to 1$.
\begin{itemize}
\item \emph{Soft} radiation at wide angles has energy of order $\omega\equiv 2E_X\simeq\sqrt{\hat s}(1-z)$.
\item Radiation collinear to an incoming parton must also be soft, i.e.\ it carries a small fraction of the
  parton momentum. It is described by the PDFs at large momentum fractions.
\item For the observed jet, radiation inside the jet cone is part of $p_5$ and unconstrained, while radiation
  outside it is restricted to be soft. For a narrow jet ($R\ll1$) this gives rise to two further
  modes~\cite{Becher:2015hka,Becher:2016mmh,Chien:2015cka}. \emph{Collinear} radiation inside the jet has a
  virtuality set by the jet energy and opening angle, $\mu_j\sim E_JR_{\rm eff}$.
  \emph{Soft-collinear} radiation (called ``coft'' in ref.~\cite{Becher:2015hka}) is emitted at angles of order
  $R_{\rm eff}$ around the jet axis with energy of order $\omega$; it is sensitive to the jet boundary, because only
  its part outside the cone contributes to $\omega$. Its natural scale is $\mu_c\sim\omega R_{\rm eff}/2$.
\end{itemize}
Here $E_J$ and $R_{\rm eff}$ are the jet energy and the opening half-angle of the jet cone in the partonic
centre-of-mass frame, which coincides with the rest frame of the $t\bar tj$ system at leading power. For a narrow jet
of rapidity $y_J$ in this frame, $R_{\rm eff}\simeq R/\cosh y_J=R\,p_T^{j}/E_J$, so that $E_JR_{\rm eff}=p_T^jR$,
which is invariant under boosts along the beam.

The resulting factorisation theorem for the hard-scattering kernel reads
\begin{equation}
  \begin{split}
  C(z,\Phi_3,\mu_f) ={}& \int d\omega\;\delta\!\left(\omega-\sqrt{\hat s}(1-z)\right)
  \int d\omega'\;\mathrm{Tr}\Big[\bm H(\Phi_3,\mu_f)\,\bm S(\omega-\omega',\Phi_3,\mu_f)\Big]\\
  &\times\mathcal U_c\!\left(\frac{\omega' R_{\rm eff}}{2},\mu_f\right)
  J\big(p_T^jR,\mu_f\big)\;\mathcal S_{\rm NG}(\mu_c,\mu_j)\,,
  \end{split}
  \label{eq:factorisation}
\end{equation}
up to corrections suppressed by powers of $1-z$, of $\mu_c/\mu_j$ and of $R^2$. Here:
\begin{itemize}
\item $\bm H$ is the \emph{hard function}, a matrix in colour space built from the virtual amplitudes of the $2\to3$
  process after the removal of their infrared poles. Its characteristic scale is $\mu_h\sim M$.
\item $\bm S$ is the \emph{global soft function}, also a matrix in colour space. It describes soft radiation emitted
  by the five external partons. Each parton acts on soft gluons as a classical colour source moving along a
  straight line, which is represented by a \emph{Wilson line}. For the massless partons and the jet the line is
  light-like (along a direction $n$ with $n^2=0$); for the top quarks it is time-like (along their velocity
  $v$, with $v^2=1$). ``Global'' means that the energy of all soft radiation is measured, independently of its
  direction. The natural scale is $\mu_s\sim\omega$, or $M/\Nbar$ in Mellin space.
\item $\mathcal U_c$ is the coft function. Its convolution with $\bm S$ in the energies $\omega-\omega'$ and
  $\omega'$ distributes the out-of-jet soft energy between the two modes.
\item $J$ is the \emph{unmeasured jet function}~\cite{Ellis:2010rwa,Hornig:2016ahz}. Only the energy and the
  direction of the jet are constrained, not its mass or internal structure, so it does not depend on $\omega$ and
  enters as a multiplicative factor.
\item $\mathcal S_{\rm NG}$ accounts for \emph{non-global logarithms}~\cite{Dasgupta:2001sh,Becher:2016mmh}. Because
  radiation inside and outside the jet is treated differently, a soft-collinear gluon outside the jet can be emitted
  coherently by a harder gluon inside it. This produces logarithms of $\mu_j/\mu_c$ that are not captured by the
  renormalisation-group equations of the individual functions.
\end{itemize}
The trace in eq.~\eqref{eq:factorisation} runs over the colour space of the five external partons and is
discussed in section~\ref{sec:colour}. All functions are evaluated at a common scale, here chosen as $\mu_f$.
Their combined scale dependence cancels, order by order in $\as$, against the threshold limit of the
Dokshitzer--Gribov--Lipatov--Altarelli--Parisi (DGLAP) evolution of the parton luminosity in
eq.~\eqref{eq:hadronic}. This property, \emph{renormalisation-group consistency}, provides a strong check of the
ingredients, and it is used to fix the anomalous dimensions that are not known from a direct calculation
(section~\ref{sec:jet}). Related factorisation theorems for threshold resummation in the presence of an observed
jet or hadron have been studied for single-inclusive jet and hadron production~\cite{deFlorian:2007fv,Liu:2017pbb,Hinderer:2018nkb},
including the joint resummation of threshold and jet-radius logarithms in ref.~\cite{Liu:2017pbb}.

\subsection{Scale hierarchy and our scale choice}
\label{sec:hierarchy}

Equation~\eqref{eq:factorisation} involves the characteristic scales
\begin{equation}
  \mu_h\sim M\,,\qquad \mu_s\sim\omega\sim\sqrt{\hat s}(1-z)\,,\qquad \mu_c\sim\frac{\omega R_{\rm eff}}{2}\,,\qquad
  \mu_j\sim p_T^jR\,.
  \label{eq:hierarchy}
\end{equation}
Resummation consists of evaluating each function at its natural scale, where its perturbative expansion is free of
large logarithms, and evolving it to the common scale $\mu_f$ with its renormalisation-group equation, as explained
in section~\ref{sec:resummation}. Two features of eq.~\eqref{eq:hierarchy} are central to our analysis.

First, the hierarchy $\mu_h\gg\mu_s$ requires $1-z\ll1$. At a hadron collider this is realised only through the
dynamical enhancement of the previous subsection.

Second, the soft expansion around a fixed leading jet assumes that the radiation outside the jet is much softer
than the jet itself. Using eq.~\eqref{eq:hierarchy} and $R_{\rm eff}=Rp_T^j/E_J$, the two jet-related scales obey
$\mu_c/\mu_j\sim\omega/(2E_J)$, independently of $R$, so that this condition reads
\begin{equation}
  \omega\simeq M(1-z)\ll 2E_J\,.
  \label{eq:jetcondition}
\end{equation}
This condition is a requirement on the factorisation itself, and it applies whether or not the logarithms of the jet
radius are resummed. It fails once $1-z\gtrsim2E_J/M$. For a jet that carries only a small fraction of the available
energy, as is typical for a fixed transverse-momentum threshold of a few tens of GeV, it already fails at moderate
$1-z$, and the radiation can no longer be treated as a soft correction to the jet. It is for this reason that we
choose a jet threshold that grows with the invariant mass. Section~\ref{sec:highmass} discusses the value
$p_T^j>0.15\,M$ that we use.

A separate consequence of the failure of eq.~\eqref{eq:jetcondition} concerns the evolution between the two jet scales.
If $\mu_c$ exceeds $\mu_j$, evolving the jet and coft functions from their natural scales produces large corrections with
no physical basis, as section~\ref{sec:jet-hierarchy} shows for moderate $M$. Resumming the logarithms of $\mu_j/\mu_c$
would in addition require control of the non-global logarithms beyond the leading-colour approximation. In our
final predictions we therefore set $\mu_c=\mu_j=\mu_s$. The logarithms of $R$ are then kept at fixed order, the
evolution between $\mu_c$ and $\mu_j$ is absent, and the non-global factor reduces to $\mathcal S_{\rm NG}=1$. The
separate scales $\mu_c$ and $\mu_j$ are used only to diagnose the breakdown of condition~\eqref{eq:jetcondition}
in section~\ref{sec:dominance}.

\subsection{Colour space}
\label{sec:colour}

The hard and soft functions in eq.~\eqref{eq:factorisation} are matrices because soft radiation is sensitive to
the colour flow of the hard process, i.e.\ to the way the colour indices of the external partons are connected. We
use the colour-space formalism of refs.~\cite{Catani:1996vz,Becher:2009cu}. An amplitude is a vector
$|\mathcal M\rangle=\sum_I\mathcal M_I\,|c_I\rangle$ in the space of colour-singlet tensors carrying the colour
indices of the five external partons, where $\{|c_I\rangle\}$ is a basis of this space. The hard and soft functions
are matrices with elements
\begin{equation}
  H_{IJ}=\sum_{\rm spin}\mathcal M_I\,\mathcal M_J^{*}\,,\qquad S^{(0)}_{JI}=\langle c_J|c_I\rangle\,,
  \label{eq:HS}
\end{equation}
where $\bm S^{(0)}$, the lowest-order soft function, is the metric (Gram matrix) of the basis. The squared
amplitude summed over colours and spins is $\mathrm{Tr}[\bm H\bm S^{(0)}]$. For the $q\bar q\to t\bar tg$ and
$qg\to t\bar tq$ channels the colour space is four-dimensional. For $gg\to t\bar tg$ it is eleven-dimensional for
generic $N_c$. At $N_c=3$ one of the eleven basis tensors, a projector onto an irreducible representation that
does not exist for three colours, vanishes identically, so ten basis vectors remain.

As in refs.~\cite{Chargeishvili:2021msc,Chargeishvili:2022ngl}, we use bases built from Kronecker deltas, the generators
$t^a$ and the structure constants $f^{abc}$ and $d^{abc}$. For $q\bar q\to t\bar tg$, for example,
\begin{equation}
  c_1=\delta_{c_1c_2}\,t^{g}_{c_3c_4}\,,\quad c_2=\delta_{c_3c_4}\,t^{g}_{c_2c_1}\,,\quad
  c_3=i f^{xyg}\,t^x_{c_2c_1}t^y_{c_3c_4}\,,\quad c_4=d^{xyg}\,t^x_{c_2c_1}t^y_{c_3c_4}\,,
  \label{eq:basisqq}
\end{equation}
where $c_1,\dots,c_4$ are the colour indices of the quark, antiquark, top and antitop, and $g$ that of the gluon. The
basis for $qg\to t\bar tq$ follows by crossing. These bases are orthogonal, so $\bm S^{(0)}$ is a diagonal matrix,
although not the unit matrix since the vectors are not normalised. Physical results do not depend on the choice of
basis.

The construction of such orthogonal bases by hand quickly becomes cumbersome as the number of partons grows. For
processes with many gluons the bases are most naturally classified by the irreducible representations of
$\mathrm{SU}(N_c)$ that appear in the decomposition of the colour space, and the number of basis vectors grows
rapidly. The colour algebra of $t\bar t+$jet production, where this difficulty first becomes apparent for the present
programme, was one of the motivations for the program \textsc{OrthoBase}~\cite{Chargeishvili:2024pnq}, which automates
the construction of orthogonal multiplet bases using projectors onto irreducible representations. For the five-parton
process studied here the bases of ref.~\cite{Chargeishvili:2022ngl} are sufficient, and all colour algebra is carried
out numerically. Extensions of this work to higher multiplicities, such as $t\bar t+2$~jets, would benefit from
\textsc{OrthoBase} directly.

%% file: sections/ingredients.tex
\section{Ingredients of the factorisation theorem}
\label{sec:ingredients}

In this section we describe each function in eq.~\eqref{eq:factorisation}. Each one is defined in dimensional
regularisation with $d=4-2\epsilon$ and renormalised in the $\MSbar$ scheme. We give its one-loop expression and
its anomalous dimension, which governs its dependence on the renormalisation scale $\mu$ and is the input to the
resummation of section~\ref{sec:resummation}. Throughout we write
\begin{equation}
  a\equiv\frac{\as(\mu)}{4\pi}\,,
  \label{eq:a}
\end{equation}
with $\as$ the five-flavour $\MSbar$ coupling (the top quark is decoupled), and the top-quark mass is defined in
the on-shell scheme. Every function is expanded as $F=F^{(0)}+aF^{(1)}+\order{a^2}$, and every anomalous dimension
as $\gamma(\as)=\sum_{n\ge0}\gamma_n\,a^{n+1}$. The coefficients needed at NNLL are collected, with their
sources, in appendix~\ref{app:anomalous}.

\paragraph{Colour-space conventions.}
We use the colour-space notation of refs.~\cite{Catani:1996vz,Becher:2009cu}. The colour generator $\bm T_k^a$ of
parton $k$ acts on the colour index of that parton. It is $(\bm T_k^a)_{\alpha\beta}=t^a_{\alpha\beta}$ for an
outgoing quark or incoming antiquark, $(\bm T_k^a)_{\alpha\beta}=-t^a_{\beta\alpha}$ for an outgoing antiquark or
incoming quark, and $(\bm T_k^a)_{bc}=-if^{abc}$ for a gluon, where $t^a$ are the generators of the fundamental
representation. We write $\bm T_k\cdot\bm T_l=\sum_a\bm T_k^a\bm T_l^a$, and $\bm T_k^2=C_k$ is the quadratic Casimir
($C_F$ for quarks, $C_A$ for gluons). Because the amplitude is a colour singlet, it satisfies \emph{colour
conservation}, $\sum_k\bm T_k|\mathcal M\rangle=0$. In the basis of section~\ref{sec:colour} each colour operator is
represented by a matrix acting on the vector of colour coefficients of the amplitude.

\subsection{The hard function}
\label{sec:hard}

\paragraph{Definition.}
The hard function encodes the virtual corrections to the $2\to3$ scattering amplitudes. One-loop amplitudes
contain ultraviolet (UV) and infrared (IR) divergences. After UV renormalisation the IR poles in $\epsilon$ remain.
They arise from virtual gluons that are soft or collinear to an external parton, and they cancel against the real
radiation that is described in the factorisation theorem by the soft and jet functions and by the PDFs. The
structure of these poles is universal~\cite{Catani:1998bh,Catani:2000ef}: they are removed by a multiplicative
renormalisation factor $\bm Z$, a matrix in colour space, whose form is established to the order needed
here~\cite{Aybat:2006mz,Becher:2009cu,Becher:2009qa,Gardi:2009qi,Becher:2009kw,Ferroglia:2009ep,Ferroglia:2009ii,Mitov:2009sv}.
We define the IR-subtracted amplitude as
\begin{equation}
  |\mathcal C(\mu)\rangle = \lim_{\epsilon\to0}\bm Z^{-1}(\epsilon,\mu)\,|\mathcal M_{\rm ren}(\epsilon)\rangle\,,
  \qquad
  \bm Z = 1 + a\left[\frac{\bm\Gamma_0'}{4\epsilon^2}+\frac{\bm\Gamma_0}{2\epsilon}\right]+\order{a^2}\,,
  \label{eq:Z}
\end{equation}
where $\bm\Gamma_0$ is the one-loop coefficient of the anomalous-dimension matrix given in
eq.~\eqref{eq:Gamma0} below, and $\bm\Gamma_0'=\partial\bm\Gamma_0/\partial\ln\mu$. The hard function is
\begin{equation}
  \bm H(\mu)=\sum_{\rm hel}|\mathcal C(\mu)\rangle\langle\mathcal C(\mu)|\,,\qquad
  \bm H(\mu_h)=\big(4\pi\as(\mu_h)\big)^3\Big[\bm H^{(0)}+a(\mu_h)\,\bm H^{(1)}(\mu_h)+\order{a^2}\Big]\,,
  \label{eq:Hdef}
\end{equation}
where the sum runs over the helicities of all external particles. The three powers of the coupling of the squared
Born amplitude are factored out and evaluated at the hard scale. In components, with
$|\mathcal C\rangle=\sum_I\mathcal C_I|c_I\rangle$, the matrix elements are
$H_{IJ}=\sum_{\rm hel}\mathcal C_I\mathcal C_J^*$, as in eq.~\eqref{eq:HS}. At one loop, $\bm H^{(1)}$ is the sum of
the interference terms $|\mathcal C^{(0)}\rangle\langle\mathcal C^{(1)}|$ and
$|\mathcal C^{(1)}\rangle\langle\mathcal C^{(0)}|$.

\paragraph{Computation.}
We obtain the colour-resolved tree-level and one-loop amplitudes from the program
\textsc{Recola}~\cite{Actis:2012qn,Actis:2016mpe}, which evaluates the one-loop tensor integrals with the library
\textsc{Collier}~\cite{Denner:2016kdg}. For every phase-space point, helicity configuration and colour structure,
\textsc{Recola} provides the finite part and the coefficients of $\epsilon^{-2}$ and $\epsilon^{-1}$ of the
UV-renormalised one-loop amplitude. We project the amplitudes onto the colour bases of section~\ref{sec:colour},
subtract the IR poles with eq.~\eqref{eq:Z}, and assemble $\bm H^{(0)}$ and $\bm H^{(1)}$ at $\mu=M$.

Two details deserve comment. First, one-loop programs normalise the IR poles with an $\epsilon$-dependent
prefactor that differs from the factor $(4\pi)^\epsilon e^{-\epsilon\gamma_E}$ of the $\MSbar$ scheme at
$\order{\epsilon^2}$. Multiplied by the double pole, this difference leaves a finite term proportional to the
$\epsilon^{-2}$ coefficient. For the normalisation used by \textsc{Recola} the $\MSbar$ finite part is obtained by
adding $\pi^2/12$ times the $\epsilon^{-2}$ coefficient. We confirmed this on the one-loop quark form factor, whose
$\MSbar$ finite part is reproduced to $10^{-15}$. Second, eq.~\eqref{eq:Z} predicts the pole coefficients of the
one-loop amplitude completely, so the computed poles test both the amplitudes and $\bm\Gamma_0$, including its
imaginary parts. We perform this test at every phase-space point and for every helicity configuration, and
report its outcome in section~\ref{sec:hard-num}.

\paragraph{Anomalous dimension.}
The scale dependence of the hard function follows from that of $\bm Z$,
\begin{equation}
  \frac{d\bm H}{d\ln\mu}=\bm\Gamma_H\,\bm H+\bm H\,\bm\Gamma_H^\dagger\,,
  \label{eq:RGEH}
\end{equation}
where $\bm\Gamma_H$ is the IR anomalous dimension of the amplitude. In this equation the Born coupling factored out
in eq.~\eqref{eq:Hdef} is held fixed. For amplitudes with massless and massive coloured particles $\bm\Gamma_H$ is
known to two loops~\cite{Becher:2009kw,Ferroglia:2009ep,Ferroglia:2009ii,Mitov:2009sv}. Labelling the massless
partons by $i,j$ and the top quarks by $I,J$, its one-loop coefficient reads
\begin{align}
  \bm\Gamma_0 &= \sum_{(i,j)}\bm T_i\cdot\bm T_j\;\gamma^{\rm cusp}_0\ln\frac{\mu^2}{-s_{ij}}
  +\sum_{I,j}\bm T_I\cdot\bm T_j\;\gamma^{\rm cusp}_0\ln\frac{m_t\,\mu}{-s_{Ij}}\nonumber\\
  &\quad-\sum_{(I,J)}\bm T_I\cdot\bm T_J\;\gamma^{\rm cusp}_0(\beta_{IJ})
  +\sum_i\gamma^i_0+\sum_I\gamma^Q_0\,,
  \label{eq:Gamma0}
\end{align}
where the sums over $(i,j)$ and $(I,J)$ run over unordered pairs of distinct partons, and
$s_{kl}=2\sigma_{kl}\,p_k\cdot p_l+i0$, with $\sigma_{kl}=+1$ if both momenta are incoming or both outgoing and
$\sigma_{kl}=-1$ otherwise. The $+i0$ prescription fixes the imaginary parts: $\ln(-s_{kl})=\ln|s_{kl}|-i\pi$ for
$\sigma_{kl}=+1$.

The terms of eq.~\eqref{eq:Gamma0} have a simple physical interpretation. The \emph{cusp anomalous dimension}
$\gamma^{\rm cusp}$, with $\gamma^{\rm cusp}_0=4$, governs the soft radiation emitted coherently by a pair of colour
charges, represented by two Wilson lines that meet at the hard vertex at an angle (the ``cusp'')~\cite{Korchemsky:1987wg}.
When a massless parton is involved it multiplies a logarithm of the scale, which generates the Sudakov double
logarithms. For two massive partons the radiation depends on their relative velocity, encoded in the \emph{cusp
angle} $\beta_{IJ}$ defined by $\cosh\beta_{IJ}=-s_{IJ}/(2m_t^2)$. For the two outgoing top quarks
$\beta_{34}=b-i\pi$, with $\cosh b=v_3\cdot v_4$ and $v_k=p_k/m_t$, and the velocity-dependent cusp anomalous
dimension is at one loop
\begin{equation}
  \gamma^{\rm cusp}_0(\beta_{34})=\gamma^{\rm cusp}_0\,\beta_{34}\coth\beta_{34}
  =\gamma^{\rm cusp}_0\,(b-i\pi)\coth b\,.
  \label{eq:cuspbeta}
\end{equation}
Its imaginary part is the Coulomb phase of the top-quark pair, which grows like $1/v_{\rm rel}$ as the relative
velocity $v_{\rm rel}$ of the top quarks goes to zero. Our formalism does not resum the corresponding Coulomb
terms~\cite{Beneke:2011mq}. The non-cusp terms are $\gamma^q_0=-3C_F$, $\gamma^g_0=-\beta_0$ and $\gamma^Q_0=-2C_F$,
with $\beta_0=\tfrac{11}{3}C_A-\tfrac43T_Fn_l$, $T_F=\tfrac12$ and $n_l=5$. Colour conservation implies
$\bm\Gamma_0'=-\gamma^{\rm cusp}_0\sum_iC_i$, with the sum over the massless partons, which is proportional to the
unit matrix. At two loops $\bm\Gamma_H$ contains, in addition to the two-loop analogues of these terms, a genuine
three-parton correlation involving both top quarks and one massless
parton~\cite{Ferroglia:2009ep,Ferroglia:2009ii,Mitov:2009sv}, which we include in our NNLL evolution. The imaginary
parts of $\bm\Gamma_H$ and its non-diagonal colour structure are essential: through eq.~\eqref{eq:RGEH} they mix the
colour components of the hard function.

\subsection{The soft function and its anomalous dimension}
\label{sec:soft}

\paragraph{Definition.}
Soft gluons cannot resolve the internal structure of the hard process. They couple to each energetic or massive
external parton only through its colour charge and its direction of motion. This coupling is described by a
\emph{Wilson line}, the path-ordered exponential
$\bm W_k=\mathcal P\exp\big[ig_s\int_0^\infty ds\,n_k\cdot A^a(sn_k)\bm T_k^a\big]$ of the gluon field along the
classical trajectory of the parton, where for incoming partons the path runs from $-\infty$ to $0$ instead. For a
massless parton of direction $n_k$ the trajectory is light-like ($n_k^2=0$), and for a top quark of velocity
$v_k=p_k/m_t$ it is time-like ($v_k^2=1$, with $n_k$ replaced by $v_k$). The soft function is the squared matrix
element of the product $\bm W$ of the five Wilson lines, summed over soft final states $X_s$, with the total energy
of the soft radiation fixed in the partonic centre-of-mass frame:
\begin{equation}
  \bm S(\omega,\mu)=\sum_{X_s}\langle 0|\bm W^\dagger|X_s\rangle\langle X_s|\bm W|0\rangle\,
  \delta\big(\omega-2\,n_0\cdot p_{X_s}\big)\,,\qquad n_0=(1,\vec 0)\,.
  \label{eq:Sdef}
\end{equation}
The soft function is \emph{global}: the energy of all soft radiation is measured, independently of its direction.
In the small-$R$ factorisation the observed jet enters $\bm W$ as a single light-like Wilson line along the jet
axis. Soft radiation inside the jet cone, a region of size $R^2$, does not in fact contribute to $\omega$. The
leading effect of this restriction is supplied by the coft function of section~\ref{sec:jet}, and the remainder is
a power correction in $R$.

The variable $\omega=2E_{X_s}$ is related to the threshold variable by $\omega\simeq\sqrt{\hat s}(1-z)$. The
convolution structure is simplest in Laplace space, so we use
\begin{equation}
  \widetilde{\bm s}(L,\mu)=\int_0^\infty d\omega\;\exp\!\left(-\frac{\omega}{e^{\gamma_E}\kappa}\right)\bm S(\omega,\mu)\,,
  \qquad L=\ln\frac{\kappa^2}{\mu^2}\,,
  \label{eq:Laplace}
\end{equation}
where $\kappa$ is the Laplace variable conjugate to $\omega$. At large $N$, Mellin moments with respect to $z$
coincide with the Laplace transform~\eqref{eq:Laplace} at $\kappa=\sqrt{\hat s}/\Nbar$, up to corrections of order
$1/N$~\cite{Becher:2006nr,Becher:2007ty}. Large logarithms of $\Nbar$ thus appear as large values of $|L|$ when
$\mu\sim M$.

\paragraph{One-loop result.}
At tree level $\widetilde{\bm s}^{(0)}=\bm S^{(0)}$, the colour metric of eq.~\eqref{eq:HS}. At one loop the
virtual corrections are scaleless and vanish in dimensional regularisation. The real emission of one soft gluon
gives, before renormalisation, a sum over pairs of Wilson lines (``dipoles''),
\begin{equation}
  \begin{split}
  \widetilde{\bm s}^{(1)}_{JI}&=-\sum_{k,l}\langle c_J|\bm T_k\cdot\bm T_l|c_I\rangle\,\mathcal K_{kl}(L)\,,\\
  \mathcal K_{kl}&=\frac{e^{-\epsilon\gamma_E}(4\pi)^\epsilon}{\pi}\,\Gamma(-2\epsilon)\,e^{-\epsilon L}
  \int d\Omega_{2-2\epsilon}(n)\,\frac{p_k\cdot p_l}{(p_k\cdot n)(p_l\cdot n)}\,,
  \end{split}
  \label{eq:s1}
\end{equation}
where $n=(1,\hat n)$ is the direction of the emitted gluon. In contrast to eq.~\eqref{eq:Gamma0}, the sum runs over
ordered pairs and includes $k=l$; the diagonal terms are non-zero only for the massive top quarks. The factor
$\Gamma(-2\epsilon)$, which arises from the energy integral, has a pole, so the angular integral is needed through
$\order{\epsilon}$ to obtain the finite part. For two massless partons it has, in addition, collinear singularities
along $n_k$ and $n_l$.

The angular integrals depend non-trivially on the directions and velocities of all five particles, and we
evaluate them numerically. Each dipole is first split into two terms, each singular along at most one light-like
direction. From each singular term we subtract a function with the same collinear limit, whose $d$-dimensional
integral is known in closed form. The remainder is integrable. For its $\order{\epsilon}$ part we use, in a frame
aligned with the singular direction, the measure
$d\Omega_{2-2\epsilon}\propto\sin^{1-2\epsilon}\!\theta\,d\theta\,\sin^{-2\epsilon}\!\phi\,d\phi\,
\sin^{-1-2\epsilon}\!\chi\,d\chi$, and expand it in $\epsilon$, treating the endpoint singularities in $\chi$ by a
further subtraction. After $\MSbar$ renormalisation the one-loop soft function is an exact quadratic polynomial in $L$,
\begin{equation}
  \widetilde{\bm s}^{(1)}(L)=\bm M_0+\bm M_1\,L+\bm M_2\,L^2\,,
  \label{eq:spoly}
\end{equation}
whose colour matrices $\bm M_k$ depend on the Born kinematics. We tested this calculation against the
$t\bar t$ soft integrals of ref.~\cite{Ahrens:2010zv} and against Drell--Yan production. Its numerical precision is
discussed in section~\ref{sec:soft-num}.

\paragraph{Anomalous dimension.}
The soft function satisfies the renormalisation-group equation (RGE)
\begin{equation}
  \frac{d\widetilde{\bm s}}{d\ln\mu}=-\bm\Gamma_s^\dagger\,\widetilde{\bm s}-\widetilde{\bm s}\,\bm\Gamma_s\,,\qquad
  \bm\Gamma_s=\bm\Gamma_H-\gamma_{s,\bm1}\,\bm 1\,,
  \label{eq:RGES}
\end{equation}
so that the colour-mixing terms cancel in $\mathrm{Tr}[\bm H\widetilde{\bm s}]$. The term $\gamma_{s,\bm1}$,
proportional to the unit matrix, depends on $\ln(\kappa/\mu)$ and on the observable. It is compensated by the jet,
coft and PDF anomalous dimensions, as required by renormalisation-group consistency. The calculation of the soft
function provides a check of eq.~\eqref{eq:RGES}, which we have carried out: the full one-loop $\mu$ dependence of
$\mathrm{Tr}[\bm H\widetilde{\bm s}]$ combined with the jet and coft functions and the PDFs cancels. Because
$\bm T_k\cdot\bm T_l$ is self-adjoint with respect to the colour metric, the imaginary parts of $\bm\Gamma_0$ drop out
of the combination $\bm\Gamma_0^\dagger\bm S^{(0)}+\bm S^{(0)}\bm\Gamma_0$ that the real-emission soft function
probes. This check therefore tests only the real part of the anomalous dimension. The imaginary parts are tested
by the pole check of the hard amplitudes.

\paragraph{Relation to ref.~\cite{Chargeishvili:2022ngl}.}
In our earlier work~\cite{Chargeishvili:2022ngl} the one-loop soft anomalous-dimension matrices for the
$q\bar q\to t\bar tg$ and $gg\to t\bar tg$ channels were computed directly from the eikonal diagrams with the
program \textsc{FormSoft}, which is based on \textsc{Form}~\cite{Vermaseren:2000nd}. Universality implies that,
in the normalisation of that reference (units of $\as/\pi$), they must equal $\bm\Gamma_0/4$ of
eq.~\eqref{eq:Gamma0} up to a multiple of the unit matrix. The unit-matrix terms include logarithms of an auxiliary
gauge vector used in the calculation, which do not affect the colour mixing. We have compared the expressions
provided with ref.~\cite{Chargeishvili:2022ngl} with $\bm\Gamma_0/4$ for random phase-space points, for $N_c=3$ and
$N_c=4$. The two agree to better than $10^{-12}$, but only after three corrections, which refer to version~2 of ref.~\cite{Chargeishvili:2022ngl} and have been incorporated in its version~3:
\begin{enumerate}
\item The massive dipole is expressed in ref.~\cite{Chargeishvili:2022ngl} through
  $L_\beta=\frac{1-2m_t^2/s}{\beta_v}\big[\ln\frac{1-\beta_v}{1+\beta_v}+i\pi\big]$ with
  $\beta_v=\sqrt{1-4m_t^2/s}$. Comparison with eq.~\eqref{eq:cuspbeta} shows that $L_\beta=-\beta_{34}\coth\beta_{34}$
  provided $s$ is the invariant mass of the top-quark pair, $s_{t\bar t}=(p_3+p_4)^2$. In
  ref.~\cite{Chargeishvili:2022ngl} $s$ is the partonic centre-of-mass energy $\hat s$, which coincides with
  $s_{t\bar t}$ only for $t\bar t$ production without an additional jet. This correction changes both the real and
  the imaginary part of $L_\beta$.
\item The dipole formed by the two incoming partons contributes a logarithm written as $\ln(-\tfrac12)$ and
  evaluated in ref.~\cite{Chargeishvili:2022ngl} as $\ln\tfrac12+i\pi$. The $+i0$ prescription of
  eq.~\eqref{eq:Gamma0}, which also produces the imaginary part of $L_\beta$, requires $\ln\tfrac12-i\pi$.
\item The imaginary parts of the two dipoles formed by an outgoing top quark and the outgoing jet,
  for which $\sigma_{35}=\sigma_{45}=+1$, are missing. The term $i\pi\,(\bm T_3\cdot\bm T_5+\bm T_4\cdot\bm T_5)$, in
  units of $\as/\pi$, has to be added.
\end{enumerate}
Apart from the argument of $L_\beta$, the real parts of all logarithms and all colour factors of
ref.~\cite{Chargeishvili:2022ngl} are correct. The imaginary parts do not drop out of physical predictions: they enter
the evolution of the hard function through eq.~\eqref{eq:RGEH} and the traces of section~\ref{sec:resummation}. The
comparison covers the $q\bar q$ and $gg$ channels, for which matrices were given; the $qg$ channel is related to
$q\bar q$ by crossing. In the present work we do not use the matrices of ref.~\cite{Chargeishvili:2022ngl}. We use
the general expression~\eqref{eq:Gamma0} and its two-loop extension. Our implementation reproduces the closed-form
two-loop anomalous-dimension matrices for $q\bar q\to t\bar t$ and $gg\to t\bar t$ of ref.~\cite{Ferroglia:2009ii} for
generic kinematics. For $m_t\to0$ the non-diagonal part of its matrices approaches that of the corresponding massless
five-parton amplitudes, with deviations of order $m_t^2/s_{kl}$.

\subsection{The jet and soft-collinear functions}
\label{sec:jet}

\paragraph{Unmeasured jet function.}
The collinear radiation inside the jet is described by the jet function of an \emph{unmeasured} jet: a jet whose
energy and direction are fixed but whose mass and internal structure are not measured~\cite{Ellis:2010rwa}. For a jet
of energy $E_J$ and opening half-angle $R_{\rm eff}$ the renormalised one-loop result
is~\cite{Ellis:2010rwa,Hornig:2016ahz}
\begin{equation}
  \begin{split}
  J^{(1)}_q &= 2\left[\frac{C_F}{2}L_J^2+\frac32C_FL_J+C_F\Big(\frac{13}{2}-\frac{3\pi^2}{4}\Big)\right],\\
  J^{(1)}_g &= 2\left[\frac{C_A}{2}L_J^2+\frac{\beta_0}{2}L_J+C_A\Big(\frac{67}{9}-\frac{3\pi^2}{4}\Big)
  -\frac{23}{9}T_Fn_l\right],
  \end{split}
  \label{eq:J1}
\end{equation}
for a quark and a gluon jet respectively, with $L_J=\ln\big(\mu^2/(E_JR_{\rm eff})^2\big)$. At this order the jet
contains at most two partons, and the anti-$k_T$, Cambridge/Aachen and $k_T$ algorithms give the same result. For a
narrow jet the scale $2E_J\tan(R/2)$ of ref.~\cite{Ellis:2010rwa} becomes $E_JR_{\rm eff}$, so the natural scale is
$\mu_j=E_JR_{\rm eff}=p_T^jR$, the jet transverse momentum times the jet radius~\cite{Hornig:2016ahz}. The jet function
satisfies the local RGE
\begin{equation}
  \frac{d\ln J}{d\ln\mu}=2C_J\,\gamma^{\rm cusp}(\as)\,\ln\frac{\mu}{E_JR_{\rm eff}}+\gamma^J(\as)\,,
  \label{eq:RGEJ}
\end{equation}
with $C_J=C_F$ or $C_A$, $\gamma^J_0=6C_F$ for quarks and $\gamma^J_0=2\beta_0$ for gluons. The two-loop
coefficient $\gamma^J_1$ was computed for quark jets in ref.~\cite{Liu:2021xzi}. For gluon jets it was obtained in
ref.~\cite{Cal:2024yjz} from renormalisation-group consistency.

\paragraph{Soft-collinear (``coft'') function.}
Soft radiation emitted at angles of order $R_{\rm eff}$ around the jet axis resolves the jet boundary: it becomes
part of the jet if it falls inside the cone, and contributes to the energy $\omega$ if it falls outside. Such
radiation cannot resolve the directions of the other partons, which all lie at large angles from the jet. By colour
conservation their Wilson lines combine into a single line in the direction $\bar n_J$ opposite to the jet, with
colour charge $-\bm T_J$. The coft function is therefore the matrix element of two Wilson lines, along $n_J$ and
$\bar n_J$, with the energy outside the cone measured~\cite{Becher:2015hka,Becher:2016mmh,Chien:2015cka}. It is
proportional to the unit matrix in colour space and depends only on the Casimir $C_J$ of the jet parton. With its
Laplace transform defined as in eq.~\eqref{eq:Laplace}, its one-loop result is
\begin{equation}
  \widetilde{\mathcal U}_c^{(1)}(\ell)=-C_J\Big(\ell^2+\frac{\pi^2}{2}\Big)\,,\qquad
  \ell=\ln\frac{\kappa^2R_{\rm eff}^2}{4\mu^2}\,,
  \label{eq:coft}
\end{equation}
which identifies the natural scale $\mu_c=\kappa R_{\rm eff}/2$, as in eq.~\eqref{eq:hierarchy}. Its RGE reads
\begin{equation}
  \frac{d\ln\widetilde{\mathcal U}_c}{d\ln\mu}=-2C_J\,\gamma^{\rm cusp}(\as)\,\ln\frac{2\mu}{\kappa R_{\rm eff}}+\gamma^c(\as)\,,
  \label{eq:RGEc}
\end{equation}
with $\gamma^c_0=0$, since eq.~\eqref{eq:coft} contains no single logarithm.

\paragraph{Renormalisation-group consistency.}
Adding eqs.~\eqref{eq:RGEJ} and~\eqref{eq:RGEc}, the cusp terms combine into $2C_J\gamma^{\rm cusp}\ln(\kappa/(2E_J))$.
The dependence on the jet radius cancels, as it must, since the global soft function knows nothing about $R$. The
remaining terms must combine with those of the hard, soft and PDF anomalous dimensions to give a $\mu$-independent
cross section. For the non-cusp terms associated with the jet parton this requires
\begin{equation}
  \gamma^J+\gamma^c+\gamma^S_J=-2\gamma^{i}\,,\qquad i=q,g\,,
  \label{eq:consistency}
\end{equation}
where $\gamma^i$ is the contribution of the jet parton to $\bm\Gamma_H$ and $\gamma^S_J$ is the non-cusp part of
$\gamma_{s,\bm1}$ in eq.~\eqref{eq:RGES} associated with the jet direction. At one loop $\gamma^S_J=\gamma^c_0=0$ and
$\gamma^J_0=-2\gamma^i_0$. At two loops, to our knowledge, neither $\gamma^c_1$ nor the separation of the soft non-cusp
term into contributions of the five directions is known for this observable. When we evolve the jet and coft
functions separately, we assume $\gamma^S_J=0$ also at two loops and fix $\gamma^c_1$ by eq.~\eqref{eq:consistency}.
This is an assumption, and it is not expected to be exact: the threshold soft function of the Drell--Yan process has
a non-zero two-loop non-cusp anomalous dimension~\cite{Becher:2007ty}, which suggests a non-zero $\gamma^S_J$ here.
For our final predictions the assumption is irrelevant. With $\mu_c=\mu_j=\mu_s$ (section~\ref{sec:hierarchy}) the
soft, coft and jet functions are evaluated at a common scale, so only the sum of their anomalous dimensions enters.
This sum is fixed unambiguously by $\bm\Gamma_H$ and the PDF evolution. The assumption enters only where the scales are
separated, namely in the diagnostic studies of section~\ref{sec:dominance} and in the variations of $\mu_c$ and $\mu_j$.

\subsection{Non-global logarithms}
\label{sec:ngl}

The observable treats radiation inside and outside the jet differently. As a result, the soft-collinear radiation
at the jet boundary can be emitted coherently by a gluon that is itself inside the jet. This gives rise to
non-global logarithms (NGLs) of $\mu_j/\mu_c$~\cite{Dasgupta:2001sh}, which are not captured by the RGEs above. Their
leading tower, the terms $\as^nL^n$ with $L=\ln(\mu_j/\mu_c)$, is formally of NLL accuracy, and it is known
numerically in the limit of a large number of colours, $N_c\to\infty$. For a narrow jet a boost along the jet axis
maps the cone onto a hemisphere, so that the result for hemisphere energy flows
applies~\cite{Dasgupta:2001sh,Becher:2016mmh}. We use the parametrisation of ref.~\cite{Dasgupta:2001sh},
\begin{equation}
  \mathcal S_{\rm NG}(\mu_c,\mu_j)=\exp\!\left[-C_JC_A\frac{\pi^2}{3}\,\frac{1+(a_1t)^2}{1+(b_1t)^{c_1}}\,t^2\right],\qquad
  t=\int_{\mu_c}^{\mu_j}\frac{d\mu}{\mu}\,\frac{\as(\mu)}{2\pi}\,,
  \label{eq:ngl}
\end{equation}
with $a_1=0.85\,C_A$, $b_1=0.86\,C_A$ and $c_1=1.33$. The Casimir $C_J$ replaces the factor $C_F$ of the original
quark result, which is an approximation for gluon jets. Note that $t$ is defined with $\as/(2\pi)$, following
ref.~\cite{Dasgupta:2001sh}. The function is defined for $\mu_c<\mu_j$, i.e.\ when the
hierarchy~\eqref{eq:jetcondition} holds, and we set $\mathcal S_{\rm NG}=1$ otherwise. For the central scale choice
$\mu_j=\mu_c$ the argument $t$ vanishes and NGLs play no role.
For observables that restrict radiation in part of phase space but leave it unrestricted along the beams, Glauber
phases give rise to super-leading logarithms~\cite{Forshaw:2006fk,Becher:2021zkk}, and for massive final states the
Coulomb phase of the heavy-quark pair provides an additional source~\cite{Banerjee:2025kkq}. Our observable constrains
the soft radiation in all directions outside the jet, so these terms do not arise at the accuracy considered here.

\subsection{Parton distributions}
\label{sec:pdfs}

The PDFs complete the factorisation theorem. We use PDFs evolved with the complete DGLAP equations at NNLO, evaluated
at $\mu_f$. In the factorisation theorem, however, only the threshold limit of their evolution enters: the large-$z$
limit of the splitting functions, which corresponds to large $N$ in Mellin space. This limit concerns the partonic
variable $z$, and it is relevant independently of the momentum fractions at which the PDFs are probed. In this limit
the evolution is diagonal in flavour~\cite{Korchemsky:1988si,Becher:2007ty,Ahrens:2010zv},
\begin{equation}
  \frac{d\ln\widetilde f_i(N,\mu)}{d\ln\mu}=-2C_i\,\gamma^{\rm cusp}(\as)\ln\Nbar+2\gamma^{\phi}_i(\as)
  +\order{1/N}\,,
  \label{eq:pdfrge}
\end{equation}
with $C_q=C_F$, $C_g=C_A$, $\gamma^{\phi}_{q,0}=3C_F$ and $\gamma^{\phi}_{g,0}=\beta_0$. Equation~\eqref{eq:pdfrge} is
used only to evolve the partonic kernel between $\mu_s$ and $\mu_f$ (section~\ref{sec:rge-solution}). Its accuracy for
the moderate values of $N$ relevant at the LHC is one of the questions examined in section~\ref{sec:dominance}.

%% file: sections/resummed_formula.tex
\section{The resummed cross section}
\label{sec:resummation}

The functions of section~\ref{sec:ingredients} are free of large logarithms only when each is evaluated at its own
natural scale. The factorisation theorem~\eqref{eq:factorisation}, however, requires them at a common scale. In this
section we solve their renormalisation-group equations to evolve them from their natural scales to the common scale.
This resums the threshold logarithms~\cite{Sterman:1986aj,Catani:1989ne}. We then assemble the resummed partonic
kernel, define the logarithmic accuracy, specify the scale choices, match the result to the exact NLO cross section,
and discuss logarithms of the top-quark mass.

\subsection{Solution of the renormalisation-group equations}
\label{sec:rge-solution}

\paragraph{Evolution functions.}
A scalar RGE $d\ln F/d\ln\mu=\Gamma(\mu)$ is solved by
\begin{equation}
  F(\mu)=F(\mu_0)\exp\!\left[\int_{\ln\mu_0}^{\ln\mu}d\ln\mu'\;\Gamma(\mu')\right].
  \label{eq:RGEsol}
\end{equation}
The jet, coft and PDF anomalous dimensions have the form $\Gamma(\mu)=A\,\gamma^{\rm cusp}(\as)\ln(\mu/Q)+\gamma(\as)$,
with a constant $A$ and a scale $Q$. Because $\as$ depends on $\mu'$, the exponent is not simply a polynomial in
$\ln(\mu/\mu_0)$. The integral of the cusp term produces the \emph{Sudakov} double logarithms, the terms
$\as^n\ln^{n+1}$ in the exponent, while the integral of $\gamma$ produces single logarithms. We denote
\begin{equation}
  A_{\gamma}(\nu,\mu)=-\int_{\ln\nu}^{\ln\mu}d\ln\mu'\;\gamma\big(\as(\mu')\big)\,,
  \label{eq:AS}
\end{equation}
and $A_\Gamma\equiv A_{\gamma^{\rm cusp}}$~\cite{Becher:2006mr,Becher:2006nr,Becher:2007ty}. These integrals are often
expanded in closed form in terms of the ratio $\as(\mu)/\as(\nu)$. We instead evaluate them by numerical quadrature,
with $\as(\mu)$ taken from the PDF set at all logarithmic orders. The two approaches differ only by terms beyond the
formal accuracy.

\paragraph{Hard evolution.}
The solution of the matrix RGE~\eqref{eq:RGEH} is
\begin{equation}
  \bm H(\mu_s)=\bm U\,\bm H(\mu_h)\,\bm U^\dagger\,,\qquad
  \bm U\equiv\bm U(\mu_h,\mu_s)=\mathcal P\exp\!\left[\int_{\ln\mu_h}^{\ln\mu_s}d\ln\mu\;\bm\Gamma_H(\mu)\right],
  \label{eq:U}
\end{equation}
where $\mathcal P$ denotes path ordering: in the product, matrices closer to $\mu_s$ stand to the left. Path ordering
is needed because $\bm\Gamma_H(\mu)=a(\mu)\bm\Gamma_0+a^2(\mu)\bm\Gamma_1+\ldots$, and $\bm\Gamma_0$ and $\bm\Gamma_1$
do not commute, so that $\bm\Gamma_H$ at different scales does not commute either. In the literature $\bm U$ is often
constructed by diagonalising $\bm\Gamma_0$ and treating the two-loop terms
perturbatively~\cite{Kidonakis:1997gm,Ahrens:2010zv}. We instead solve $d\bm U/d\ln\mu=\bm\Gamma_H(\mu)\bm U$ numerically
for every Born configuration, with a high-order adaptive Runge--Kutta method, which avoids the diagonalisation of
complex matrices of dimension up to ten. The explicit dependence of $\bm\Gamma_H$ on $\ln\mu$ arises from all dipoles
with at least one massless parton in eq.~\eqref{eq:Gamma0}. By colour conservation it is proportional to the unit
matrix and multiplies $-\gamma^{\rm cusp}\sum_iC_i$. Its integral produces the Sudakov double logarithms of the hard
evolution. At NNLL we include the three-loop cusp coefficient in this term only; the corresponding $\order{a^3}$ terms
multiplying the kinematic logarithms in $\bm\Gamma_H$ are formally of higher order.

\paragraph{Evolution to a common scale.}
We evolve all functions to the soft scale $\mu_s$. The hard function is evolved with eq.~\eqref{eq:U}, and the jet and
coft functions from $\mu_j$ and $\mu_c$ with eqs.~\eqref{eq:RGEJ} and~\eqref{eq:RGEc}. The product then has to be
related to the parton luminosity, which is evaluated at the factorisation scale $\mu_f$. Since the hadronic cross
section~\eqref{eq:hadronic} does not depend on $\mu_f$, the kernel at $\mu_f$ follows from the kernel at $\mu_s$ by
undoing the evolution of the two PDFs between $\mu_s$ and $\mu_f$. In Mellin space, and at leading power, this
evolution is multiplicative and given by eq.~\eqref{eq:pdfrge}. Hence the $\mu_f$ dependence of the kernel compensates
that of the PDFs only at leading power.

Collecting the exponents, all dependence on $N$ in the evolution factors takes the form $\Nbar^{-2\eta}$, and the
scalar evolution factors combine into
\begin{equation}
  e^{E}\,\Nbar^{-2\eta}\,,\qquad
  \eta\equiv\sum_{i=1,2}C_i\,A_\Gamma(\mu_s,\mu_f)-C_J\,A_\Gamma(\mu_c,\mu_s)\,,
  \label{eq:eta}
\end{equation}
\begin{align}
  E={}&2\sum_{i=1,2}A_{\gamma^\phi_i}(\mu_s,\mu_f)
  +\int_{\ln\mu_c}^{\ln\mu_s}d\ln\mu\left[-2C_J\gamma^{\rm cusp}\ln\frac{2\mu}{MR_{\rm eff}}+\gamma^c\right]\nonumber\\
  &+\int_{\ln\mu_j}^{\ln\mu_s}d\ln\mu\left[2C_J\gamma^{\rm cusp}\ln\frac{\mu}{p_T^jR}+\gamma^J\right],
  \label{eq:E}
\end{align}
where the sums run over the two incoming partons. For our final scale choice $\mu_c=\mu_j=\mu_s$ the last two terms
of $E$ and the second term of $\eta$ vanish, and only the PDF evolution remains. For $\mu_s<\mu_f$ the parameter $\eta$
is then negative. With the scales of section~\ref{sec:scales} it ranges from $\eta\approx-0.09$ ($q\bar q$ at $M=1$~TeV)
to $\eta\approx-0.30$ ($gg$ at $M=4$~TeV).

\subsection{The resummed kernel}
\label{sec:kernel}

\paragraph{Mellin space.}
Combining these elements, the resummed hard-scattering kernel, for one partonic channel and one Born configuration,
reads in Mellin space
\begin{align}
  \widetilde C_N(\mu_f)={}&\big(4\pi\as(\mu_h)\big)^3\,e^{E}\;\mathcal S_{\rm NG}(\mu_c,\mu_j)\;
  \bigg\{\mathrm{Tr}\Big[\bm U\big(\bm H^{(0)}+a(\mu_h)\bm H^{(1)}(\mu_h)\big)\bm U^\dagger\bm S^{(0)}\Big]\nonumber\\
  &+a(\mu_s)\,\mathrm{Tr}\Big[\bm U\bm H^{(0)}\bm U^\dagger\,\widetilde{\bm s}^{(1)}(L_s+\partial_\eta)\Big]\nonumber\\
  &+\mathrm{Tr}\Big[\bm U\bm H^{(0)}\bm U^\dagger\bm S^{(0)}\Big]\Big(a(\mu_c)\widetilde{\mathcal U}_c^{(1)}(\ell_c+\partial_\eta)
  +a(\mu_j)J^{(1)}\Big)\bigg\}\,f_N(\eta)\,,
  \label{eq:kernelN}
\end{align}
with $L_s=\ln(M^2/\mu_s^2)$, $\ell_c=\ln\big(M^2R_{\rm eff}^2/(4\mu_c^2)\big)$, and $J^{(1)}$ evaluated with
$L_J=\ln\big(\mu_j^2/(p_T^jR)^2\big)$. As in eq.~\eqref{eq:Hdef}, the factor $(4\pi\as)^3$ contains the three powers of
the coupling of the Born process. Each matching correction is multiplied by the coupling at its own scale. We
retain the one-loop matching corrections linearly, without their products, which are of higher order.

Two features of eq.~\eqref{eq:kernelN} require explanation. First, the logarithm $L=\ln(\kappa^2/\mu_s^2)$ of the
Laplace-space soft function, eq.~\eqref{eq:spoly}, is replaced by $L_s+\partial_\eta$. This is possible because
$\kappa=M/\Nbar$, so $L=L_s-2\ln\Nbar$, and because $\partial_\eta\Nbar^{-2\eta}=-2\ln\Nbar\,\Nbar^{-2\eta}$. The
derivative thus generates the $\Nbar$ dependence of the soft function. It is understood that $\eta$ is treated as a
variable, the derivatives are taken, and only then $\eta$ is set to the value~\eqref{eq:eta}. The same holds for the
coft function. Second, the factor $\Nbar^{-2\eta}$ is replaced by
\begin{equation}
  f_N(\eta)=\frac{e^{-2\gamma_E\eta}\,\Gamma(N-\eta)}{\Gamma(N+\eta)}
  =\Nbar^{-2\eta}\left[1+\order{1/N}\right].
  \label{eq:fN}
\end{equation}
The two expressions agree at leading power. We use $f_N(\eta)$ because it is the exact Mellin transform of a simple
function of $z$, see eq.~\eqref{eq:kernelz} below, so that the Mellin-space and momentum-space versions of the resummed
formula are identical~\cite{Becher:2006nr,Becher:2007ty}. It is one of several choices that differ by terms
suppressed by $1/N$. We examine the size of such terms in section~\ref{sec:lp-validity}.

\paragraph{Momentum space.}
Since $\int_0^1dz\,z^{N-1}\,z^{-\eta}(1-z)^{2\eta-1}=\Gamma(N-\eta)\Gamma(2\eta)/\Gamma(N+\eta)$, the kernel in $z$ space
is obtained from eq.~\eqref{eq:kernelN} by the replacement
\begin{equation}
  f_N(\eta)\;\longrightarrow\;\frac{e^{-2\gamma_E\eta}}{\Gamma(2\eta)}\,z^{-\eta}(1-z)^{2\eta-1}\,.
  \label{eq:kernelz}
\end{equation}
Inserting it into eq.~\eqref{eq:hadronic} gives the hadronic cross section as a one-dimensional $z$ integral over the
luminosity. For $\eta>0$ the integral converges at $z\to1$. For $\eta<0$, which is our case, the factor
$(1-z)^{2\eta-1}$ is not integrable and the integral is defined by analytic continuation in $\eta$. In practice we
subtract from the smooth function $h(z)=z^{-\eta-1}\mathcal L(\tau/z)$ the first two terms of its Taylor expansion
around $z=1$ and integrate them analytically:
\begin{equation}
  \begin{split}
  \int_\tau^1dz\,(1-z)^{2\eta-1}h(z)={}&\int_\tau^1dz\,(1-z)^{2\eta-1}\big[h(z)-h(1)+h'(1)(1-z)\big]\\
  &+h(1)\frac{(1-\tau)^{2\eta}}{2\eta}-h'(1)\frac{(1-\tau)^{2\eta+1}}{2\eta+1}\,.
  \end{split}
  \label{eq:subtraction}
\end{equation}
The remaining integral converges for $\eta>-1$, which covers all cases encountered here. Because $\mu_s$ is a fixed
scale at the level of the hadronic cross section, no coupling is evaluated at $M/\Nbar$ or at $M(1-z)$, and the
Landau-pole ambiguity of formulations in which the soft scale follows $N$ does not
arise~\cite{Catani:1996yz,Becher:2006nr,Bonvini:2010tp,Bonvini:2012az}. The form~\eqref{eq:kernelz} is used for our
undamped results, and the form~\eqref{eq:kernelN} in section~\ref{sec:damping}, where we modify the kernel at small $N$.

\paragraph{Fixed-order expansion.}
Setting all scales equal, $\mu_h=\mu_s=\mu_c=\mu_j=\mu_f=\mu$, gives $\bm U=\bm 1$, $E=\eta=0$, and $\mathcal S_{\rm NG}=1$,
and eq.~\eqref{eq:kernelN} reduces to the threshold approximation to the NLO kernel,
\begin{equation}
  \begin{split}
  \widetilde C^{\rm thr}_N(\mu)={}&\big(4\pi\as(\mu)\big)^3\Big\{\mathrm{Tr}[\bm H^{(0)}\bm S^{(0)}]
  +a\Big(\mathrm{Tr}[\bm H^{(1)}\bm S^{(0)}]+\mathrm{Tr}[\bm H^{(0)}\widetilde{\bm s}^{(1)}(L)]\\
  &+\mathrm{Tr}[\bm H^{(0)}\bm S^{(0)}]\big(\widetilde{\mathcal U}_c^{(1)}(\ell)+J^{(1)}\big)\Big)\Big\}\,,
  \end{split}
  \label{eq:thr}
\end{equation}
with $L=\ln(M^2/\mu^2)-2\ln\Nbar$ and $\ell=L+\ln(R_{\rm eff}^2/4)$. Here $\bm H^{(1)}$ is taken at the scale $\mu$, and
the $\partial_\eta$ derivatives of $f_N$ at $\eta=0$ amount to the replacement $\ln\Nbar\to\gamma_E+\psi(N)$, with $\psi$
the digamma function. Equation~\eqref{eq:thr} contains the terms of the NLO correction that are leading-power in the
joint expansion in $1-z$ and $R$, together with some terms of order $1/N$ induced by $f_N$. Terms suppressed by powers
of $1-z$ or by $R^2$ are not included. Its comparison with the exact NLO correction is the subject of
sections~\ref{sec:dominance} and~\ref{sec:lp-validity}.

\subsection{Logarithmic accuracy}
\label{sec:counting}

In Mellin space the logarithm of the resummed kernel has the structure
$\ln\widetilde C_N=L\,g_1(\as L)+g_2(\as L)+\as\,g_3(\as L)+\ldots$, with $L=\ln\Nbar$. The functions $g_k$ collect
the series (``towers'') of terms $\as^nL^{n+1}$ (leading logarithms, LL), $\as^nL^n$ (NLL), $\as^nL^{n-1}$ (NNLL), and so
on. In the counting $\as L\sim1$, the accuracy of eq.~\eqref{eq:kernelN} is determined by the loop orders of its
ingredients, as summarised in table~\ref{tab:orders}. We use the conventions of
refs.~\cite{Becher:2007ty,Ahrens:2010zv}, in which NNLL includes the one-loop matching corrections, i.e.\ the one-loop
terms of the functions at their natural scales. NLL$'$ denotes NLL evolution combined with one-loop matching.
\begin{table}[t]
  \centering
  \begin{tabular}{lccc}
    \hline
    accuracy & $\gamma^{\rm cusp}$ & $\bm\Gamma_H$ (other terms), $\gamma^\phi$, $\gamma^J$, $\gamma^c$ & matching\\
    \hline
    NLL      & 2-loop & 1-loop & tree\\
    NLL$'$   & 2-loop & 1-loop & 1-loop\\
    NNLL     & 3-loop & 2-loop & 1-loop\\
    \hline
  \end{tabular}
  \caption{Loop orders of the ingredients of eq.~\eqref{eq:kernelN} at the logarithmic accuracies considered. The
  velocity-dependent cusp anomalous dimension counts among the ``other terms''. At NLL and NLL$'$ the two-loop term of
  $\bm\Gamma_H$ is approximated by $(\gamma^{\rm cusp}_1/\gamma^{\rm cusp}_0)$ times the part of $\bm\Gamma_0$ without
  the non-cusp terms. The running coupling is the same at all orders.}
  \label{tab:orders}
\end{table}
At NNLL we use the three-loop cusp anomalous dimension~\cite{Moch:2004pa}, the two-loop hard anomalous dimension
including the velocity-dependent cusp~\cite{Korchemsky:1987wg,Kidonakis:2009ev} and the three-parton
correlation~\cite{Ferroglia:2009ep,Ferroglia:2009ii,Mitov:2009sv}, and the two-loop large-$z$ PDF anomalous
dimensions~\cite{Becher:2007ty,Ahrens:2010zv}. The two-loop jet and coft non-cusp coefficients of section~\ref{sec:jet} enter only
if $\mu_c$ or $\mu_j$ differs from $\mu_s$.

\subsection{Scale choices and uncertainties}
\label{sec:scales}

\paragraph{Hard and factorisation scales.}
We set $\mu_h=\mu_f=M$, the same value as the renormalisation and factorisation scales of the NLO calculation.

\paragraph{Soft scale.}
In Mellin space the natural soft scale is $M/\Nbar$. The relevant values of $N$ are selected by the luminosity, as
explained in section~\ref{sec:hadronic}, so the soft scale has to be fixed at the level of the hadronic cross section.
We follow the criterion of refs.~\cite{Becher:2007ty,Ahrens:2010zv}. We consider the one-loop soft correction to the
hadronic cross section, summed over partonic channels and integrated over the Born phase space. With the coupling in
front of it held fixed, it is a quadratic polynomial in $\ln(M^2/\mu_s^2)$, and we choose $\mu_s$ at its extremum. There
the correction is stationary with respect to $\mu_s$, which indicates that the soft logarithms are minimised. We
evaluate the polynomial once for each $M$ with all available Born points and keep the resulting value fixed in all
further calculations. At $\sqrt S=13.6$~TeV with $p_T^j>0.15\,M$ it gives
\begin{equation}
  \mu_s=331,\;410,\;480,\;542,\;594,\;669~\text{GeV}\qquad\text{for}\qquad M=1,\;1.5,\;2,\;2.5,\;3,\;4~\text{TeV},
  \label{eq:musvalues}
\end{equation}
i.e.\ $\mu_s\approx M/3$ to $M/6$. These values reflect the steepness of the luminosity: a steeper luminosity selects
larger $N$ and hence a lower soft scale relative to $M$.

\paragraph{Jet and coft scales.}
As discussed in section~\ref{sec:hierarchy}, we set $\mu_c=\mu_j=\mu_s$ in our final predictions. The logarithms of $R$
in the one-loop jet and coft functions are then treated at fixed order. In section~\ref{sec:dominance} we also use the
natural scales $\mu_j=p_T^jR$ and $\mu_c=\mu_{\rm lumi}R_{\rm eff}/2$. Here $\mu_{\rm lumi}$ is the effective value of
$M/\Nbar$ selected by the luminosity, determined separately for each partonic channel from the extremum of the coft
correction.

\paragraph{Uncertainty estimate.}
We vary $\mu_h$, $\mu_s$, $\mu_c$ and $\mu_j$ separately by factors of $2$ and $1/2$ around their central values. The
scale $\mu_f$ is varied by the same factors, together with the renormalisation scale of the NLO calculation and
consistently in all parts of the matched result. The upper uncertainty is the sum in quadrature of the positive
deviations, one for each scale, and the lower uncertainty that of the negative deviations. For the central choice
$\mu_c=\mu_j=\mu_s$, the variation of $\mu_s$ moves all three scales together, while the variations of $\mu_c$ and
$\mu_j$ separate them. The latter thus activate the jet and coft evolution, including the two-loop coefficients of
section~\ref{sec:jet} and, for $\mu_c<\mu_j$, the non-global factor. They probe the sensitivity to the jet-radius
logarithms that we otherwise treat at fixed order.

\subsection{Matching to NLO}
\label{sec:matching}

The resummed cross section describes the leading-power terms to all orders but misses the terms of the exact NLO
result that are power-suppressed. We combine the two with additive matching,
\begin{equation}
  \frac{d\sigma^{\rm NLO+NNLL}}{dM}=\frac{d\sigma^{\rm NLO}}{dM}
  +\left[\frac{d\sigma^{\rm NNLL}}{dM}-\frac{d\sigma^{\rm thr}}{dM}\right],
  \label{eq:matching}
\end{equation}
where $d\sigma^{\rm thr}$ is the NNLL result expanded through $\order{\as^4}$, i.e.\ eq.~\eqref{eq:thr} at $\mu=\mu_f$
inserted into eq.~\eqref{eq:hadronic}. The bracket starts at $\order{\as^5}$, so the matched result reproduces the exact
NLO cross section and adds the leading-power terms of all higher orders. The NLO cross section is computed with
\textsc{MadGraph5\_aMC@NLO}~\cite{Alwall:2014hca}, and the resummed and expanded terms with our own implementation
(section~\ref{sec:implementation}), with the same Born points, so that their statistical fluctuations are strongly
correlated in the bracket. As a check we also consider multiplicative matching,
$d\sigma^{\rm NLO}\times d\sigma^{\rm NNLL}/d\sigma^{\rm thr}$. It differs from eq.~\eqref{eq:matching} by
$(d\sigma^{\rm NLO}-d\sigma^{\rm thr})(d\sigma^{\rm NNLL}/d\sigma^{\rm thr}-1)$, a term of $\order{\as^5}$ which
involves the power-suppressed part of the NLO correction.

\subsection{Logarithms of the top-quark mass}
\label{sec:masslogs}

The hard and soft functions of section~\ref{sec:ingredients} retain the full dependence on $m_t$. If all invariants
involving a top quark are much larger than $m_t^2$, they contain large logarithms of $m_t$. For boosted top quarks such
logarithms can be resummed by a further factorisation~\cite{Mitov:2006xs,Ferroglia:2012ku}. The massive hard function
factorises into a massless hard function and, for each top quark, a hard-collinear matching coefficient $C_D$ with
natural scale $\mu_{dh}\sim m_t$. The massive soft function factorises into a massless soft function and, for each top
quark, a soft-collinear function $\widetilde s_D$ with natural scale $\mu_{ds}\sim m_t/\Nbar$. This factorisation has
been used for the $t\bar t$ invariant-mass distribution at large
$M_{t\bar t}$~\cite{Ferroglia:2012ku,Pecjak:2016nee,Czakon:2018nun}.

In our process large $M$ does not imply boosted top quarks in this sense. A large part of the cross section comes from
top-quark pairs of moderate invariant mass, recoiling against a hard jet, in particular through $qg\to q+g^*$ with
$g^*\to t\bar t$. We find that $\sqrt{s_{34}}<6m_t$ for 58\% of the leading-order cross section at $M=4$~TeV and for 81\%
at $M=2$~TeV. The factorisation of refs.~\cite{Mitov:2006xs,Ferroglia:2012ku} does not describe such configurations. We
therefore do not use it to construct a full resummation of the mass logarithms. We use it only to estimate the size of
the effect, with one global factor per value of $M$.

Equation~\eqref{eq:kernelN} evaluates $C_D$ at $\mu_h$ and $\widetilde s_D$ at $\mu_s$, at fixed order, as part of the
massive functions. To estimate the effect of moving them to their natural scales, we consider, for each top quark,
\begin{equation}
  R_t=\frac{C_D(\mu_{dh})\,\widetilde s_D(\mu_{ds})\,
  \exp\!\Big[\int_{\ln\mu_{dh}}^{\ln\mu_h}d\ln\mu\;\Gamma_{C}(\mu)+\int_{\ln\mu_{ds}}^{\ln\mu_s}d\ln\mu\;\Gamma_{s}(\mu)\Big]}
  {1+a(\mu_h)\,C_D^{(1)}(\mu_h)+a(\mu_s)\,\widetilde s_D^{(1)}(\mu_s)}\,.
  \label{eq:Rfactor}
\end{equation}
The numerator contains $C_D$ and $\widetilde s_D$ at their natural scales, evolved to $\mu_h$ and $\mu_s$ with their
anomalous dimensions $\Gamma_C$ and $\Gamma_s$. The denominator removes the one-loop terms already contained in the
massive functions, in the linear form in which they appear in eq.~\eqref{eq:kernelN}. By construction
$R_t=1+\order{\as^2}$. The one-loop functions are~\cite{Ferroglia:2012ku}
\begin{equation}
  C_D^{(1)}=C_F\Big[L_m^2+L_m+4+\frac{\pi^2}{6}\Big]\,,\qquad
  \widetilde s_D^{(1)}=C_F\Big[-L_D^2-2L_D-\frac{5\pi^2}{6}\Big]\,,
  \label{eq:CDsD}
\end{equation}
with $L_m=\ln(\mu^2/m_t^2)$ and $L_D=\ln\big(m_t^2/(\Nbar^2\mu^2)\big)$. We use the two-loop non-cusp anomalous
dimensions of ref.~\cite{Ferroglia:2012ku} and set $\mu_{dh}=m_t$, $\mu_{ds}=m_t\mu_s/M$, which corresponds to the
identification $\Nbar=M/\mu_s$ in $\widetilde s_D$.

With this identification and $\mu_h=M$, the logarithms in eq.~\eqref{eq:CDsD} satisfy $L_D=-L_m=\ln(m_t^2/M^2)$. If the
two couplings were equal, the double logarithms would cancel between $C_D^{(1)}$ and $\widetilde s_D^{(1)}$, leaving
$C_F[3\ln(M^2/m_t^2)+4-2\pi^2/3]$, with $3C_F=\gamma^\phi_{q,0}$. The numerical result for the factor is
$R_t^2=1.003$, 1.006, 1.007, 1.006, 1.004 and 0.998 at $M=1$, 1.5, 2, 2.5, 3 and 4~TeV. This small value is a numerical
finding: the couplings at $\mu_h$ and $\mu_s$ differ, single logarithms remain, and these effects compensate only in
the complete expression~\eqref{eq:Rfactor}. Varying $\mu_{dh}$ and $\mu_{ds}$ simultaneously by factors of 2 changes
$R_t^2$ by about $-7\%$ and $+2\%$. The largest change arises from lowering $\mu_{ds}$. We conclude that there is no
indication of a large effect of the mass logarithms on $d\sigma/d\Mttj$ in this kinematic regime. This statement concerns the logarithms contained in the leading-power kernel. Terms outside it do depend on $M/m_t$: section~\ref{sec:offdiag} shows that the difference between the threshold and exact $K$ factors grows with this ratio at fixed $\tau$.

\paragraph{Comparison with $t\bar t$ production.}
This conclusion does not contradict the use of this factorisation for $t\bar t$ production. In
refs.~\cite{Pecjak:2016nee,Czakon:2018nun} the simultaneous resummation of soft and small-mass logarithms is applied at
large $M_{t\bar t}$ in the boosted-soft limit, $\hat s,|t_1|,|u_1|\gg m_t^2$, and matched to the soft resummation with the
exact mass dependence and to the fixed-order result. These analyses combine the two types of logarithms in one prediction
and do not isolate the effect of the mass logarithms alone, so we cannot quote a number to compare with
$R_t^2$. The kinematic difference, however, is plain. In $t\bar t$ production at large $M_{t\bar t}$ both top quarks
have energies of the order of $M_{t\bar t}/2$ in the partonic centre-of-mass frame. In our case the available energy is shared
with the jet, and much of the cross section has a light pair. Weighted with the leading-order cross section, the average of the
smaller of the two top-quark energies is $1.6\,m_t$, $2.5\,m_t$ and $4.3\,m_t$ at $M=1$, 2 and 4~TeV, and it is below
$2m_t$ for 88\%, 30\% and 8\% of the cross section. Thus at least one top quark is not strongly boosted for a large part of the
cross section, even at $M=4$~TeV. A second difference is the choice of the hard scale. Ref.~\cite{Czakon:2018nun} shows that
in the boosted regime the relevant hard scale of the $M_{t\bar t}$ distribution is $H_T$ rather than $M_{t\bar t}$, as the
$t$- and $u$-channel diagrams enhance forward and backward scattering. We have not investigated this for our process. Our
global factor $R_t^2$ remains an estimate, and a full treatment of the mass logarithms at high $M$, in which a separate
scale is assigned to each configuration, would require a dedicated study.

%% file: sections/implementation.tex
\section{Numerical implementation and validation}
\label{sec:implementation}

In this section we describe how the resummed and fixed-order cross sections are evaluated, and the checks we
performed on each component. The resummed and threshold-expanded cross sections are computed with a dedicated
\textsc{Python} implementation, which relies on \textsc{SciPy}~\cite{Virtanen:2019joe} for the quasi-random numbers
and the integration of differential equations. The exact NLO cross section is computed with
\textsc{MadGraph5\_aMC@NLO}~\cite{Alwall:2014hca}.

\subsection{Set-up}
\label{sec:setup}

Unless stated otherwise we consider proton--proton collisions at $\sqrt S=13.6$~TeV with $m_t=172.5$~GeV and
the NNPDF3.1 NNLO parton distributions with $\as(m_Z)=0.118$~\cite{NNPDF:2017mvq}, accessed through
\textsc{LHAPDF}~\cite{Buckley:2014ana} (set number 303600). The top quark is treated as stable, with vanishing width,
and five massless quark flavours are active. Jets are reconstructed from all light partons with the anti-$k_T$
algorithm~\cite{Cacciari:2008gp} with $R=0.4$, as implemented in \textsc{FastJet}~\cite{Cacciari:2011ma}, without
rapidity restrictions. We require at least one jet with $p_T^j>0.15\,M$ and consider $M=1$--4~TeV.

The NLO cross section is a histogram: we evaluate it in windows $M\pm10$~GeV ($M\pm40$~GeV at 4~TeV) and divide by the
width of the window. The resummed and threshold-expanded cross sections are instead computed at the centre of the
window. For a spectrum falling like $M^{-n}$ the two differ by a relative amount
$n(n+1)(\Delta M/M)^2/6$, which is at most about 0.2\% at 4~TeV and negligible at the other masses. Since only the
difference of the resummed and expanded results enters the matching~\eqref{eq:matching}, and that difference is
itself a correction of order 20\% or less, the effect on our predictions can be neglected. In
section~\ref{sec:dominance} we also use lower collider energies, lower masses and a fixed jet threshold of 50~GeV.

\subsection{Born phase space and Monte Carlo integration}
\label{sec:born}

The resummed kernel~\eqref{eq:kernelN} depends on the Born configuration $\Phi_3$, so the hadronic cross section
requires an integral over the three-body phase space at fixed $\sqrt{\hat s}=M$. By rotational invariance around the
beam axis this integral is four-dimensional. We parametrise it by the invariant mass $s_{34}$ of the top-quark pair,
the polar angle $\theta_5$ of the jet in the partonic centre-of-mass frame, and the polar and azimuthal angles of the
top quark in the rest frame of the pair. The jet energy follows from $s_{34}=M^2-2ME_5$. At this order the jet is a
single massless parton, so that the requirement on $p_T^j$ is a constraint on $E_5\sin\theta_5$ alone.

The integration points are drawn from scrambled Sobol sequences. A Sobol sequence~\cite{Sobol:1967} is a deterministic set of points
constructed to fill the unit hypercube more evenly than random points do. For smooth integrands the error then
decreases faster with the number of points $n$ than the $n^{-1/2}$ of plain Monte Carlo integration. Applying a random
scrambling~\cite{Owen:1995,Owen:1997,Owen:1998} preserves this property and, in addition, makes different sequences statistically independent, which
allows for an unbiased error estimate as described below. We use the implementation in \textsc{SciPy}~\cite{Virtanen:2019joe,Roy:2023}, with the direction numbers of ref.~\cite{Joe:2008} and linear matrix scrambling with a digital shift~\cite{Matousek:1998}.

At high $M$ the integrand is far from uniform, and its largest values must be sampled efficiently. In the $qg$ channel
the process $qg\to q+g^*$, followed by the splitting $g^*\to t\bar t$, produces a rise like $1/s_{34}$ towards the
threshold $s_{34}=4m_t^2$, while soft jets produce a rise like $1/E_5$. In importance sampling, one generates points
from a density $p(x)$ that follows the shape of the integrand $f(x)$, and averages the weights $f/p$; these are then
nearly constant, which reduces the variance. To describe both enhancements we use a multi-channel density in the
sense of ref.~\cite{Kleiss:1994qy}, the equal mixture
\begin{equation}
  p(s_{34})=\frac12\,\frac{1}{2ME_5\,\ln(E_5^{\rm max}/E_5^{\rm min})}
  +\frac12\,\frac{1}{s_{34}\,\ln(s_{34}^{\rm max}/s_{34}^{\rm min})}\,,
  \label{eq:multichannel}
\end{equation}
where the first term is uniform in $\ln E_5$, with $E_5^{\rm min}=p_{T,\rm min}^j$, and the second is uniform in $\ln s_{34}$, with
$s_{34}^{\rm min}=4m_t^2$. A fifth quasi-random variable selects the term from which $s_{34}$ is generated, while the weight always contains the full density $p(s_{34})$. The
jet direction is sampled uniformly in the variable $y_5={\rm artanh}\cos\theta_5$, within the range allowed by
the jet requirement, and the polar angle of the top quark uniformly in $y_3={\rm artanh}(\beta\cos\theta_3)$ with
$\beta=(1-4m_t^2/s_{34})^{1/2}$. These substitutions flatten the peaks of the matrix elements at small angles to the
beams and at small angles between the top quark and the axis of the pair. Even so the distribution of weights is not
flat: at $M=4$~TeV the 1\% of points with the largest weights carry 31\% of the leading-order cross section, compared
with 7\% at 1~TeV and 17\% at 2~TeV.

For every value of $M$ we use 80 independently scrambled sequences of 256 points for each of the partonic channels
$q\bar q$, $qg$ and $gg$, including the sub-channels with interchanged beams and with antiquarks. This amounts to
$61\,440$ Born points for each value of $M$. With $X_i$ the estimate of a quantity from the sequence $i$, the central
value is the mean of the $X_i$ and its statistical uncertainty is
\begin{equation}
  \Delta X=\Bigl[\frac{1}{80\cdot79}\sum_{i=1}^{80}(X_i-\bar X)^2\Bigr]^{1/2}\,.
  \label{eq:statistics}
\end{equation}
All scale variations, logarithmic orders and the threshold-expanded cross section are evaluated with the same Born
points, so that their fluctuations are strongly correlated. The difference that enters the
matching~\eqref{eq:matching} is therefore determined considerably more precisely than the individual cross sections:
for the NNLL result the relative uncertainty is 0.13\% at 1~TeV and 1.4\% at 4~TeV, whereas the uncertainty on
$d\sigma^{\rm NNLL}-d\sigma^{\rm thr}$, normalised to the same NNLL cross section, is 0.02\% and 0.4\%.

\subsection{Hard function}
\label{sec:hard-num}

For each Born point and helicity configuration, the tree-level and one-loop amplitudes are computed with
\textsc{Recola}~\cite{Actis:2012qn,Actis:2016mpe} and \textsc{Collier}~\cite{Denner:2016kdg} for all colour flows
and projected onto the orthogonal bases of section~\ref{sec:colour}. We performed the following checks.

\begin{itemize}
\item \emph{Tree level.} The tree-level hard functions agree with an independent implementation of the tree-level
  amplitudes.

\item \emph{Normalisation of the poles.} The finite part of a one-loop amplitude depends on the factor taken out in
  front of the $\epsilon$ expansion, as explained in section~\ref{sec:hard}. We determined the convention of
  \textsc{Recola} from the one-loop massless quark form factor, whose result is known in closed form.

\item \emph{Infrared poles.} The coefficients of $\epsilon^{-2}$ and $\epsilon^{-1}$ of the one-loop amplitude are
  predicted by eq.~\eqref{eq:Z}. This check involves the full anomalous dimension, including its imaginary part, and
  therefore the corrections of section~\ref{sec:hard} to the anomalous dimension of ref.~\cite{Chargeishvili:2022ngl}.
  We evaluated it point by point in production. On a dedicated sample of 768 Born points for each of 8 sequences at
  high mass, the relative deviation per helicity configuration has a median of about $10^{-12}$, and is below
  $3.3\times10^{-6}$ for 99\% of the configurations. It reaches $2.2\times10^{-2}$ at rare points for $\epsilon^{-1}$
  and $2.7\times10^{-4}$ for $\epsilon^{-2}$. These largest deviations occur for individual helicity configurations at
  4~TeV. After summing over colour and helicities with the weights that enter the cross section, the relative
  deviation does not exceed $4.7\times10^{-7}$ for $\epsilon^{-1}$ and $9.5\times10^{-8}$ for $\epsilon^{-2}$, which shows
  that these outliers are irrelevant for the physical result. As a safeguard the code stops when a deviation of a single helicity configuration exceeds
  a tolerance. The tolerance is $10^{-2}$ by default; for the production at 4~TeV, in which the isolated points
  mentioned above occur, and for the later runs at other top-quark masses and jet thresholds, it is $5\times10^{-2}$. In production only this safeguard applies: the weighted deviation quoted above was evaluated on the dedicated sample, and not on the production points themselves.

\item \emph{Scale dependence.} The dependence of $\bm H^{(1)}$ on the renormalisation scale, obtained by running
  \textsc{Recola} at different values of $\mu$, agrees with the prediction of eq.~\eqref{eq:RGEH}.

\item \emph{Comparison with \textsc{MadLoop}.} We compared the squared amplitudes summed over colour and helicities
  with \textsc{MadLoop}~\cite{Hirschi:2011pa}, the one-loop provider of our NLO calculation, as a standalone program
  with $\Gamma_t=0$. The comparison uses 108 Born points of our production samples, for all three partonic channels at
  $M=1$, 2 and 4~TeV. In each case we selected the points with the largest weights, random points, and those with the
  largest pole deviation. With the prefactor conventions of the two programs, which differ by
  $\Gamma(1+\epsilon)$ against $1/\Gamma(1-\epsilon)$, the finite parts are related by
  \begin{equation}
    c_0^{\textsc{Recola}}=c_0^{\textsc{MadLoop}}-\frac{\pi^2}{12}\,c_{-2}^{\textsc{MadLoop}}\,,
    \label{eq:mlconversion}
  \end{equation}
  where $c_{-k}$ is the coefficient of $\epsilon^{-k}$ in units of $(\as/2\pi)$ times the Born cross section.
  The results are given in table~\ref{tab:madloop}. The ratio of the tree-level matrix elements is constant to
  $10^{-13}$. The finite parts agree with a median absolute difference of about $10^{-12}$. The largest difference,
  $2.4\times10^{-4}$, or $1.5\times10^{-5}$ relative to the finite part, occurs at one of the points at 4~TeV selected for their large
  pole deviation, at which \textsc{MadLoop} itself estimates its accuracy to be $6\times10^{-6}$. This is a stringent test of
  the common origin of the hard function and of the virtual correction in the fixed-order cross section, which is
  essential for the matching~\eqref{eq:matching}. It concerns the squared amplitudes summed over colour and does not test the individual entries of the hard matrix, whose colour structure is tested by the pole check.
\end{itemize}

\begin{table}[t]
  \centering
  \begin{tabular}{cccc}
    \hline
    channel & $\max|\Delta c_{-2}|$ & $\max|\Delta c_{-1}|$ & $\max|\Delta c_0|$\\
    \hline
    $q\bar q$ & $2.2\times10^{-6}$ & $1.0\times10^{-5}$ & $5.4\times10^{-5}$\\
    $gg$      & $1.1\times10^{-5}$ & $4.7\times10^{-6}$ & $2.4\times10^{-4}$\\
    $qg$      & $1.5\times10^{-9}$ & $3.1\times10^{-8}$ & $2.4\times10^{-8}$\\
    \hline
  \end{tabular}
  \caption{Largest absolute difference between \textsc{Recola} and \textsc{MadLoop} for the coefficients of the
  one-loop amplitude interfered with the Born amplitude, summed over colour and helicities, in units of $\as/2\pi$
  times the Born cross section, at 108 Born points. The finite parts are compared using
  eq.~\eqref{eq:mlconversion}.}
  \label{tab:madloop}
\end{table}

The hard function is computed at $\mu=M$. For other values of $\mu_h$ the one-loop matrix is obtained from
eq.~\eqref{eq:RGEH}.

\subsection{Soft function}
\label{sec:soft-num}

The angular integrals of the one-loop soft function, eq.~\eqref{eq:s1}, are evaluated on a product grid in the polar
and azimuthal angles of the emitted gluon and in the angle that parametrises the additional dimensions, with
$80\times48\times24$ nodes. A change of variables concentrates the polar nodes near the axis of each dipole term.
This is necessary because, for boosted top quarks, the integrand is sharply peaked around their directions, within
angles of order $m_t/E_t$.

We tested the convergence of the grid by increasing the number of nodes by factors of three and six in each direction,
at 28 Born points of the $qg$ and $gg$ channels at $M=2$ and 4~TeV, namely the four points with the largest weights
and three random points per channel and mass. These include top-quark energies up to $E_t/m_t=11.3$, close to the
kinematic limit $M/(2m_t)\approx11.6$ at 4~TeV. The change in the one-loop soft function does not exceed $1.2\times10^{-4}$
in units of its tree-level value. We did not repeat the test for the $q\bar q$ channel, which contains the same
dipole integrals and contributes at most 3\% of the cross section. We also verified that the poles of the bare soft
function cancel against those predicted by the soft anomalous dimension, and that the full one-loop product of hard,
soft, jet and coft functions and PDFs is independent of $\mu$ at $\order{\as}$, as required by
renormalisation-group consistency.

\subsection{Resummed and expanded cross sections}
\label{sec:res-num}

The evolution matrix $\bm U$ of eq.~\eqref{eq:U} is obtained by solving its differential equation numerically with
an eighth-order Runge--Kutta method, with relative and absolute tolerances of $10^{-10}$ and $10^{-12}$. The scalar
evolution integrals, section~\ref{sec:rge-solution}, are computed with 64-point Gauss--Legendre quadrature in
$\ln\mu$. The running coupling is taken from the PDF set.

The hadronic convolution in $z$ is performed as in eq.~\eqref{eq:subtraction}. We use a Gauss--Legendre rule with 160
nodes in a variable $s$ defined by $1-z=(1-\tau)s^5$, which concentrates the nodes near $z=1$. The subtraction
terms at $z=1$ are determined from the luminosity and from its derivative with respect to $z$, computed by a central
difference with step $10^{-4}$. Very close to $z=1$ the difference in the subtracted integrand suffers from numerical
cancellations. We therefore freeze the smooth function that multiplies $(1-z)^{2\eta+1}$ at its value at $1-z=2\times10^{-3}$
for smaller $1-z$. The derivatives with respect to $\eta$ in eq.~\eqref{eq:kernelN} are computed by five-point
finite differences with step $2\times10^{-3}$. Two checks of this part of the calculation are essential.

\begin{itemize}
\item \emph{Expansion of the resummed formula.} In the limit $\as\to0$ the resummed kernel with widely separated
  scales, expanded to $\order{\as}$, must reproduce the fixed-order threshold kernel at the factorisation scale. We
  test this by rescaling $\as$ and extrapolating to zero (the one-loop hard coefficient being set to zero in this test,
  since it is independent of the evolution). The relative deviation at complex values of $N$ is below
  $5\times10^{-6}$ at one random phase-space point in each of the three channels, at $M=700$~GeV and at 4~TeV. This tests the formula and its implementation, and is not a scan of the phase space.

\item \emph{Comparison at lowest order.} Our leading-order cross section agrees with that of
  \textsc{MadGraph5\_aMC@NLO} with the same cuts. The ratios of the two at $M=1$, 1.5, 2, 2.5, 3 and 4~TeV are
  1.0005(18), 0.9952(32), 1.0002(43), 0.9963(59), 0.9903(70) and 0.9841(99), where the uncertainties combine the
  statistical uncertainties of both calculations. The values at high mass lie somewhat below one, in the direction expected from the large weights discussed above, so that the uncertainty at 4~TeV may be somewhat underestimated. The fractions of the three partonic channels, compared at 1, 2 and 4~TeV, agree to within a few per mille. For instance, the fractions at 4~TeV are 0.013, 0.188 and 0.799 for $q\bar q$, $gg$
  and $qg$, against 0.014, 0.185 and 0.801.
\end{itemize}

For the modifications of the kernel at small $N$ in section~\ref{sec:damping}, the inverse Mellin transform with
respect to $z$ is computed numerically on a contour in the complex $N$ plane. Tests of this procedure with the
closed-form expression~\eqref{eq:kernelz} yield a relative precision of $2\times10^{-8}$ or better, as detailed in
appendix~\ref{app:mellin}.

\subsection{Fixed-order NLO cross section}
\label{sec:nlo-num}

The NLO cross section is computed with \textsc{MadGraph5\_aMC@NLO} version 3.8.1~\cite{Alwall:2014hca} with the
one-loop Standard Model with massless bottom quarks. It evaluates the one-loop amplitudes with
\textsc{MadLoop}~\cite{Hirschi:2011pa} and subtracts the infrared divergences of the real emission with the FKS
method~\cite{Frixione:1995ms,Frederix:2009yq}. In this method, the real-emission matrix element is combined with
counter-events, which share the kinematics of a Born configuration and cancel its singular limits. We require a
relative accuracy of 0.5\% for the integrated cross section in the generation window. At $M=3$~TeV we average two independent runs with a required accuracy of 1\%, which agree with each other within 0.5\%. We use the fixed scales
$\mu_R=\mu_F=M$ in each window and estimate the scale uncertainty from the seven-point variation of $\mu_R$ and $\mu_F$
by factors of 2 around $M$, excluding opposite variations. PDF uncertainties are obtained by reweighting to the 100
replicas of the NNPDF3.1 set. The jet and $\Mttj$-window requirements are implemented as user cuts.

Generating events only with $\hat s$ near the window would improve the efficiency considerably, and earlier runs of
our calculation restricted the Born $\sqrt{\hat s}$ to values above $0.75\,(M-\Delta M)$. Such a restriction is not
harmless, however. Although $\hat s\ge M^2$ holds for every physical real-emission event in the window, the
restriction acts on the partonic energy of the underlying Born configuration, which may be smaller than that of the
real-emission event. We therefore lifted the restriction in all final results, keeping only a generation bound
$\sqrt{\hat s}>0.01\,(M-\Delta M)$ as a technical cut. To quantify the effect, table~\ref{tab:taucut} compares
the runs with and without the restriction.

\begin{table}[t]
  \centering
  \begin{tabular}{ccc}
    \hline
    $M$ [TeV] & earlier run & no restriction / earlier run\\
    \hline
    1   & restricted & 1.016(8)\\
    1.5 & not restricted & 1.001(8)\\
    2   & restricted & 1.027(5)\\
    2.5 & not restricted & 0.992(11)\\
    3   & restricted & 1.033(8)\\
    4   & restricted & 1.026(10)\\
    \hline
  \end{tabular}
  \caption{Ratio of the NLO cross section without any restriction on the Born partonic energy to the one from an earlier run, in the window
  around each value of $M$. The earlier runs at 1.5 and 2.5~TeV did not have the restriction and serve as a control of the statistical
  uncertainties, given in brackets.}
  \label{tab:taucut}
\end{table}

The restriction thus removes about 2--3\% of the cross section at the masses at which it was applied, and has no visible effect in the
controls. At $M=4$~TeV the same comparison in the runs with accuracy of 1\% that also contain $M_{t\bar tX}$ gives 1.036 for $\Mttj$, compatible with the table, 1.032 for events with one jet and 1.092 for the invariant mass $M_{t\bar tX}$ of section~\ref{sec:candidates}, which suggests that the restriction mainly affects events with hard additional radiation, although we have not studied the mechanism in detail. Details of the set-up are collected in appendix~\ref{app:mg5}.

For the studies in section~\ref{sec:lp-validity} we additionally record the distribution in $z=M^2/\hat s$ in each
window. Here $\hat s$ is computed from the momenta of the incoming partons, and for counter-events these are the momenta
of the corresponding Born configuration. Because the real-emission events and their counter-events have different
values of $z$, the content of an individual bin near $z=1$ is sensitive to the details of the subtraction. We
quantify this effect on the Drell--Yan process in section~\ref{sec:lp-validity}.

%% file: sections/threshold_dominance.tex
\section{Dominance of the threshold region}
\label{sec:dominance}

Threshold resummation improves a fixed-order prediction only if the partonic cross section is dominated by
configurations close to threshold, and if the factorisation theorem~\eqref{eq:factorisation} applies to them. In
this section we test the first condition, and the part of the second that concerns the jet, by comparing the
leading-power approximation with the exact NLO correction. We first explain the test, then apply it to moderate
invariant masses at $\sqrt S=13$~TeV and to lower collider energies, and finally to the high-mass region that is the
focus of our predictions.

\subsection{A diagnostic: the threshold approximation at NLO}
\label{sec:diagnostic}

The NLO correction to $d\sigma/dM$ consists of terms that are singular at $z\to1$, the logarithmically enhanced
plus distributions and the $\delta(1-z)$ terms, and of terms that remain finite or vanish as $z\to1$. The first
class is reproduced exactly by the expansion~\eqref{eq:thr} of the resummed formula; the second class is not. We
define the relative NLO correction and its leading-power approximation as
\begin{equation}
  \delta_{\rm NLO}=\frac{d\sigma^{\rm NLO}/dM}{d\sigma^{\rm LO}/dM}-1\,,\qquad
  \delta_{\rm thr}=\frac{d\sigma^{\rm thr}/dM}{d\sigma^{\rm LO}/dM}-1\,,
  \label{eq:deltas}
\end{equation}
where $d\sigma^{\rm thr}$ is eq.~\eqref{eq:hadronic} evaluated with the kernel~\eqref{eq:thr}, and all quantities
are computed at $\mu=M$. If the threshold region dominates, the ratio $\delta_{\rm thr}/\delta_{\rm NLO}$ is close
to one. Resummation then exponentiates terms that constitute the bulk of the correction, and its predictions for
higher orders are expected to be reliable. If the ratio differs substantially from one, terms suppressed by powers
of $1-z$ are as important as the threshold logarithms. Questions of this kind have been studied for threshold
resummation in colour-singlet production~\cite{Becher:2007ty,Bonvini:2010tp}. It is a necessary
but not a sufficient condition: a ratio close to one could also arise from a cancellation between regions of $z$
in which the approximation fails. We therefore also compare the distributions in $z$ in
section~\ref{sec:lp-validity}.

A useful measure of the typical distance from threshold is the effective value of $\Nbar$ selected by the
luminosity. One such measure is the position of the saddle point of the Mellin inversion integral~\cite{Bonvini:2010tp}. Here we use a definition adapted to the form of our kernel,
\begin{equation}
  \ln\Nbar_{\rm eff}=-\frac{1}{2}\,\frac{F'(0)}{F(0)}\,,\qquad
  F(\eta)=\int_\tau^1\frac{dz}{z}\,\mathcal L\!\left(\frac{\tau}{z}\right)\frac{e^{-2\gamma_E\eta}}{\Gamma(2\eta)}\,
  z^{-\eta}(1-z)^{2\eta-1}\,,
  \label{eq:Neff}
\end{equation}
for each partonic channel. Because $F(\eta)$ is the convolution of the luminosity with the momentum-space form
of $\Nbar^{-2\eta}$, eq.~\eqref{eq:kernelz}, $\ln\Nbar_{\rm eff}$ is formally the average of $\ln\Nbar$ weighted by the
luminosity. Equivalently, $M/\Nbar_{\rm eff}$ is the scale at which the double logarithm of the one-loop soft correction is stationary, in analogy with the criterion for the soft scale in refs.~\cite{Becher:2007ty,Ahrens:2010zv}. This is the
scale $\mu_{\rm lumi}$ introduced in section~\ref{sec:scales}. The threshold logarithms are large only if
$\Nbar_{\rm eff}\gg1$.

\subsection{Moderate invariant masses and the collider energy}
\label{sec:lowmass}

\begin{figure}[t]
  \centering
  \includegraphics[width=0.72\textwidth]{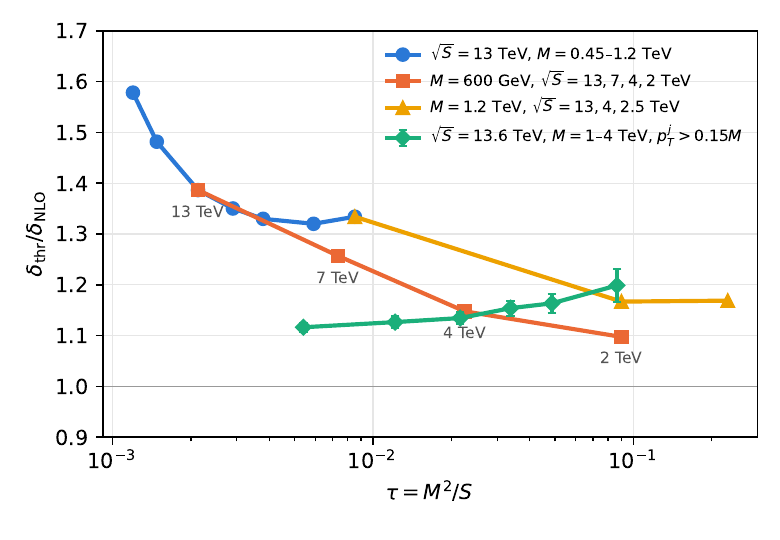}
  \caption{Ratio of the leading-power approximation to the exact NLO correction, eq.~\eqref{eq:deltas}, as a
  function of $\tau=M^2/S$, for $\mu=M$ and $R=0.4$. Blue circles: $\sqrt S=13$~TeV, $M=0.45$--1.2~TeV,
  $p_T^j>50$~GeV. Orange squares and yellow triangles: $M=600$~GeV and $M=1.2$~TeV at several collider energies (as
  labelled for the squares), $p_T^j>50$~GeV. Green diamonds: $\sqrt S=13.6$~TeV, $M=1$--4~TeV, $p_T^j>0.15\,M$,
  with statistical uncertainties. For the other points the statistical uncertainties are comparable to the symbol
  size or smaller.}
  \label{fig:dominance}
\end{figure}

We first consider $M=0.45$--1.2~TeV at $\sqrt S=13$~TeV with a fixed jet threshold $p_T^j>50$~GeV. This is the
region where most events are recorded and where the sensitivity to the top-quark mass is
largest~\cite{Alioli:2013mxa}. Figure~\ref{fig:dominance} shows $\delta_{\rm thr}/\delta_{\rm NLO}$ as a function of
$\tau=M^2/S$. At $\sqrt S=13$~TeV the leading-power approximation overestimates the NLO correction by 58\% at
$M=450$~GeV and by 32--33\% at $M=1$--1.2~TeV. The reason becomes clear from $\Nbar_{\rm eff}$. At
$M=600$~GeV we find $\Nbar_{\rm eff}=1.0$, 1.8 and 1.3 for the $q\bar q$, $gg$ and $qg$ channels. These values
are not large: the luminosity falls too slowly with $\tau/z$ to suppress the region away from threshold.

To confirm this interpretation we keep $M=600$~GeV fixed and lower the collider energy, which increases $\tau$
and makes the luminosity steeper. The ratio decreases steadily, from 1.39 at 13~TeV to 1.26, 1.15 and 1.10 at 7,
4 and 2~TeV. At $\sqrt S=2$~TeV we find $\Nbar_{\rm eff}=7.4$, 6.6 and 5.2 for the three channels. For $M=1.2$~TeV
the ratio falls from 1.33 at 13~TeV to 1.17 at 4~TeV, and does not decrease further at 2.5~TeV ($\tau=0.23$). The ratio is therefore not a function of $\tau$ alone: at $\tau\approx0.09$ it is 1.10 for $M=600$~GeV and 1.17 for $M=1.2$~TeV. The steepening of the luminosity accounts for the decrease of the ratio, but only at values of $\tau$ well above those relevant for $M\lesssim1$~TeV at the LHC, and it does not by itself bring the ratio to one. This is in line with the Drell--Yan analysis of ref.~\cite{Becher:2007ty}, in which the dynamical enhancement
remains effective down to $\tau\approx0.2$ and weakens at much smaller $\tau$. Our range $M=1$--4~TeV at $\sqrt S=13.6$~TeV
corresponds to $\tau=0.005$--0.09. That analysis also finds that, even far from threshold, the terms leading at $z\to1$ dominate the
Drell--Yan cross section, and suggests, assuming that this holds for other processes, that they provide useful approximations to
higher orders. The comparison of this section tests this assumption for $t\bar tj$.

\subsection{The jet and the inverted scale hierarchy}
\label{sec:jet-hierarchy}

A second, independent problem arises from the jet. Condition~\eqref{eq:jetcondition} requires the soft radiation
outside the jet to be much softer than the jet, i.e.\ $\mu_c\ll\mu_j$. With the natural scales of
section~\ref{sec:scales} the ratio of the two scales is
\begin{equation}
  \frac{\mu_c}{\mu_j}=\frac{\mu_{\rm lumi}R_{\rm eff}/2}{E_JR_{\rm eff}}=\frac{M}{2E_J\,\Nbar_{\rm eff}}\,,
  \label{eq:ratio}
\end{equation}
independent of $R$. For a fixed jet threshold of 50~GeV and $M=600$~GeV the jet typically carries only
15--23\% of the available energy $M$, as the leading-order matrix elements favour soft jets. Together with
$\Nbar_{\rm eff}\approx1$--2, eq.~\eqref{eq:ratio} gives $\mu_c/\mu_j\approx1.4$--2.6. The hierarchy assumed by the
factorisation theorem is inverted. Evolving the jet and coft functions between these scales with their RGEs then
generates large factors that have no physical basis: the evolution resums logarithms of $\mu_j/\mu_c$ that would
be large only if the hierarchy were the opposite one. At $M=600$~GeV and $\sqrt S=13$~TeV, the matched NLL$'$ result
with the natural jet and coft scales exceeds the NLO cross section by a factor 1.52, compared with 1.28 when
$\mu_c=\mu_j=\mu_s$. The difference is largely due to this evolution, although changing the scales also changes the logarithms in the fixed-order functions.

\subsection{The high-mass region}
\label{sec:highmass}

Both problems are reduced at large invariant masses with a jet threshold that grows with $M$. We therefore focus on
$M=1$--4~TeV at $\sqrt S=13.6$~TeV with $p_T^j>0.15\,M$. With this requirement the jet carries on average 31--35\%
of $M$ at $M=1$~TeV and 37--45\% at 4~TeV, depending on the channel.

The value 0.15 is a compromise between rate and hard jet, and it does not follow from a calculation. Its order of magnitude is set by
eq.~\eqref{eq:jetcondition}: with $1-z\sim1/\Nbar$ and $\Nbar\approx M/\mu_s$, the condition requires $E_J$ to be well above
$M/(2\Nbar)\approx0.17\,M$, $0.12\,M$ and $0.08\,M$ at $M=1$, 2 and 4~TeV. A higher threshold would improve this only marginally, because the
matrix elements already favour an energetic jet once the threshold is imposed, while it would reduce the rate quickly: $p_T^j>0.2\,M$ keeps 62--63\% of the leading-order cross section at $p_T^j>0.15\,M$ at all three masses, and $p_T^j>0.3\,M$ only 21--24\%, whereas the average of $\mu_c/\mu_j$ changes by less than 7\% and 17\%, respectively. The dependence of our results on this choice is discussed in section~\ref{sec:threshdep}.

Table~\ref{tab:hierarchy} gives the corresponding scale ratios, with the value $\Nbar_{\rm eff}$ of each channel. For $M\ge2$~TeV radiation at the
scale selected by the luminosity is softer than the jet for essentially the whole cross section. At $M=1$~TeV the condition is
marginal: $\mu_c>\mu_j$ for about 16\% of the leading-order cross section, namely 21\% of the $qg$ channel and 10\% of the $gg$ channel, and for most configurations in the
$q\bar q$ channel, which however contributes only 2.7\% of the cross section.

These ratios must be read with care. Since we set $\mu_c=\mu_j=\mu_s$ in our final predictions, no evolution takes place between the two scales at central values, and the problem of section~\ref{sec:jet-hierarchy} does not arise there. What the ratios measure is whether eq.~\eqref{eq:jetcondition} holds, and hence whether the leading-power expansion can be expected to apply, which we test directly in section~\ref{sec:lp-validity}. They do not imply that the logarithms of the jet radius are under control. These stay at fixed order and do not become small for a hard jet: at $\mu_c=\mu_s$ the argument $\ell$ of the coft function, eq.~\eqref{eq:coft}, is close to $\ln(R_{\rm eff}^2/4)\approx-3.2$ at all masses, so that the double logarithm alone contributes about $10\,C_J\,\as/(4\pi)$. At $\order{\as}$ these terms are replaced by the exact result through the matching. At higher orders they are a limitation of our result. Section~\ref{sec:shower} shows that the sensitivity to the jet radius is indeed sizeable. Only the variations of $\mu_c$ and $\mu_j$ in our uncertainty estimate activate the evolution between the two scales.

\begin{table}[t]
  \centering
  \begin{tabular}{ccccc}
    \hline
    $M$ [TeV] & $\mu_s$ [GeV] & $q\bar q$ & $gg$ & $qg$\\
    \hline
    1 & 331 & 1.29 (71\%) & 0.69 (10\%) & 0.88 (21\%)\\
    2 & 480 & 0.61 (8\%)  & 0.36 (0\%)  & 0.45 (1\%)\\
    4 & 669 & 0.20 (0\%)  & 0.18 (0\%)  & 0.21 (0\%)\\
    \hline
  \end{tabular}
  \caption{Soft scale and Born-weighted average of the ratio $\mu_c/\mu_j$ of eq.~\eqref{eq:ratio} for each
  partonic channel at $\sqrt S=13.6$~TeV with $p_T^j>0.15\,M$. In parentheses: fraction of the leading-order cross
  section of the channel with $\mu_c>\mu_j$.}
  \label{tab:hierarchy}
\end{table}

The green diamonds in figure~\ref{fig:dominance} show $\delta_{\rm thr}/\delta_{\rm NLO}$ in this region. The ratio is
1.116(9), 1.127(11), 1.135(11), 1.154(15), 1.164(18) and 1.199(32) at $M=1$, 1.5, 2, 2.5, 3 and 4~TeV. This is closer to one than at moderate masses at 13~TeV. Nevertheless, the ratio does not approach one as $M$ increases, and
even rises slowly, although $\tau$ increases by a factor of 16 over this range and the comparison at fixed $M=600$~GeV suggests that
the approximation should improve. Section~\ref{sec:offdiag} shows the reason: at fixed $\tau$ the difference between the threshold and exact $K$ factors grows with $M/m_t$, so that the comparison at fixed $M$ does not carry over to the comparison at fixed $\sqrt S$. As explained in section~\ref{sec:diagnostic}, a ratio close to one is a
necessary condition only. In section~\ref{sec:lp-validity} we find that it can arise from a cancellation between different regions of
$z$, so that this ratio should not be used on its own to rate the approximation.

%% file: sections/leading_power_validity.tex
\section{Validity of the leading-power approximation at high mass}
\label{sec:lp-validity}

In section~\ref{sec:highmass} we found that the threshold expansion overestimates the exact NLO correction even at
large invariant masses, where the luminosity is steep. A resummation that exponentiates this expansion can only be
trusted if we understand the origin of the difference. In this section we therefore ask which terms the leading-power
approximation misses. We first test three physically motivated candidates, then determine directly where in $z$ the
difference arises, and finally validate the method of this comparison on Drell--Yan production, for which the exact
result is known analytically.

\subsection{The size of the difference}
\label{sec:overshoot}

We work at $\order{\as}$ with all scales equal to $M$, and measure the difference in units of the leading-order cross
section,
\begin{equation}
  \Delta\equiv\delta_{\rm thr}-\delta_{\rm NLO}\,,
  \label{eq:Delta}
\end{equation}
with $\delta_{\rm thr}$ and $\delta_{\rm NLO}$ from eq.~\eqref{eq:deltas}. It amounts to $\Delta=0.105$, 0.118, 0.129, 0.149, 0.161 and
0.197 at $M=1$, 1.5, 2, 2.5, 3 and 4~TeV, with statistical uncertainties of about 0.01--0.03. The separate runs used for the histograms of $z$ below give values that agree within these uncertainties, for example 0.193 at 4~TeV. Any candidate explanation must yield a negative contribution to $\delta_{\rm NLO}-\delta_{\rm
thr}$ of this size.

\subsection{Candidates for the missing terms}
\label{sec:candidates}

Before examining the exact result, it is instructive to estimate terms known to be absent from the leading-power
kernel. Each estimate below contains no free parameter, and each is an estimate only: it retains the term that is
enhanced by a large logarithm or by a large luminosity ratio, and is not a complete calculation of the power
corrections.

\paragraph{Non-soft fragmentation of the top quark.}
The threshold expansion contains logarithms of $m_t$ only through the soft limit of the emission from the top quark. The
collinear emission of a gluon that carries a finite fraction of the momentum of a boosted top quark is described by the
perturbative fragmentation function~\cite{Mele:1990cw}
\begin{equation}
  D(x)=\delta(1-x)+\frac{\as}{2\pi}\,C_F\left[\frac{1+x^2}{1-x}\left(\ln\frac{\mu^2}{m_t^2}-2\ln(1-x)-1\right)\right]_+\,.
  \label{eq:fragmentation}
\end{equation}
The soft limit of this function, in which $(1+x^2)/(1-x)\to2/(1-x)$, is part of our kernel through the product of
$C_D$ and $\widetilde s_D$ of section~\ref{sec:masslogs}. The remainder,
\begin{equation}
  r(x)=-\frac{\as}{2\pi}\,C_F\,(1+x)\left(\ln\frac{\mu^2}{m_t^2}-2\ln(1-x)-1\right)\,,
\end{equation}
is suppressed by a power of $1-x$ but enhanced by $\ln(M^2/m_t^2)$. Since the top quark and the gluon are not
recombined in $\Mttj$, such an emission changes the invariant mass, with $1-z=f_i(1-x)$ and
$f_i=1-s_{jk}/M^2$, where $j,k$ are the other two particles of the Born configuration. Convolving $r(x)$
with the luminosity, summing over the two top quarks and averaging over the Born phase space gives the contribution to
$\delta_{\rm NLO}-\delta_{\rm thr}$ listed in table~\ref{tab:candidates}. It grows with $M$, as expected from the
logarithm, but it is positive when integrated over all $z$. It therefore does not explain the total difference; it makes it larger. This sign arises from the endpoint at $z=1$. At $z<0.995$ the estimated contribution to the threshold minus exact result is positive, see section~\ref{sec:offdiag}, and it is overcompensated at $z=1$.

\paragraph{Non-soft initial-state collinear emission.}
For an invariant-mass observable with $\mu_F=M$, the collinear emission from an incoming parton leaves the universal
remnant familiar from Drell--Yan production~\cite{Altarelli:1979ub},
\begin{equation}
  \frac{\as}{2\pi}\,P_{ab}(x)\,\ln\frac{(1-x)^2}{x}\,,
  \label{eq:isrremnant}
\end{equation}
convolved with the luminosity in which the parton $b$ replaces $a$ on that leg. The threshold expansion retains only
its soft limit $2C_a[\ln(1-x)^2/(1-x)]_+$. It therefore lacks the regular part of the diagonal splitting functions as
well as the off-diagonal terms, which involve new partonic channels. The sum of these terms is given in the second row of
table~\ref{tab:candidates}. The diagonal terms are positive. The off-diagonal terms are negative, and become important
at high mass, where the $gg$ Born channel receives a large contribution from incoming quarks whose luminosity is much
larger than that of gluons. The sum falls with $M$, as opposed to the difference $\Delta$, and has the wrong sign at 1 and
2~TeV.

\begin{table}[t]
  \centering
  \begin{tabular}{lccc}
    \hline
    & $M=1$~TeV & 2~TeV & 4~TeV\\
    \hline
    non-soft top fragmentation & $+0.033$ & $+0.074$ & $+0.145$\\
    non-soft initial-state collinear & $+0.258$ & $+0.084$ & $-0.021$\\
    \hline
    observed $\delta_{\rm NLO}-\delta_{\rm thr}$ & $-0.105$ & $-0.129$ & $-0.197$\\
    \hline
  \end{tabular}
  \caption{Estimates of terms absent from the leading-power kernel, as contributions to $\delta_{\rm NLO}-\delta_{\rm
  thr}$ in units of the leading-order cross section, compared with the exact difference. Parameters as in
  section~\ref{sec:setup}, $\mu=M$. The estimates use 20 of the sequences of section~\ref{sec:born}.}
  \label{tab:candidates}
\end{table}

\paragraph{Migration of a second jet.}
Our observable uses the leading jet only. At NLO a second parton may form a separate jet, which is then excluded from
$\Mttj$, whereas the kernel does not distinguish it from any other radiation outside the leading jet. To test whether this
matters, we compare $\Mttj$ with the invariant mass $M_{t\bar tX}$ of the top quarks and all jets above the threshold.
Replacing $\Mttj$ by $M_{t\bar tX}$ changes the $K$ factor by $-0.05$, $+0.04$, $+0.11$ and $+0.19$ at $M=1$, 2, 3 and 4~TeV. The statistical uncertainties of the individual $K$ factors are about 0.03 at $M\le3$~TeV and 0.003 at 4~TeV.
Events with two jets account for only 4.8\%, 2.9\%, 1.9\% and 1.3\% of the cross
section in the $\Mttj$ window, so the large change at 4~TeV is a result of migration. Because the spectrum falls steeply, a window in $M_{t\bar tX}$ receives
events with a hard second jet from values of $\Mttj$ at which the cross section is much larger. At 4~TeV this has the same size as $\Delta$. The threshold result exceeds the exact $K$ factor for $M_{t\bar tX}$ by $+0.17$, $+0.11$, $+0.03$ and $-0.004(3)$ at the four masses, whereas the excess for $\Mttj$ remains between $+0.12$ and $+0.19$. This agreement in the total may be accidental, since it can hide compensating differences between ranges of $z$, and since the shift has the opposite sign at 1~TeV. It nevertheless shows that the emission of a hard second jet is an effect of
the required size at high mass. This is not a leading-power effect, since a jet above the threshold $0.15\,M$ is not soft, and it would correspond to $z$ well below one.

\paragraph{Top-pair invariant mass.}
A difference that originates from boosted top-quark dynamics, or from the peak of $g^*\to t\bar t$, would be concentrated
at a specific value of the invariant mass $\sqrt{s_{34}}$ of the pair. We therefore decomposed $\Delta$ in bins of
$\sqrt{s_{34}}$. Using 40 of the sequences of section~\ref{sec:born}, the ratio
$(K_{\rm thr}-K_{\rm NLO})/(K_{\rm NLO}-1)$ lies between 0.18 and 0.23 in all bins between 345 and 1600~GeV at 4~TeV, and between 0.11
and 0.14 at 2~TeV. At 1~TeV it rises from 0.06 in the lowest bin to 0.22 at 700--1000~GeV. The difference is thus spread over the
whole range, nearly uniformly at the two larger masses, and is not tied to a particular kinematic configuration of the pair.

\subsection{Distribution of the correction in $z$}
\label{sec:zdensity}

The most direct information comes from comparing the distributions in $z=M^2/\hat s$. For $z<1$ only real emission
contributes, and the prediction of the threshold kernel can be written in closed form. For a Born point the expansion
of the kernel to $\order{\as}$ has the form $A_0+A_1\partial_\eta+A_2\partial_\eta^2$, acting on
\begin{equation}
  g(\eta,z)=G(z)\,\frac{e^{-2\gamma_E\eta}}{\Gamma(2\eta)}\,z^{-\eta}(1-z)^{2\eta-1}\,,\qquad
  G(z)=\frac1z\,\mathcal L\!\left(\frac\tau z\right)\,,
\end{equation}
see eq.~\eqref{eq:kernelz}, at $\eta=0$. Using $1/\Gamma(2\eta)=2\eta\,e^{2\gamma_E\eta}\,[1+\order{\eta^2}]$ one finds for $z<1$
\begin{equation}
  \partial_\eta g\big|_{\eta=0}=\frac{2\,G(z)}{1-z}\,,\qquad
  \partial_\eta^2 g\big|_{\eta=0}=\frac{4\,G(z)}{1-z}\,\bigl[2\ln(1-z)-\ln z\bigr]\,,
  \label{eq:zdensity}
\end{equation}
while $A_0$ and the endpoint contributions of the plus distributions are located at $z=1$. Equation~\eqref{eq:zdensity},
integrated over bins of $z$, gives the threshold density $d\delta_{\rm thr}/dz$ in the same units as
$\delta_{\rm thr}$. The exact density $d\delta_{\rm NLO}/dz$ follows from a histogram of $z$ in the
NLO calculation, as described in section~\ref{sec:nlo-num}, normalised to the leading-order cross section in the
window. Both densities are negative over most of the range, since at fixed $M$ an emission requires a larger $\hat s$ and
the first term of the expansion is dominated by the logarithm $\ln(1-z)$. Their integrals are compensated by the
positive contributions at $z=1$.

\begin{figure}[t]
  \centering
  \includegraphics[width=\textwidth]{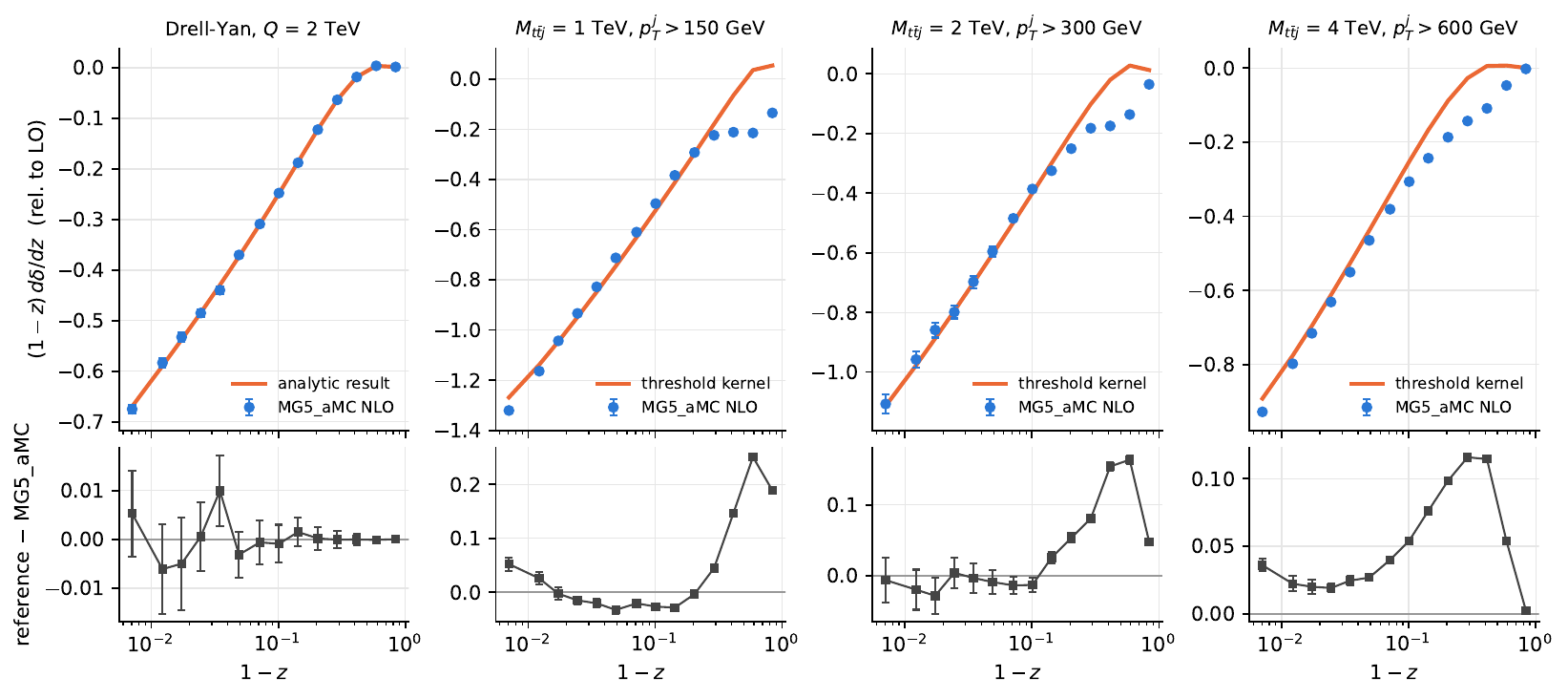}
  \caption{Contribution of real emission at $z<1$ to the NLO correction, normalised to the leading-order cross section and
  multiplied by $1-z$. Upper panels: histogram from \textsc{MadGraph5\_aMC@NLO} (points with statistical uncertainties) and
  reference curve. The reference is the analytic coefficient function for Drell--Yan production (left) and the threshold
  kernel, eq.~\eqref{eq:zdensity}, for $t\bar tj$. Lower panels: reference minus histogram. Parameters as in
  section~\ref{sec:setup}, $\mu=M$. The panels have different vertical scales.}
  \label{fig:zdens}
\end{figure}

Figure~\ref{fig:zdens} and table~\ref{tab:zgroups} compare the two densities. At $M=1$ and 2~TeV, from $z\approx0.8$ up to $z=0.98$ the threshold kernel follows the exact density closely: the differences are at most
0.02 per range in units of the leading-order cross section, out of a density of several tenths per range. In the range $0.98<z<0.995$ the two agree at 2~TeV and differ by $+0.046(10)$ at 1~TeV. Below
$z\approx0.8$ the kernel no longer describes the exact result: the exact density stays negative while the kernel tends to zero and becomes positive. The integrated difference for $z<0.8$ is $+0.222$ at 1~TeV and
$+0.172$ at 2~TeV, whereas the total difference $\Delta$ amounts to $+0.105$ and $+0.128$, so the contribution of the region far from threshold alone exceeds the total difference.

At $M=4$~TeV the agreement is less good, and the small statistical uncertainties of this run make this clear. The kernel lies above the exact density in every range, by 3--4\% of its content for $z>0.95$, by 10\% for $0.9<z<0.95$ and by 31\% for $0.8<z<0.9$.
The region $z<0.8$ contributes $+0.122$ and the region $0.8<z<0.995$ a comparable $+0.139$. Thus at this mass the difference does not arise only far from threshold. It nevertheless grows smoothly with the distance from threshold, in a way
that is similar in terms of the emitted energy $E=(1-z)M/2$ at all masses: the deviation reaches about 10\% at $E\approx100$--200~GeV, which corresponds to $1-z\approx0.2$--0.3, 0.1--0.15 and 0.05--0.1 at $M=1$, 2 and 4~TeV.
The disagreement is therefore organised by an emitted energy rather than by a fixed value of $z$. The larger deviation close to threshold at 4~TeV is not a consequence of the change in the mix of partonic channels; section~\ref{sec:offdiag} studies its origin. Table~\ref{tab:zchannels}, obtained by summing the individual integration jobs of the same NLO runs, shows that $0.8<z<0.995$ contributes $+0.158$ in the $gg$ channel and $+0.139$ in the $qg$ channel at 4~TeV, relative to the leading-order cross section of each channel, compared with $-0.022$ and $+0.005$ at 1~TeV. The deviation thus grows with $M$ within each channel by a similar amount, although the $qg$ share rises from 44\% to 80\%.

\begin{table}[t]
  \centering
  \begin{tabular}{ccccccc}
    \hline
    $M$ [TeV] & channel & LO share & total & $z<0.8$ & $0.8<z<0.995$ & $0.995<z<1$\\
    \hline
    1 & $gg$ & 0.532 & $+0.150$ & $+0.274$ & $-0.022$ & $-0.102$\\
      & $qg$ & 0.440 & $+0.057$ & $+0.165$ & $+0.005$ & $-0.114$\\
    2 & $gg$ & 0.361 & $+0.187$ & $+0.228$ & $-0.004$ & $-0.037$\\
      & $qg$ & 0.619 & $+0.101$ & $+0.143$ & $-0.025$ & $-0.018$\\
    4 & $gg$ & 0.188 & $+0.345$ & $+0.162$ & $+0.158$ & $+0.026$\\
      & $qg$ & 0.799 & $+0.164$ & $+0.115$ & $+0.139$ & $-0.091$\\
    \hline
  \end{tabular}
  \caption{Threshold minus exact result per Born channel, in units of the leading-order cross section of the same channel. The $q\bar q$ channel contributes less than 3\% of the cross section and is omitted. In the exact result the assignment to a channel follows the Born process of the FKS subtraction, which is also the assignment in the kernel.}
  \label{tab:zchannels}
\end{table}

\begin{table}[t]
  \centering
  \input{sections/tab_zgroups}
  \caption{Contribution of each range of $z$ to the $K$ factor of $t\bar tj$ production, from the threshold kernel and from
  the exact NLO calculation, in units of the leading-order cross section. Numbers in brackets are the statistical
  uncertainties of the last digits of the exact result. The range $0.995<z<1$ contains the Born term, the virtual correction and
  the contributions at $z=1$.}
  \label{tab:zgroups}
\end{table}

\paragraph{The bin next to threshold.}
The last range of table~\ref{tab:zgroups}, $0.995<z<1$, does not agree perfectly: the threshold result is smaller
by $0.110(12)$, $0.025(32)$ and $0.068(5)$ at 1, 2 and 4~TeV. This range should be reproduced by a correct leading-power expression up
to terms that vanish for $R\to0$, as the formula is obtained in the small-$R$ limit. To test this, we repeated the comparison
for $R=1$ at 1 and 2~TeV. The difference grows to $0.257(16)$ and $0.230(43)$, respectively. If we describe the dependence by
$a+bR^2$ using only these two values, the coefficient $b$ is $-0.18$ and $-0.24$ at 1 and 2~TeV, compatible
within the uncertainties of about $0.03$ and $0.06$. At $R=0.4$ the term proportional to $R^2$ then gives $-0.03$ and $-0.04$. This
accounts for the difference at 2~TeV but not at 1~TeV, where $a=-0.08$ remains. Assuming a similar coefficient at 4~TeV, a remainder of about $-0.03$ is left there. These are two-point fits at a value $R=1$ outside
the range of validity of the small-$R$ expression, so they indicate the order of magnitude only.

We tested a further possibility at $M=1$~TeV, the migration of events across the jet threshold. A soft emission outside the jet
lowers the transverse momentum of the jet and can move an event below the threshold, an effect not included in the factorisation
formula. Its size should follow the leading-order density at the threshold. With $p_T^j>300$~GeV in place of 150~GeV,
the fraction of the leading-order cross section within 10~GeV of the threshold rises from 8.8\% to 12.9\%, and the difference
in the last range is $0.123(20)$, compared with $0.110(12)$. These two values are compatible with no change as well as with a growth
in proportion to this density, so this comparison does not decide the question. The density, in units of $1/M$, is nearly the
same at $M=1$, 2 and 4~TeV (8.8, 8.5 and 10.0 within 1\% of $M$), whereas the remainder is $-0.08$, $+0.02(3)$ and about $-0.03$. These values do not follow the density, nor do they show a clear trend with $M$. The origin of the
remainder, which amounts to at most about 1.5\% of the content of this range, remains open.

\subsection{Initial states absent from the kernel}
\label{sec:offdiag}

The channels of table~\ref{tab:zchannels} are labelled by the Born process. The real emission of a given FKS configuration can, however, start from a different initial state. For example, $qg\to t\bar tgq$ with a soft or collinear final-state quark has the underlying Born process $gg\to t\bar tg$, the incoming quark having turned into a gluon. Such off-diagonal contributions have no $1/(1-z)$ enhancement, since a soft quark, unlike a soft gluon, does not produce a soft singularity. They are therefore absent from the leading-power kernel. To measure them, we record the initial state of every event and counter-event in addition to its Born channel. The classes add up to the $z$ histogram, and the bin $0.995<z<1$, which contains the Born and virtual terms, is populated almost entirely by the diagonal initial state, as it should be.

Table~\ref{tab:offdiag} shows the result for $0.8<z<0.995$, at the two masses of the earlier tables and additionally for $M=1$~TeV at $\sqrt S=3.4$~TeV. This third point has the same $\tau=0.0865$ as $M=4$~TeV at $\sqrt S=13.6$~TeV, and the same $p_T^j/M$, so that the two differ only through $M/m_t$ and through the running of $\as$ and the PDFs.

\begin{table}[t]
  \centering
  \input{sections/tab_offdiag}
  \caption{Threshold minus exact result for $0.8<z<0.995$, per Born channel and in units of the leading-order cross section of that channel. Off-diagonal entries are the initial states absent from the kernel, for which the threshold result vanishes; for the Born channel $qg$ these are $gg$, $qq$ and $q\bar q$. The diagonal estimate is the sum of the regular initial-state terms and the fragmentation estimate, see the text. Entries marked $*$ are the total of the higher-precision run minus the off-diagonal part of a separate run with 1\% accuracy; the corresponding direct values, $+0.023(155)$ and $-0.051(48)$, are compatible but much less precise.}
  \label{tab:offdiag}
\end{table}

The off-diagonal terms are well described by the remnant~\eqref{eq:isrremnant} of the collinear subtraction. For a Born channel $ab$ and an initial state $b'b$ it gives, at $\mu=M$,
\begin{equation}
  \frac{\as(M)}{2\pi}\,P_{ab'}(z)\,\ln\frac{(1-z)^2}{z}\;\frac{\mathcal L_{b'b}(\tau/z)}{\mathcal L_{ab}(\tau)}\,,
  \label{eq:offdiagpred}
\end{equation}
summed over the two legs and over flavours. This expression has no free parameter and does not involve the hard matrix element of the process. Only the logarithm in it is universal, and non-logarithmic terms are not included. In the range considered it agrees with the exact result to within about 15\% in the two main channels, over a variation by a factor of 3.5 of the $gg$ term. It fails for $z<0.8$, where it falls short of the exact result by a factor of about 2 to 8, and in the small $q\bar q$ Born channel, where the exact result has the opposite sign.

Three conclusions follow. First, the off-diagonal terms depend on $\tau$ and not on $M$. At fixed $\tau=0.0865$ they are $+0.286(3)$ and $+0.260(12)$ for the $gg$ Born channel at $M=1$ and 4~TeV, compared with $+0.081(1)$ at $\tau=0.0054$. Their growth follows from the luminosity ratio in eq.~\eqref{eq:offdiagpred}. Second, the good agreement of the total at $M=1$~TeV in table~\ref{tab:zgroups} is in part a cancellation. In the $gg$ Born channel the diagonal part, $-0.087(18)$, compensates the off-diagonal part, $+0.081(1)$. The diagonal part itself differs from the exact result by about 2\% of the content of this range. Third, at $M=4$~TeV the deviation close to threshold is entirely off-diagonal in the sum over channels. Weighted with the leading-order shares, the off-diagonal terms give about $+0.14$ of the total $+0.139$, and the diagonal terms of the $gg$ and $qg$ channels compensate each other. Close to threshold the terms remain small compared with the density: for $0.95<z<0.995$ they amount to about 1\% of it at 1~TeV and 3\% at 4~TeV.

The diagonal part in the $gg$ channel is nearly the same at all three points. In the $qg$ channel it moves from $-0.074(7)$ to $+0.026(9)$ between $M=1$ and 4~TeV at fixed $\tau$. This is the only clear dependence on $M$ at fixed $\tau$ in table~\ref{tab:offdiag}. Two known terms contribute to the diagonal part: the regular part of the diagonal splitting functions, combined with the difference between the kernel and the form~\eqref{eq:isrremnant}, which is negative, and non-soft emission from the top quarks, estimated as in section~\ref{sec:candidates} and positive. The second contains $\ln(M^2/m_t^2)$. Their sum is the estimate in table~\ref{tab:offdiag}. It reproduces the three values in the $qg$ channel to within about 0.025, but in the $gg$ channel it differs by 0.04 at 1~TeV and predicts an increase of $0.08$ at 4~TeV that is not seen, and as noted in section~\ref{sec:candidates} the fragmentation estimate does not describe the last bin. We therefore regard it as an indication of the size of these terms and not as an explanation of the data. 

To separate the dependence on $\tau$ from that on $M$ directly, we add a fourth point, $M=4$~TeV at $\sqrt S=54.4$~TeV, which has the same $\tau$ as $M=1$~TeV at $\sqrt S=13.6$~TeV. Such a collider energy serves only as a numerical test. Table~\ref{tab:taum} shows the result. The two channels behave differently. The $gg$ channel is small at the lower value of $\tau$ and large at the higher value, at either mass, in accordance with its off-diagonal part. The $qg$ channel is small at $M=1$~TeV and about $+0.14$ at $M=4$~TeV, at either value of $\tau$. The difference $\Delta$ between the threshold and exact $K$ factors, summed over channels and over all $z$, grows by more than a factor of 2 from $M=1$ to 4~TeV at both values of $\tau$, whereas it decreases slightly with increasing $\tau$ at fixed $M$. A steeper luminosity alone therefore does not reduce it appreciably, and the growth with $M$ explains why $\delta_{\rm thr}/\delta_{\rm NLO}$ does not approach one in figure~\ref{fig:dominance}, although $\tau$ rises.

\begin{table}[t]
  \centering
  \input{sections/tab_taum}
  \caption{Separation of the dependence on $\tau$ and on $M$ at fixed $p_T^j/M=0.15$. Threshold minus exact result for $0.8<z<0.995$ per Born channel, in units of the leading-order cross section of the channel, and the difference $\Delta=K_{\rm thr}-K$ in units of the total leading-order cross section. The point at $\sqrt S=54.4$~TeV serves only as a numerical test.}
  \label{tab:taum}
\end{table}

The comparison at fixed $\tau$ leaves two possible origins of the dependence on $M$: the ratio $M/m_t$, and the running of $\as$ and of the PDFs between the two scales. To decide between them we keep $M=1$~TeV and $\sqrt S=3.4$~TeV and lower the top-quark mass to $m_t=172.5/4=43.125$~GeV. This point has the same $\tau$, $p_T^j/M$ and $M/m_t$ as $M=4$~TeV at $\sqrt S=13.6$~TeV, and it is again a purely numerical test. We fixed the expectation beforehand: for the $qg$ channel in $0.8<z<0.995$, the initial-state and fragmentation estimates together with the off-diagonal terms give $+0.131$, compared with the value $+0.024(7)$ at the physical top-quark mass, which would persist if $M/m_t$ played no role. The result is $+0.136(6)$, in agreement with the expectation. In the $gg$ channel the outcome does not confirm the estimate. The same estimate gives $+0.234$ there, the result is $+0.196(12)$, and the value at the physical mass is $+0.181(18)$ in the run with initial-state separation and $+0.232(28)$ in an earlier run, which are compatible with each other. The result is therefore about three standard deviations from the estimate and shows no clear change with $m_t$. The agreement in the $qg$ channel also rests on a cancellation between a negative initial-state term and a positive fragmentation term of similar size, so that it carries limited weight on its own.

Table~\ref{tab:mtscale} shows the comparison for all channels, in each range of $z$. Lowering $m_t$ moves the $M=1$~TeV result onto that at $M=4$~TeV. The remaining differences are consistent with the overall coupling: when the ratio $\as(1~\text{TeV})/\as(4~\text{TeV})=1.15$ is applied, $K-1$, $K_{\rm thr}-1$ and $\Delta$ agree to within 1\%, and the ranges of $z$ agree within two standard deviations. This comparison was not part of the expectation fixed beforehand, although the ratio is calculated and not fitted. We conclude that, summed over channels, the dependence on $M$ at fixed $\tau$ and $p_T^j/M$ is a dependence on $M/m_t$. It also provides a useful internal check of both calculations. The exact $K$ factor barely changes when $m_t$ is lowered at fixed $M$, whereas $K_{\rm thr}$ increases from 2.22 to 2.36, so that the growth of $\Delta$ arises in the threshold approximation. Close to threshold the kernel moves by more than the exact density: for $0.8<z<0.9$ the former changes from $-0.187$ to $-0.139$, the latter from $-0.208$ to $-0.197$.

This points to terms that depend on $m_t$ and that the leading-power kernel does not reproduce. Radiation from the top quarks, which is not part of $\Mttj$, is the natural candidate: only its soft limit is contained in the kernel, and the non-soft remainder carries $\ln(M^2/m_t^2)$, as in the fragmentation estimate of section~\ref{sec:candidates}. That estimate, however, does not describe the $gg$ channel at 4~TeV or the last bin, so the mechanism is not established by this analysis. To test it more directly, we add to the top quarks all partons outside the leading jet that lie within a distance $\Delta R$ of them, which recovers radiation from the top quarks in the invariant mass and leaves the Born level unchanged. Since the kernel has no such version, we compare only the change with $m_t$, again at $M=1$~TeV and $\sqrt S=3.4$~TeV, with a criterion fixed beforehand: if radiation from the top quarks carries the growth of $K_{\rm thr}$ by $+0.137$, the $K$ factor of the dressed mass should grow by a similar amount when $m_t$ is lowered, whereas otherwise it should stay constant, like the undressed one, for which we find $+0.012(35)$ in these runs. For $\Delta R<1$ the growth is $+0.161(33)$, and $K_{\rm thr}-K_d$ is $+0.012$ and $-0.011$ at the two masses. For $\Delta R<0.4$ the growth is $+0.052(35)$, in the range that we had noted beforehand as a partial explanation. After seeing this result, we note that a small cone may be expected to capture little of radiation that the mass of the top quark suppresses at angles below $m_t/E_t$, which is about 0.5 at the physical mass. This supports radiation from the top quarks as the origin of the dependence on $M/m_t$, without proving it: a cone of this size also collects other radiation, and at $M=4$~TeV the mass $M_{t\bar tX}$ closes the same gap through a second jet, so that these recoveries cannot be simply added. Since the onset was found to be organised by the emitted energy in section~\ref{sec:zdensity}, the present result would also fit an onset at energies of order $m_t$; we have not tested this separately.

\begin{table}[t]
  \centering
  \input{sections/tab_mtscale}
  \caption{Scaling test. Threshold minus exact result per range of $z$, and $K$ factors, in units of the total leading-order cross section. $M=1$~TeV is at $\sqrt S=3.4$~TeV and $M=4$~TeV at $\sqrt S=13.6$~TeV, with the same $\tau$ and $p_T^j/M$. The last column shows the $M=4$~TeV values multiplied by $\as(1~\text{TeV})/\as(4~\text{TeV})$.}
  \label{tab:mtscale}
\end{table}

\subsection{Validation on Drell--Yan production}
\label{sec:dy}

The comparison of figure~\ref{fig:zdens} relies on the assumption that the histogram of $z$ produced by the NLO program
represents the partonic coefficient function bin by bin. This is not obvious. In the FKS method, the real-emission
contribution diverges for $z\to1$ and is rendered finite only in combination with counter-events, and these
contributions are not assigned to the same value of $z$. To quantify the resulting effect we applied the same
procedure to the production of a virtual photon of mass $Q=2$~TeV in proton--proton collisions at $\sqrt S=13.6$~TeV. In
this case the exact coefficient functions are known~\cite{Altarelli:1979ub,Hamberg:1990np}. For $\mu_R=\mu_F=Q$ and $z<1$
they read
\begin{align}
  C_{q\bar q}(z)&=\frac{\as}{4\pi}\,C_F\left[8(1+z^2)\,\frac{\ln(1-z)}{1-z}-4(1+z^2)\,\frac{\ln z}{1-z}\right]\,,\\
  C_{qg}(z)&=\frac{\as}{4\pi}\,T_F\left[2\bigl(z^2+(1-z)^2\bigr)\ln\frac{(1-z)^2}{z}+1+6z-7z^2\right]\,.
\end{align}
Table~\ref{tab:dy} compares the integrals of these functions with the luminosity to the histogram from
\textsc{MadGraph5\_aMC@NLO}, obtained with the window $Q\pm10$~GeV imposed at generation level, as for $t\bar tj$, a
target accuracy of 0.2\% and no restriction on $\hat s$.

\begin{table}[t]
  \centering
  \input{sections/tab_dy}
  \caption{Contribution of each range of $z$ to the NLO correction of Drell--Yan production ($\gamma^*$ only) in units
  of the leading-order cross section, for $Q=2$~TeV and $\sqrt S=13.6$~TeV. Numbers in brackets are the statistical uncertainties
  of the last digits.}
  \label{tab:dy}
\end{table}

The agreement is excellent in every range, up to the differences of at most 0.0012 that are compatible with
the statistical uncertainties. The $K$ factor from the histogram, 1.2452(24), also agrees with the analytic result of 1.248 within
1.2 standard deviations. We conclude that the histogram of $z$ does represent the coefficient function bin by bin, and that
the method is validated for initial-state radiation. This process has no final-state radiation and no jet, so that it does not test the treatment of final-state counter-events. Subject to this limitation, we regard the differences found for $t\bar tj$ in figure~\ref{fig:zdens} as physical. Two practical remarks are in order. It is important
to impose the window at generation level: without it, the statistical uncertainty of a bin of $z$ is an order of magnitude larger. We also
note that the luminosity in this comparison enters as $\int dz\,\mathcal L_{\rm c}(\tau/z)\,C(z)$ in terms of
$\mathcal L_{\rm c}(y)=y\mathcal L(y)$, and that a factor of $z$ in this convolution is clearly visible in the comparison.

\subsection{Interpretation}
\label{sec:lp-interpretation}

The kernel is defined only up to powers of $z$, which are formally suppressed by $1-z$. The results of this section
do not suggest that this freedom explains the difference. Multiplying the kernel by $z$ raises $K_{\rm thr}$, to 2.14 at 1~TeV and
2.26 at 4~TeV from 2.00 and 2.19, and therefore increases the difference from the exact result instead of reducing it. This holds although $p=1$ corresponds to the form of the convolution that the exact Drell--Yan coefficient function takes in terms of $\mathcal L(y)$.

We also caution against using $\delta_{\rm thr}/\delta_{\rm NLO}$ alone as a measure of the quality of the kernel. At $R=1$
we find $\Delta=-0.05$ and $+0.01$ at 1 and 2~TeV, close to zero, although the differences in the individual regions of $z$ are as
large as for $R=0.4$. They merely compensate each other in the integral. The same happens at $M=1$~TeV with $p_T^j>300$~GeV, where $\Delta=+0.01$ while the region $z<0.8$ gives $+0.135$ and the last
range $-0.123$.

Taken together, these results give a consistent picture of two properties of the difference: initial states absent from the kernel, which depend on $\tau$, and a dependence on $M/m_t$ of the sum over channels, whose mechanism we have not established (section~\ref{sec:offdiag}). The threshold kernel is a good description of exact real emission over a wide range close to threshold, including the logarithms and
the colour structure that are the purpose of the calculation. The difference $\Delta$ originates mostly from emitted energies beyond 100--200~GeV, in the sum over ranges entirely so at 1 and 2~TeV, through a cancellation with the bin next to threshold and in part between initial states, and about one half at 4~TeV, where $z\lesssim0.8$ is complemented by a smaller and gradually growing deviation at larger $z$. This is the region
where the kernel extrapolates a soft expansion to hard emission, and where the exact result contains the terms studied above. At $\order{\as}$, this mismatch is removed by matching to the exact
NLO. At higher orders it is not: the resummed correction also contains this extrapolation, and its size is not controlled by the threshold expansion. The next section quantifies this.

%% file: sections/tab_zgroups.tex
\begin{tabular}{ccccccc}
\hline
$z$ range & \multicolumn{2}{c}{1~TeV} & \multicolumn{2}{c}{2~TeV} & \multicolumn{2}{c}{4~TeV}\\
 & thr. & exact & thr. & exact & thr. & exact\\
\hline
0--0.500 & $+0.032$ & $-0.120(1)$ & $+0.014$ & $-0.058(3)$ & $+0.002$ & $-0.016(0)$\\
0.5--0.700 & $-0.044$ & $-0.109(1)$ & $-0.018$ & $-0.089(3)$ & $+0.000$ & $-0.059(1)$\\
0.700--0.800 & $-0.096$ & $-0.101(1)$ & $-0.060$ & $-0.088(2)$ & $-0.022$ & $-0.066(1)$\\
0.800--0.900 & $-0.288$ & $-0.271(2)$ & $-0.210$ & $-0.219(4)$ & $-0.118$ & $-0.170(1)$\\
0.900--0.950 & $-0.439$ & $-0.422(3)$ & $-0.348$ & $-0.342(7)$ & $-0.238$ & $-0.266(1)$\\
0.950--0.980 & $-0.801$ & $-0.783(4)$ & $-0.665$ & $-0.662(12)$ & $-0.503$ & $-0.523(2)$\\
0.980--0.995 & $-1.644$ & $-1.689(10)$ & $-1.425$ & $-1.404(26)$ & $-1.134$ & $-1.174(4)$\\
0.995--1 & $+5.277$ & $+5.388(12)$ & $+4.802$ & $+4.827(32)$ & $+4.199$ & $+4.267(5)$\\
\hline
total & $1.997$ & $1.892$ & $2.091$ & $1.963$ & $2.186$ & $1.993$\\
\hline
\end{tabular}

%% file: sections/tab_offdiag.tex
\begin{tabular}{llccc}
\hline
Born & & $M=1$~TeV & $M=1$~TeV & $M=4$~TeV\\
 & & $\sqrt S=13.6$~TeV & $\sqrt S=3.4$~TeV & $\sqrt S=13.6$~TeV\\
\hline
$gg$ & off-diagonal, exact & $+0.081(1)$ & $+0.286(3)$ & $+0.260(12)$\\
 & off-diagonal, eq.~\eqref{eq:offdiagpred} & $+0.076$ & $+0.273$ & $+0.257$\\
 & diagonal, threshold minus exact & $-0.087(18)$ & $-0.105(18)$ & $-0.103(16)^*$\\
 & diagonal, estimate & $-0.130$ & $-0.110$ & $-0.033$\\
 & all initial states & $-0.022(13)$ & $+0.232(28)$ & $+0.158(11)$\\
\hline
$qg$ & off-diagonal, exact & $+0.044(2)$ & $+0.100(1)$ & $+0.113(7)$\\
 & off-diagonal, eq.~\eqref{eq:offdiagpred} & $+0.048$ & $+0.107$ & $+0.100$\\
 & diagonal, threshold minus exact & $-0.083(16)$ & $-0.074(7)$ & $+0.026(9)^*$\\
 & diagonal, estimate & $-0.059$ & $-0.052$ & $+0.020$\\
 & all initial states & $+0.005(19)$ & $+0.024(7)$ & $+0.139(5)$\\
\hline
\end{tabular}

%% file: sections/tab_taum.tex
\begin{tabular}{lcccc}
\hline
 & \multicolumn{2}{c}{$\tau=0.0054$} & \multicolumn{2}{c}{$\tau=0.0865$}\\
 & $M=1$~TeV & $M=4$~TeV & $M=1$~TeV & $M=4$~TeV\\
 & $\sqrt S=13.6$~TeV & $\sqrt S=54.4$~TeV & $\sqrt S=3.4$~TeV & $\sqrt S=13.6$~TeV\\
\hline
$gg$, $0.8<z<0.995$ & $-0.022(13)$ & $+0.014(12)$ & $+0.232(28)$ & $+0.158(11)$\\
$qg$, $0.8<z<0.995$ & $+0.005(19)$ & $+0.148(39)$ & $+0.024(7)$ & $+0.139(5)$\\
$\Delta$, all channels and $z$ & $+0.105$ & $+0.239$ & $+0.080$ & $+0.193$\\
\hline
\end{tabular}

%% file: sections/tab_mtscale.tex
\begin{tabular}{lcccc}
\hline
 & $M=1$~TeV & $M=1$~TeV & $M=4$~TeV & $M=4$~TeV\\
 & $m_t=172.5$~GeV & $m_t=43.125$~GeV & $m_t=172.5$~GeV & $\times\,\as$ ratio\\
\hline
$z<0.8$ & $+0.085(1)$ & $+0.138(1)$ & $+0.122(1)$ & $+0.140$\\
$0.8<z<0.9$ & $+0.021(1)$ & $+0.058(1)$ & $+0.052(1)$ & $+0.059$\\
$0.9<z<0.95$ & $+0.006(2)$ & $+0.030(1)$ & $+0.028(1)$ & $+0.032$\\
$0.95<z<0.98$ & $+0.002(3)$ & $+0.020(2)$ & $+0.020(2)$ & $+0.023$\\
$0.98<z<0.995$ & $+0.037(7)$ & $+0.036(4)$ & $+0.040(4)$ & $+0.046$\\
\hline
$K-1$ & $1.141$ & $1.138$ & $0.993$ & $1.144$\\
$K_{\rm thr}-1$ & $1.221$ & $1.358$ & $1.186$ & $1.366$\\
$\Delta$ & $+0.080$ & $+0.220$ & $+0.193$ & $+0.222$\\
\hline
\end{tabular}

%% file: sections/tab_dy.tex
\begin{tabular}{cccc}
\hline
$z$ range & \textsc{MG5\_aMC} & analytic & difference\\
\hline
0.000--0.500 & $+0.0020(2)$ & $+0.0019$ & $+0.0000$\\
0.500--0.700 & $-0.0135(5)$ & $-0.0133$ & $-0.0002$\\
0.700--0.800 & $-0.0367(8)$ & $-0.0369$ & $+0.0002$\\
0.800--0.900 & $-0.1299(15)$ & $-0.1303$ & $+0.0004$\\
0.900--0.950 & $-0.2166(22)$ & $-0.2155$ & $-0.0011$\\
0.950--0.980 & $-0.4079(38)$ & $-0.4067$ & $-0.0012$\\
0.980--0.995 & $-0.8582(76)$ & $-0.8585$ & $+0.0002$\\
\hline
\end{tabular}

%% file: sections/damping.tex
\section{Resummation away from threshold}
\label{sec:damping}

The analysis of section~\ref{sec:lp-validity} shows that the leading-power kernel describes the exact real-emission
density well close to threshold, but not once the emitted energy exceeds about 100--200~GeV. The matching~\eqref{eq:matching} removes this difference at
$\order{\as}$, since the threshold expansion is replaced by the exact NLO result. It does not remove it at higher
orders. The difference $d\sigma^{\rm NNLL}-d\sigma^{\rm thr}$ is also computed with the kernel~\eqref{eq:kernelz} for all
values of $z$, and therefore assigns higher-order terms to a region in which the kernel is not reliable. In this section we
modify the resummed correction such that it acts only close to threshold, and study how much the result depends on the
way this is done. This dependence is an intrinsic ambiguity of a leading-power resummation at these values of $\tau$.

\subsection{Damping in $z$ space}
\label{sec:zcut}

The obvious approach is to multiply the integrand of the kernel by a function of $z$ that vanishes for $z$ far from
one. This does not achieve the desired result. The kernel~\eqref{eq:kernelz} has a non-integrable behaviour for $\eta<0$
at $z\to1$, which is defined through its subtractions, eq.~\eqref{eq:subtraction}. Its contribution at $z=1$ is not
independent of its behaviour at $z<1$: the plus distributions at fixed order are defined such that their integrals over
$z$ from $0$ to $1$ vanish. If the part at $z<1$ is suppressed below some value $x_0$ while the part at $z=1$ is kept, one
effectively introduces a term proportional to powers of $\ln x_0$ at $z=1$. The resummation exponentiates such terms,
even though they have no physical origin, and the result depends strongly on the unphysical parameter $x_0$.

\subsection{Damping in Mellin space}
\label{sec:mellindamp}

Moment space does not have this problem, since in it the kernel is a smooth function of $N$. Threshold logarithms
correspond to the behaviour at large $N$, whereas the region far from threshold is mapped to small $N$. We therefore
modify the part of the kernel beyond $\order{\as}$ by a function $D(N)$ that tends to one for $N\to\infty$ and vanishes
for $N\to0$. For colour-singlet processes it is well known that forms of the soft terms that differ only by powers of $1/N$ can give very different results far from the physical threshold, and that this is a central part of the uncertainty of the method~\cite{Kramer:1996iq,Ball:2013bra,Bonvini:2014joa}. With $K_{\rm res}$ the resummed kernel and $K_{\rm thr}$ its
expansion to $\order{\as}$, we replace in the matching formula
\begin{equation}
  K_{\rm res}(N)-K_{\rm thr}(N)\ \longrightarrow\ D(N)\,\bigl[K_{\rm res}(N)-K_{\rm thr}(N)\bigr]\,.
  \label{eq:damped}
\end{equation}
The difference on the left starts at $\order{\as^2}$ relative to the Born cross section, so that the NLO accuracy of the
matched result is not affected. Since the factor $D(N)$ destroys the closed form in $z$, the correction relative to the
undamped result is computed from the inverse Mellin transform~\eqref{eq:Rz}, as described in appendix~\ref{app:mellin}.

We study two functional forms,
\begin{equation}
  D_{\rm c}(N)=\frac{N^2}{N^2+N_0^2}\,,\qquad
  D_{\rm r}(N)=\frac{N}{N+N_0}\,,
  \label{eq:Dforms}
\end{equation}
which differ in the location of their poles, at $N=\pm iN_0$ and at $N=-N_0$, and consequently in their shape: $D_{\rm c}$
approaches one faster at large $N$ and suppresses small $N$ more strongly. We refer to them as complex-pole and
real-pole damping. The onset $N_0$ is the parameter that defines the boundary of the threshold region. It is natural to
relate it to the energy $E$ of the emitted radiation, since $1-z\simeq2E/M$ close to threshold, and $\Nbar=Ne^{\gamma_E}$
corresponds to $1/(1-z)$. We therefore set
\begin{equation}
  N_0=\frac{M}{2E_0\,e^{\gamma_E}}\,.
  \label{eq:N0}
\end{equation}
A fixed $E_0$ rather than a fixed value of $N_0$ is suggested by the emission-energy dependence found in
section~\ref{sec:zdensity}. We use $E_0=100$~GeV as central value and vary it
between 50 and 200~GeV. For example, $N_0=2.8$ and 11 at $M=1$ and 4~TeV for $E_0=100$~GeV.

\subsection{Results}
\label{sec:dampresults}

Table~\ref{tab:damping} shows the matched NNLL cross section relative to NLO for the different forms and onsets, with
all scales at their central values. The statistical uncertainties of these ratios are at most 0.5\%.

\begin{table}[t]
  \centering
  \begin{tabular}{ccccccccc}
    \hline
    $M$ [TeV] & undamped & \multicolumn{3}{c}{$D_{\rm c}$, $E_0$ [GeV]} & \multicolumn{3}{c}{$D_{\rm r}$, $E_0$ [GeV]}\\
    & & 50 & 100 & 200 & 50 & 100 & 200\\
    \hline
    1   & 1.173 & 1.006 & 1.024 & 1.081 & 1.221 & 1.149 & 1.131\\
    1.5 & 1.187 & 1.004 & 1.017 & 1.061 & 1.370 & 1.220 & 1.164\\
    2   & 1.198 & 1.003 & 1.013 & 1.050 & 1.505 & 1.286 & 1.196\\
    2.5 & 1.206 & 1.003 & 1.012 & 1.044 & 1.622 & 1.344 & 1.224\\
    3   & 1.213 & 1.003 & 1.011 & 1.040 & 1.725 & 1.395 & 1.250\\
    4   & 1.230 & 1.003 & 1.010 & 1.040 & 1.938 & 1.502 & 1.306\\
    \hline
  \end{tabular}
  \caption{Ratio of NLO+NNLL to NLO at $\sqrt S=13.6$~TeV, $R=0.4$ and $p_T^j>0.15\,M$, for the central scales.}
  \label{tab:damping}
\end{table}

Three features stand out. First, the undamped correction of 17--23\% at this threshold (see table~\ref{tab:kscan} for its dependence on the threshold) disappears almost completely with complex-pole
damping: the matched result is within 1--2\% of NLO for $E_0=100$~GeV at all masses. Second, real-pole damping does not suppress it. At $M\ge2$~TeV and $E_0\le100$~GeV it enhances the correction, up to 94\% at 4~TeV, while at 1~TeV it lies somewhat below the undamped result for
$E_0\ge100$~GeV. Third, both forms depend strongly on $E_0$. A larger $E_0$ implies a smaller $N_0$, and $D\to1$ for $N_0\to0$, so both must ultimately approach the undamped result, the complex-pole form from below. Neither form is preferred by any
principle, and both have the same behaviour at large $N$ up to corrections of order $N_0/N$.

This strong dependence on the form becomes more transparent in momentum space. The inverse Mellin transforms of
$D-1$ are
\begin{equation}
  D_{\rm r}(N)-1\ \to\ -N_0\,z^{N_0}\,,\qquad
  D_{\rm c}(N)-1\ \to\ -N_0\,\sin\!\left(N_0\ln\frac1z\right)\,.
  \label{eq:Dinverse}
\end{equation}
Damping therefore does not simply remove the region of small $z$. It convolves the correction with a function of $\ln z$
that extends over all $z$, of definite sign for the real pole and oscillating for the complex poles. The two
forms therefore weight the correction at different values of $z$ with different signs, and the results can lie on
opposite sides of the undamped one. We do not have a more detailed explanation of the size of the effect. What matters is that
the variation of the scales does not reveal this ambiguity: the scale uncertainty of the undamped result, about
$\pm10$\%, does not cover the range of the damped results at high mass.

The scale uncertainty also behaves differently in the two cases. For complex-pole damping the band of $\pm20$\% is
essentially that of the NLO result, which is expected if the resummed correction is removed. For real-pole damping
we find upward variations of 74--226\% for $E_0\le100$~GeV. Such a band signals that the matched prediction is not
under control in this variant, and we do not use it as a prediction.

\subsection{The leading-power ambiguity in $z$}
\label{sec:zpower}

A second, independent measure of the same ambiguity is provided by powers of $z$. Since $z^p=1+\order{1-z}$, the
kernel may be multiplied by $z^p$ without changing the leading-power accuracy. The resulting variation of the
threshold $K$ factor and of the resummed correction is a direct estimate of the terms not controlled by the
factorisation theorem. Table~\ref{tab:zpower} shows the result for $p=\pm1$. The ratio to NLO changes by
$-0.23$ and $+0.19$ around the result for $p=0$ at 1~TeV, and by $-0.12$ and $+0.12$ at 4~TeV. This is the same order as the difference between the undamped result and NLO, and for $p=-1$ at
$M\le1.5$~TeV the correction even changes sign.
The threshold approximation at $\order{\as}$ is closest to the exact correction for $p=-1$, which suggests that a form of
the kernel with stronger weight at smaller $z$ would be preferred at NLO. We do not use this fact to tune the kernel,
since the exact correction at $\order{\as}$ does not determine the behaviour at higher orders.

\begin{table}[t]
  \centering
  \begin{tabular}{ccccccc}
    \hline
    & \multicolumn{3}{c}{$K_{\rm thr}$} & \multicolumn{3}{c}{NLO+NNLL/NLO (undamped)}\\
    $M$ [TeV] & $p=-1$ & $p=0$ & $p=1$ & $p=-1$ & $p=0$ & $p=1$\\
    \hline
    1   & 1.924 & 1.997 & 2.137 & 0.95 & 1.17 & 1.36\\
    1.5 & 1.948 & 2.050 & 2.173 & 0.98 & 1.19 & 1.37\\
    2   & 1.988 & 2.091 & 2.199 & 1.01 & 1.20 & 1.36\\
    2.5 & 2.025 & 2.121 & 2.217 & 1.04 & 1.21 & 1.36\\
    3   & 2.057 & 2.145 & 2.231 & 1.07 & 1.21 & 1.35\\
    4   & 2.112 & 2.186 & 2.257 & 1.11 & 1.23 & 1.35\\
    \hline
  \end{tabular}
  \caption{Effect of multiplying the kernel by $z^p$. For comparison, the exact $K$ factors are 1.89, 1.93, 1.96, 1.97, 1.98 and 1.99 at these
  masses. The soft scale $\mu_s$ is kept at its value for $p=0$.}
  \label{tab:zpower}
\end{table}

\subsection{Dependence on the jet threshold}
\label{sec:threshdep}

So far we have used $p_T^j>0.15\,M$, a choice explained in section~\ref{sec:highmass}. Table~\ref{tab:kscan} shows how our main results depend on it. Agreement between the kernel and the exact density close to threshold (section~\ref{sec:zdensity}) does not change with the threshold within the uncertainties. Neither does the excess in this range at 4~TeV, so a higher threshold does not remove it. The difference for $z<0.8$ decreases as the threshold rises, which is in line with the requirement of a hard jet but is not a test of the factorisation by itself. The ratio $\delta_{\rm thr}/\delta_{\rm NLO}$ moves closer to one as the threshold rises at 1~TeV and, less clearly, at 2~TeV, whereas at 4~TeV it does not change within its uncertainty. The resummed correction, in contrast, does depend on the threshold. The undamped result is 23--33\% for $p_T^j>0.1\,M$, 17--23\% at $p_T^j>0.15\,M$, and 13--16\% at $p_T^j>0.2\,M$. This value is therefore not a property of the process alone, and our numbers for it hold for the threshold stated. The result with complex-pole damping at $E_0=100$~GeV stays within 1--4\% of NLO throughout.

\begin{table}[t]
  \centering
  \input{sections/tab_kscan}
  \caption{Dependence on the jet threshold at $\sqrt S=13.6$~TeV, $R=0.4$ and $\mu=M$, with $\mu_s$ fixed at the values of eq.~\eqref{eq:musvalues}: exact $K$ factor, ratio of threshold to exact correction, matched NLO+NNLL over NLO without damping (undamped) and with complex-pole damping at $E_0=100$~GeV (damped), and threshold minus exact result in two ranges of $z$, in units of the leading-order cross section. Numbers in brackets are statistical uncertainties of the last digits.}
  \label{tab:kscan}
\end{table}

\subsection{Consequences for the prediction}
\label{sec:dampconsequences}

The results of this section lead us to the following presentation. A leading-power resummation at $M=1$--4~TeV and
$\sqrt S=13.6$~TeV does not yield a unique improvement of the NLO prediction. The size and even the existence of the
enhancement depend on the treatment of a region of $z$ in which the kernel is not known to be correct. It would be
misleading to quote one of these results, with its scale band, as a prediction of higher accuracy. We therefore keep the
exact NLO cross section as the central prediction. As an indication of the possible effect of threshold resummation, we show the
range between the complex-pole damped result with $E_0=100$~GeV and the undamped result. This range is not an envelope of the variants studied here and is not a bound. Real-pole damping exceeds its upper end, and the variation with $z^{-1}$ falls below its lower end at $M\le1.5$~TeV, so that the ambiguity extends in both directions. We exclude real-pole damping because of its uncontrolled scale band, which is a judgement and not a derived criterion. In the next section we compare this range with the fixed-order perturbative uncertainty and with the known
NNLO corrections.

%% file: sections/tab_kscan.tex
\begin{tabular}{cccccccc}
\hline
$M$ [TeV] & $p_T^j/M$ & $K$ & $\delta_{\rm thr}/\delta_{\rm NLO}$ & undamped & damped & $z<0.8$ & $0.8<z<0.995$\\
\hline
1 & 0.10 & 1.93 & 1.17(1) & 1.23 & 1.04 & $+0.28$ & $-0.020(10)$\\
1 & 0.15 & 1.89 & 1.12(1) & 1.17 & 1.02 & $+0.22$ & $-0.006(12)$\\
1 & 0.20 & 1.87 & 1.07(1) & 1.13 & 1.02 & $+0.18$ & $-0.029(25)$\\
\hline
2 & 0.10 & 2.05 & 1.16(1) & 1.27 & 1.02 & $+0.21$ & $+0.002(5)$\\
2 & 0.15 & 1.96 & 1.13(1) & 1.20 & 1.01 & $+0.17$ & $-0.020(30)$\\
2 & 0.20 & 1.90 & 1.11(2) & 1.14 & 1.01 & $+0.15$ & $-0.015(23)$\\
\hline
4 & 0.10 & 2.13 & 1.19(1) & 1.33 & 1.02 & $+0.14$ & $+0.141(3)$\\
4 & 0.15 & 1.99 & 1.20(3) & 1.23 & 1.01 & $+0.12$ & $+0.139(5)$\\
4 & 0.20 & 1.90 & 1.20(2) & 1.16 & 1.01 & $+0.11$ & $+0.177(34)$\\
\hline
\end{tabular}

%% file: sections/results.tex
\section{Predictions for the invariant-mass distribution}
\label{sec:results}

We now collect our predictions for $d\sigma/d\Mttj$. Following section~\ref{sec:dampconsequences}, our central
prediction is the exact NLO cross section with $\mu_R=\mu_F=M$. The possible effect of threshold resummation is indicated by a range
from the complex-pole damped result with $E_0=100$~GeV to the undamped one, which is not a bound (section~\ref{sec:dampconsequences}). We first present these results with their
uncertainties, then discuss the dependence on the NLO scale, and finally compare with the known NNLO corrections.

\subsection{The distribution}
\label{sec:spectrum}

\begin{figure}[t]
  \centering
  \includegraphics[width=0.85\textwidth]{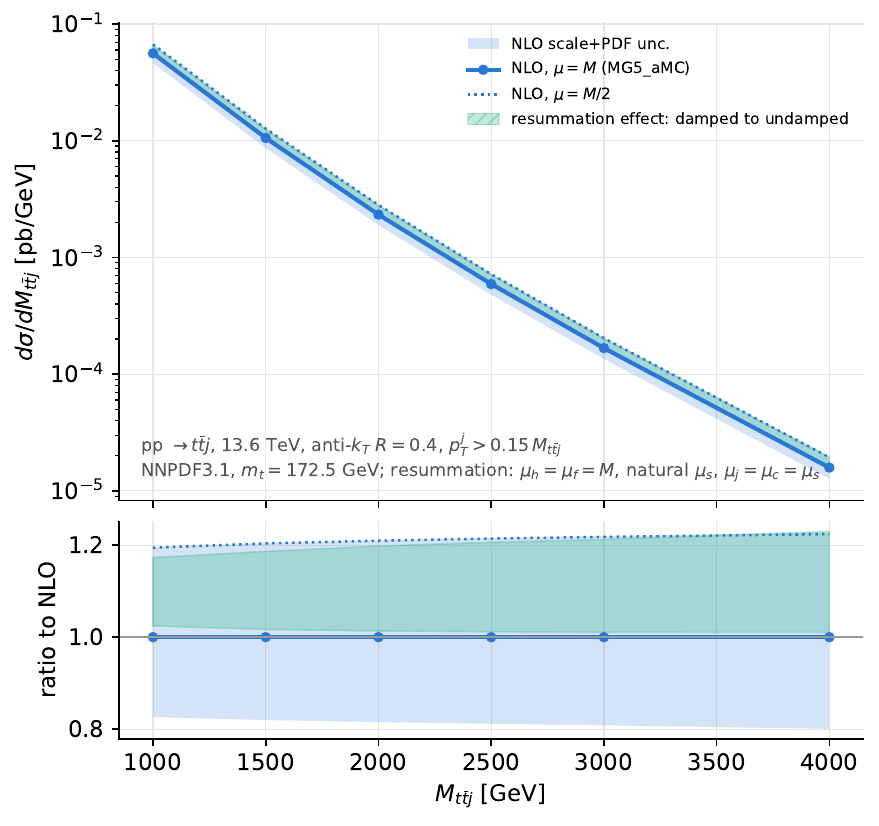}
  \caption{Distribution $d\sigma/d\Mttj$ at $\sqrt S=13.6$~TeV, for $R=0.4$ and $p_T^j>0.15\,\Mttj$ (upper panel), and ratio to NLO at $\mu=M$ (lower panel). Band around the NLO prediction:
  scale and PDF uncertainties. Hatched band: indicative range of the resummed correction, not a bound, from the complex-pole damped result with $E_0=100$~GeV to the undamped result. Also shown is NLO at
  $\mu=M/2$.}
  \label{fig:spectrum}
\end{figure}

Figure~\ref{fig:spectrum} and table~\ref{tab:spectrum} summarise our predictions.

\begin{table}[t]
  \centering
  \input{sections/tab_spectrum}
  \caption{NLO cross section with $\mu=M$, its uncertainty from scale and PDF variation, indicative range of the resummed correction relative to NLO for $p_T^j>0.15\,M$ (not an envelope of the variants of section~\ref{sec:damping}), and the NLO cross section at $\mu=M/2$ relative to $\mu=M$.}
  \label{tab:spectrum}
\end{table}

The NLO uncertainties are those of section~\ref{sec:nlo-num}. In the resummed result, we vary $\mu_h$, $\mu_s$, $\mu_c$ and
$\mu_j$ by factors of two, each separately with the others held fixed, as well as $\mu_f$, and add the positive and negative
deviations separately in quadrature (section~\ref{sec:scales}). This band is about $\pm10$\% for the undamped result and essentially that of NLO for the damped one (section~\ref{sec:damping}). We do not add it to the displayed range, since the range itself is the larger ambiguity. Neither the logarithms of the jet radius (section~\ref{sec:shower}) nor the mass logarithms (section~\ref{sec:masslogs}) are included in the table or the figure.

\subsection{Dependence on the NLO scale}
\label{sec:centralscale}

The size of the resummed correction depends strongly on the scale at which NLO is evaluated, and much of it can be reproduced by a change of this scale alone. Lowering the scale from $\mu=M$ to $\mu=M/2$ raises the NLO cross section by
\begin{equation}
  \begin{split}
  \frac{d\sigma^{\rm NLO}(\mu=M/2)}{d\sigma^{\rm NLO}(\mu=M)}&=1.195,\ 1.204,\ 1.210,\ 1.214,\ 1.218\ \text{and}\ 1.224\\
  &\quad\text{at}\quad M=1,\ 1.5,\ 2,\ 2.5,\ 3\ \text{and}\ 4~\text{TeV},
  \end{split}
  \label{eq:halfscale}
\end{equation}
which is the same size as the undamped correction of 17--23\% and grows with $M$ in the same way.

Matching at $\mu_R=\mu_F=M/2$, with $\mu_h=M$ and the other scales unchanged, confirms this. Relative to NLO at this scale, the undamped correction is then only $+1.3$, $+0.1$, $-0.9$, $-1.8$, $-2.5$ and $-3.5$\% at the masses of eq.~\eqref{eq:halfscale}, with a statistical uncertainty of the resummed part of at most 0.2\%, and the damped one lies between $+0.1$ and $-0.5$\%. The range quoted in table~\ref{tab:spectrum} therefore applies to $\mu=M$ and largely closes at $\mu=M/2$. Relative to NLO at $\mu=M$, the undamped matched result is 1.17--1.23 for $\mu=M$ and 1.18--1.21 for $\mu=M/2$, whereas the damped result follows NLO, with 1.01--1.02 and 1.20--1.22.

This does not show that the undamped result is correct. Its stability comes from the region that the damping removes, where the kernel fails at $\mu=M$, the only scale at which we have tested it (section~\ref{sec:lp-validity}). The result may also indicate that $\mu=M$ lies above the natural scale of the process: the kernel introduces the lower soft scale $\mu_s\approx M/3$ to $M/6$, eq.~\eqref{eq:musvalues}, where the coupling is larger.

\subsection{Sensitivity to the top-quark mass}
\label{sec:mtsens}

The distribution in $\Mttj$ is of particular interest as a probe of the top-quark mass~\cite{Alioli:2013mxa}, and an improved prediction
was the original motivation for threshold resummation of this process~\cite{Chargeishvili:2022ngl}. It is therefore natural to ask how the ambiguity found in
section~\ref{sec:damping} compares with the sensitivity to $m_t$. We define
\begin{equation}
  S(M)=\frac{d\ln\bigl(d\sigma/d\Mttj\bigr)}{d\ln m_t}\,,
  \label{eq:Smt}
\end{equation}
evaluated from calculations at $m_t=160$ and 185~GeV with fixed scales, parton distributions and coupling. A relative change $x$ of the cross section is then equivalent to a shift
$\delta m_t/m_t=\ln(1+x)/S$. For the normalised distribution $\sigma^{-1}d\sigma/d\Mttj$, relevant to most measurements, one has $S_{\rm norm}=S-S_{\rm tot}$, where $S_{\rm tot}$ is the corresponding quantity for the cross
section integrated over the range considered.

Table~\ref{tab:mtsens} shows the leading-order result for $p_T^j>50$~GeV, from which $S_{\rm tot}=-4.02$ for $395~\text{GeV}<\Mttj<4.2$~TeV. The sensitivity is
largest just above threshold, as is well known, and $S_{\rm norm}$ changes sign near 650~GeV. Beyond 1~TeV it remains surprisingly sizeable, $S\approx-2$. For $p_T^j>0.15\,M$, the
leading-order values are $S=-2.82(2)$ at 1~TeV and $-1.72(2)$ at 2~TeV. At NLO we find $S=-4.62(3)$ at $M=600$~GeV with $p_T^j>50$~GeV, and, with
$p_T^j>0.15\,M$, $S=-2.79(8)$, $-1.65(9)$ and $-1.08(15)$ at 1, 2 and 4~TeV, the last compared with $-1.09$ at leading order. Within the uncertainties the NLO corrections therefore do not change the sensitivity at these points.

\begin{table}[t]
  \centering
  \begin{tabular}{cccc}
    \hline
    $\Mttj$ [GeV] & fraction of $\sigma$ & $S$ & $S_{\rm norm}$\\
    \hline
    395--516   & 0.087 & $-10.82(4)$ & $-6.81$\\
    516--637   & 0.209 & $-4.90(2)$  & $-0.88$\\
    637--758   & 0.201 & $-3.51(2)$  & $+0.50$\\
    758--879   & 0.159 & $-2.96(2)$  & $+1.05$\\
    879--1000  & 0.122 & $-2.69(4)$  & $+1.33$\\
    1000--1200 & 0.074 & $-2.47(2)$  & $+1.54$\\
    1200--1400 & 0.044 & $-2.30(1)$  & $+1.72$\\
    1400--1600 & 0.026 & $-2.16(2)$  & $+1.86$\\
    2500--2700 & 0.002 & $-2.09(11)$ & $+1.92$\\
    \hline
  \end{tabular}
  \caption{Leading-order sensitivity of the distribution to the top-quark mass, eq.~\eqref{eq:Smt}, at $\sqrt S=13.6$~TeV with $p_T^j>50$~GeV. Numbers in brackets: statistical
  uncertainty of the last digits.}
  \label{tab:mtsens}
\end{table}

To measure what this means for a determination of the mass, we repeat the resummed calculation at $m_t=160$ and 185~GeV with the same scales and Born points. If data followed the matched prediction but were analysed with NLO, the fitted mass would be shifted by $\delta m_t/m_t=\ln(P/{\rm NLO})/S_{\rm NLO}$, with $P$ the matched prediction. For a distribution normalised in $1<M<4$~TeV the same relation applies to the normalised quantities, with $S_{\rm NLO}-S_{\rm tot}$ and $S_{\rm tot}=-2.37$ for this range. Table~\ref{tab:mtshift} shows the result.

\begin{table}[t]
  \centering
  \begin{tabular}{cccccccc}
    \hline
    & & \multicolumn{3}{c}{undamped} & \multicolumn{3}{c}{damped}\\
    $M$ [TeV] & $S_{\rm NLO}$ & $S-S_{\rm NLO}$ & $\delta m_t$ & $\delta m_t^{\rm norm}$ & $S-S_{\rm NLO}$ & $\delta m_t$ & $\delta m_t^{\rm norm}$\\
    \hline
    1   & $-2.79(8)$  & $+0.00$ & $-9.9$  & $+3.1$ & $+0.00$ & $-1.5$ & $-1.8$\\
    1.5 & $-2.09(7)$  & $+0.02$ & $-14.1$ & $+2.5$ & $+0.00$ & $-1.4$ & $-2.0$\\
    2   & $-1.65(9)$  & $+0.03$ & $-18.9$ & $+3.3$ & $+0.00$ & $-1.4$ & $-1.5$\\
    2.5 & $-1.62(10)$ & $+0.08$ & $-19.9$ & $+4.8$ & $+0.01$ & $-1.2$ & $-1.9$\\
    3   & $-1.31(6)$  & $+0.08$ & $-25.4$ & $+4.2$ & $+0.01$ & $-1.4$ & $-1.5$\\
    4   & $-1.08(15)$ & $+0.09$ & $-33.2$ & $+5.3$ & $+0.01$ & $-1.7$ & $-1.3$\\
    \hline
  \end{tabular}
  \caption{Effect of the resummed correction on the top-quark mass: NLO sensitivity, change of the sensitivity through the correction, and mass shifts in GeV for the absolute and for the normalised distribution, for $p_T^j>0.15\,M$ and $\sqrt S=13.6$~TeV. The statistical uncertainties of the shifts are dominated by $S_{\rm NLO}$. They are about 3--15\% for the absolute distribution and larger for the normalised one, at least 20\% at 1 and 1.5~TeV where the denominator is small. Here $S_{\rm tot}$ is the NLO value.}
  \label{tab:mtshift}
\end{table}

Three points follow. First, the resummed correction hardly changes the sensitivity itself: $S$ moves by at most 0.09, comparable to the uncertainty of $S_{\rm NLO}$, and so the leading-power correction depends only weakly on $m_t$. This says nothing about the terms depending on $M/m_t$ that the kernel does not contain (section~\ref{sec:offdiag}), where a dependence on $m_t$ could reside. Second, in the absolute distribution the difference between the damped and the undamped result corresponds to shifts of 8--32~GeV, so that the ambiguity of section~\ref{sec:damping} is far larger than any useful precision. Third, normalisation removes most of it: the two prescriptions give $+2.5$ to $+5.3$~GeV and $-1.3$ to $-2.0$~GeV, a spread of about 4--7~GeV. The denominator $S_{\rm NLO}-S_{\rm tot}$ changes sign between 1 and 1.5~TeV and is small there, so these two values are the least reliable. Even this is large compared with the current precision of top-quark mass determinations from this distribution~\cite{ATLAS:2015pfy,ATLAS:2019guf,CMS:2022emx}, which rely on a much wider range of $\Mttj$. We note that normalising in $1<M<4$~TeV is not the same as normalising the full $\rho_s$ distribution, for which the region of greatest sensitivity lies below $M\approx500$~GeV, where section~\ref{sec:lowmass} shows that the kernel does not apply. These values therefore do not translate directly to an actual measurement. They do show that threshold resummation in the present leading-power form would not reduce the theoretical uncertainty of a determination of the mass from high values of $\Mttj$.

\subsection{Comparison with the NNLO corrections}
\label{sec:nnlocomparison}

The NNLO QCD corrections to $pp\to t\bar tj$ have recently been computed in the leading-colour approximation for the two-loop
amplitudes~\cite{Badger:2025ilt}. This calculation provides the only direct test of the size of the corrections beyond NLO.
A comparison with our results requires some care. Ref.~\cite{Badger:2025ilt} considers $\sqrt S=13$~TeV with jets of
$R=0.4$, $|y_j|<2.5$, and the PDF4LHC21 parton distributions. Its results are given for a phase space with $p_T^j>30$~GeV and one with
a harder jet, and its central scale is $H_T/2$, with $H_T$ the sum of the transverse masses of the top quarks and of the transverse
momentum of the leading jet. It shows the related variable $\rho_s$ in a figure, but its text gives no numbers at the small values of $\rho_s$
corresponding to our range of $M$, about 0.085--0.34. We therefore compare only orders of magnitude and not the differential cross sections.

In our Born configurations the leading-order average of $H_T/M$ is 0.76 at 1~TeV and 0.57 at 4~TeV, so that our choice $\mu=M$ corresponds to $\mu\approx1.3$--$1.8\,H_T$ in their
notation. For $\mu=H_T$, ref.~\cite{Badger:2025ilt} finds a ratio NNLO/NLO of the total cross sections of 1.17 and 1.20 in the two
phase spaces, whereas it is 1.05 and 1.08 for $\mu=H_T/2$. If these ratios for the inclusive cross section carry over to high mass,
the NNLO correction relative to NLO at $\mu=M$ would be at least of the size of the undamped resummed correction and of
eq.~\eqref{eq:halfscale}, and larger than the complex-pole damped result, which is close to NLO. The decrease of the ratio towards lower scales shows that it mainly compensates the scale choice of the NLO calculation, which favours a lower NLO scale as much as resummation. This does not establish that the
undamped result is correct at higher orders. Its agreement could be accidental, as the size of the undamped correction depends on a
region where the kernel fails, and the inclusive ratios are dominated by much lower scales than those considered here. A differential
comparison at small $\rho_s$ would provide a much more stringent test, which we leave for when these results become available in a suitable form.

%% file: sections/tab_spectrum.tex
\begin{tabular}{ccccc}
\hline
$M$ [TeV] & NLO [pb/GeV] & uncertainty & range of resummation & $\mu=M/2$\\
\hline
1 & $5.611\times10^{-2}$ & $+19\%/-17\%$ & 1.02--1.17 & 1.19\\
1.5 & $1.056\times10^{-2}$ & $+20\%/-18\%$ & 1.02--1.19 & 1.20\\
2 & $2.330\times10^{-3}$ & $+21\%/-18\%$ & 1.01--1.20 & 1.21\\
2.5 & $5.921\times10^{-4}$ & $+22\%/-19\%$ & 1.01--1.21 & 1.21\\
3 & $1.671\times10^{-4}$ & $+22\%/-19\%$ & 1.01--1.21 & 1.22\\
4 & $1.581\times10^{-5}$ & $+23\%/-20\%$ & 1.01--1.23 & 1.22\\
\hline
\end{tabular}

%% file: sections/parton_shower.tex
\section{Comparison with NLO matched to a parton shower}
\label{sec:shower}

A resummation of threshold logarithms is not the only way to include higher orders. A parton shower generates
multiple emissions to all orders in the soft and collinear approximation, with exact kinematics for each emission and a
realistic treatment of the jet. It does not resum the threshold logarithms to the accuracy of the formula of
section~\ref{sec:resummation}, but it does describe effects that this formula treats only at fixed order. These include the
logarithms of the jet radius. In this section we compare the fixed-order result with NLO matched to a parton shower.

\subsection{Set-up}
\label{sec:ps-setup}

We use the \textsc{MC@NLO} method~\cite{Frixione:2002ik} as implemented in \textsc{MadGraph5\_aMC@NLO}~\cite{Alwall:2014hca} with
\textsc{Pythia8}~\cite{Bierlich:2022pfr}, with the settings listed in appendix~\ref{app:mg5}. The comparison is at the parton level
and with stable top quarks. It therefore compares perturbative radiation only. Hadronisation, multiple-parton interactions and top-quark
decays are not included. We reconstruct anti-$k_T$ jets from all partons other than the top quarks. The observable is $\Mttj$ from the two top quarks and the leading jet,
and we require $p_T^j>0.15\,\Mttj$. The result is averaged over five independent samples of 50\,000 events each at each mass, in a window of
$\pm2.5\%$ around $M$. This is wider than the windows of the fixed-order calculation in order to reduce the statistical uncertainty, and
the effect of this difference on the ratio to NLO is small. We estimate the shower uncertainty by varying the renormalisation
scale of the initial- and final-state shower by factors of 2, and by switching off the global recoil of final-state emissions.

\subsection{Results}
\label{sec:ps-results}

\begin{figure}[t]
  \centering
  \includegraphics[width=0.68\textwidth]{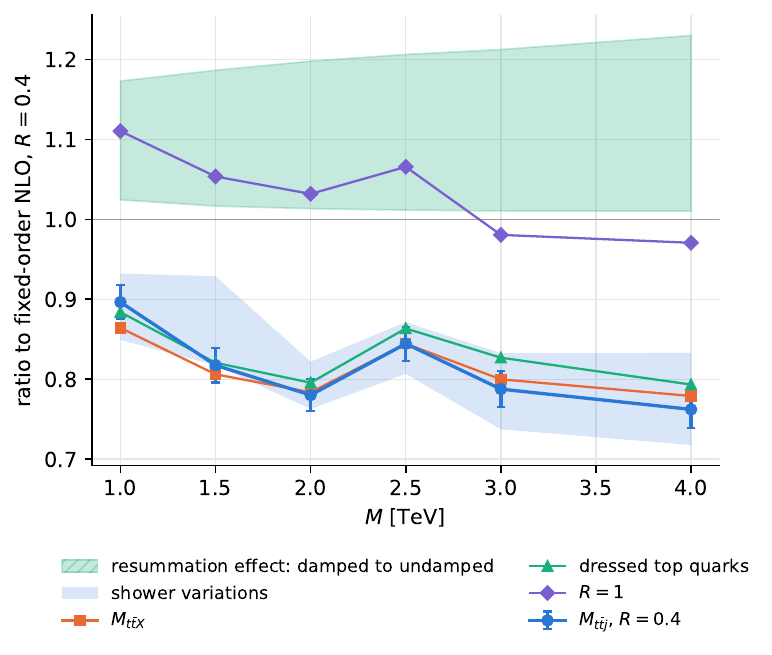}
  \caption{NLO matched to a parton shower relative to fixed-order NLO at $R=0.4$. Points: $\Mttj$ at $R=0.4$ with statistical uncertainties; band: shower variations. Other curves: the variants of the observable
  described in section~\ref{sec:ps-origin}, each relative to the same fixed-order result. Hatched band: indicative range of the resummed correction of section~\ref{sec:results} relative to NLO, not a bound.}
  \label{fig:ps}
\end{figure}

Figure~\ref{fig:ps} and table~\ref{tab:ps} show the results. At $R=0.4$ the shower reduces the cross section by 10--24\% relative to fixed-order NLO. It is smallest at 1~TeV, 10\%, and lies between 16\% and 24\% at higher masses without a monotonic trend, and we do not interpret this in detail. The shower variations change the central shower result by between $-6\%$ and $+14\%$. At most masses the downward and upward changes are each below 10\%. As a check of the samples we compared them, in a window of $\pm10\%$ around $M$, with those generated with the restriction on $\hat s$ of section~\ref{sec:nlo-num}. The ratio averages 1.008(7) over the six masses, which is smaller than the 2--3\% found at fixed order. These shower samples have a softer generation-level jet threshold and a wider slice in $\Mttj$, but we have not established why the effect differs.

\begin{table}[t]
  \centering
  \input{sections/tab_ps}
  \caption{Ratio of NLO matched to a parton shower to fixed-order NLO at $R=0.4$, in a window of $\pm2.5\%$ around $M$, with the statistical uncertainty of the last digits in brackets, the range of the shower
  variations relative to the central shower result, and the variants of the observable of section~\ref{sec:ps-origin}.}
  \label{tab:ps}
\end{table}

\subsection{Origin of the reduction}
\label{sec:ps-origin}

The total rate is, by construction, preserved by the
unitarity of the shower, so a reduction of the cross section differential in $\Mttj$ reflects migration: events leave the window around $M$ and, since the spectrum falls steeply, fewer
enter it from higher values. Emission of radiation outside the jet lowers $\Mttj$, as this radiation does not enter its definition. We
examine three variants of the observable, each of which addresses one possible source of this radiation.

\begin{itemize}
\item \emph{All jets.} We replace $\Mttj$ by the invariant mass $M_{t\bar tX}$ of the top quarks and all jets above the threshold. This
  tests the role of a separate hard jet.
\item \emph{Dressed top quarks.} Partons within $\Delta R<0.4$ of a top quark are added to it before the jets are clustered. This tests
  radiation from the top quarks, which is excluded from the jet but which would be recovered by a reconstruction of the top quark from its decay
  products together with nearby radiation.
\item \emph{Large jet radius.} We use $R=1$, which recovers wide-angle radiation from the jet.
\end{itemize}

The last columns of table~\ref{tab:ps} show that the first two variants have little effect. The ratio for $M_{t\bar tX}$ differs from that for $\Mttj$ by at most 4\%, and dressing of the top quarks changes the ratio by at most 5\%. The change to $R=1$ has a much larger effect: the ratio rises to 0.97--1.11. For comparison, at fixed order the cross section at $M=2.5$~TeV changes by only a factor of 1.035 between $R=0.4$ and $R=1$. The shower result for $R=1$ at this mass, relative to the fixed-order result for $R=1$, is thus 1.03. We have this fixed-order comparison at one mass only. These findings are consistent with radiation from the jet at angles larger than $R$. We therefore interpret the reduction tentatively as an all-order effect of logarithms of $R$; the small change from dressing with $\Delta R<0.4$ does not exclude radiation from the top quarks, as the fixed-order test of section~\ref{sec:offdiag}, which uses a different definition of the dressing, finds that this cone captures little of it. A jet of small radius loses a fraction of its energy to out-of-cone radiation, at a rate
governed by $\as\ln(1/R^2)$, and the steeply falling spectrum converts this into a sizeable reduction of the rate, as has been
studied for inclusive jet cross sections~\cite{Dasgupta:2014yra,Dasgupta:2016bnd,Kang:2016mcy}. Our resummed formula sets $\mu_c=\mu_j=\mu_s$
and therefore treats these logarithms at fixed order, so it cannot capture this effect.

\subsection{Discussion}
\label{sec:ps-discussion}

The reduction of 10--24\% is of the same size as the upper end of the resummation range of section~\ref{sec:results}, and has the opposite sign (figure~\ref{fig:ps}). We stress that the two effects are not additive, and that a
consistent combination of threshold and small-$R$ resummation is not available for this process. An approach to resum the logarithms of $R$ with
$\mu_c\ne\mu_j$ exists in the formula of section~\ref{sec:resummation}, but the hierarchy needed for it is only marginal at 1~TeV, table~\ref{tab:hierarchy}, and it would require control of the non-global logarithms beyond the leading-colour approximation.

%% file: sections/tab_ps.tex
\begin{tabular}{cccccc}
\hline
$M$ [TeV] & NLO+PS/NLO & shower variations & $M_{t\bar tX}$ & dressed & $R=1$\\
\hline
1 & $0.896(21)$ & $+4\%/-5\%$ & 0.864 & 0.884 & 1.110\\
1.5 & $0.817(22)$ & $+14\%/0\%$ & 0.806 & 0.821 & 1.054\\
2 & $0.780(20)$ & $+5\%/-2\%$ & 0.784 & 0.796 & 1.032\\
2.5 & $0.844(21)$ & $+3\%/-4\%$ & 0.844 & 0.863 & 1.066\\
3 & $0.788(22)$ & $+6\%/-6\%$ & 0.800 & 0.827 & 0.981\\
4 & $0.762(24)$ & $+9\%/-6\%$ & 0.779 & 0.793 & 0.971\\
\hline
\end{tabular}

%% file: sections/conclusions.tex
\section{Conclusions}
\label{sec:conclusions}

We have presented the threshold resummation of the invariant-mass distribution of $t\bar t$+jet production at NNLL
accuracy, matched to the exact NLO cross section. The calculation combines the one-loop hard functions from
\textsc{Recola}, the one-loop soft function with the full colour structure, the jet and soft-collinear functions of a small-radius jet,
and the two-loop anomalous dimensions. Three corrections to the one-loop soft anomalous dimension of
ref.~\cite{Chargeishvili:2022ngl} were necessary to obtain agreement with the universal infrared structure of the amplitudes. We
validated the hard functions against \textsc{MadLoop} at production phase-space points.

Our main result concerns the applicability of the method at the LHC. At invariant masses below about 1~TeV with a fixed jet threshold, it does
not apply: the luminosity is too flat and the jet too soft. At $M=1$--4~TeV with $p_T^j>0.15\,M$, the leading-power kernel describes the exact real-emission density
well close to threshold, to within a few per cent at $z>0.95$, but fails once the emitted energy exceeds about 100--200~GeV. Consequently the resummed correction beyond NLO depends on the
treatment of a region where the kernel is not reliable. It ranges from 1--2\% to 17--23\%, with an upper end of 13--16\% for $p_T^j>0.2\,M$ and 23--33\% for $p_T^j>0.1\,M$, and further with other prescriptions, and even changes
sign under a variation of the kernel that is suppressed by a power of $1-z$ at low masses. These values hold at $\mu=M$; matched at $\mu=M/2$, where NLO itself is 20--22\% higher, the correction is only $+1.3$ to $-3.5$\%. We therefore do not regard resummation as a controlled improvement of
the NLO prediction in this regime, and we retain NLO as our central prediction with a range for the possible effect of resummation.

Several statements may be useful beyond this analysis. The Drell--Yan validation shows that the histogram of $z$ from a fixed-order program, with the window imposed at generation level,
can be used to test leading-power approximations bin by bin in a process with a complicated final state, and to locate where they fail. This test is more
informative than the ratio of the expanded correction to the exact one, which we find can be close to one through a cancellation of regions of $z$. A lower limit on the partonic energy
at generation level, which is a tempting way to improve efficiency in this kind of calculation, can change the result by 2--3\%. The logarithms of the top-quark mass contained in the leading-power kernel do not
appear to have a large effect in this observable, since the top quarks are not strongly boosted for a large part of the cross section, although our treatment of them is
an estimate. This statement does not extend to the terms absent from the kernel, see below. Matching to a parton shower yields a reduction of 10--24\% at $R=0.4$, which suggests, on the basis of one fixed-order comparison, that logarithms of the jet radius may be as relevant as the
threshold logarithms for this observable.

It is instructive to compare with threshold resummation for $t\bar tH$, $t\bar tW^\pm$ and $t\bar tZ$, which uses the same threshold variable. There too the size of the correction depends strongly on the central scale, whereas the resummed result does not. For $t\bar tZ$ at 13~TeV, ref.~\cite{Kulesza:2018tqz} finds a ratio of NLO+NNLL to NLO of 1.29 for a scale equal to the invariant mass and 1.14 for half of it, with NLO+NNLL cross sections that differ by 1\%. We find a similar pattern for the undamped result (section~\ref{sec:centralscale}), although the correction nearly vanishes at $\mu=M/2$. In our case, however, this stability coexists with a kernel that fails beyond emitted energies of 100--200~GeV, and it disappears after damping. Scale stability alone is therefore not sufficient to establish that a threshold kernel is reliable.
The validation in these works compares the NLO expansion of the resummation formula with exact NLO without the $qg$ channel, which for $t\bar tZ$ at $\mu_F=M/2$ agree to about 1\% in the total cross section~\cite{Broggio:2017kzi}. Our results show that such integrated agreement can arise from a cancellation between regions of $z$ and between initial states, and that the initial states absent from the kernel grow with $\tau$. Consistently, the high-mass tail of $t\bar tW^\pm$ is dominated by the $qg$ channel~\cite{Broggio:2019ewu}. We cannot conclude that the kernels of these processes fail in the same way, since their final state contains no jet. The histogram of $z$ described above could, however, be applied to them directly.

Some questions remain open. A difference between the threshold kernel and the exact result of up to about 1.5\% in the range next to threshold persists after
accounting for terms proportional to $R^2$, and its origin is unknown. For the larger deviation at moderate $1-z$ at $M=4$~TeV we have identified two contributions. Initial states absent from the kernel depend on $\tau$ and agree to within about 15\% with a parameter-free expression from the collinear limit, for $z>0.8$ and in the main channels. A second contribution depends on $M/m_t$ at fixed $\tau$, as a numerical test at a lowered top-quark mass shows, and it is responsible for a growth of the total difference with $M$ that a steeper luminosity does not remove. Radiation from the top quarks beyond its soft limit is the natural candidate, and a test with a dressed invariant mass supports it, but this mechanism is not fully established. We do not know the reason for the size of the dependence on the damping function. Finally, a
differential comparison with the NNLO calculation of ref.~\cite{Badger:2025ilt}, at small $\rho_s$, would show whether the correction at high mass has the size of the undamped result, although it could not by itself separate resummation from a lower NLO scale.

For improvements we see three directions: the inclusion of next-to-leading-power terms, for which the off-diagonal initial states provide a first and simple target; a matching to NNLO, for which the two-loop amplitudes would have to be combined with
the two-loop hard and soft functions in the kernel; and the resummation of logarithms of $R$ together with those at threshold.
Next-to-leading-power effects have so far been studied mainly for colour-singlet final
states~\cite{Laenen:2008ux,Bonocore:2015esa,Beneke:2018gvs,Bahjat-Abbas:2019fqa}, including phenomenological
applications matched to NNLO~\cite{Bhattacharya:2021hae,Bhattacharya:2025rqk}, and more recently for Higgs plus jet
production~\cite{Pal:2023vec}. The leading logarithms of channels that first appear at next-to-leading power, the
analogue of our off-diagonal initial states, have been resummed for deep-inelastic
scattering~\cite{Vogt:2010cv,Almasy:2010wn,Beneke:2020ibj} and for Drell--Yan and Higgs
production~\cite{LoPresti:2014ihe}.

%% file: sections/appendix_anomalous.tex
\section{Anomalous dimensions}
\label{app:anomalous}

In this appendix we list the perturbative coefficients used at NNLL, in the normalisation
$\gamma(\as)=\sum_{n\ge0}\gamma_n\,a^{n+1}$ with $a=\as/(4\pi)$. Here $T_F=\tfrac12$, $C_A=3$, $C_F=\tfrac43$ and $n_l=5$.

\paragraph{Cusp anomalous dimension~\cite{Moch:2004pa}.}
\begin{align}
  \gamma^{\rm cusp}_0&=4\,,\qquad
  \gamma^{\rm cusp}_1=\Big(\frac{268}{9}-\frac{4\pi^2}{3}\Big)C_A-\frac{80}{9}T_Fn_l\,,\nonumber\\
  \gamma^{\rm cusp}_2&=C_A^2\Big(\frac{490}{3}-\frac{536\pi^2}{27}+\frac{44\pi^4}{45}+\frac{88}{3}\zeta_3\Big)
  +C_AT_Fn_l\Big(-\frac{1672}{27}+\frac{160\pi^2}{27}-\frac{224}{3}\zeta_3\Big)\nonumber\\
  &\quad+C_FT_Fn_l\Big(-\frac{220}{3}+64\zeta_3\Big)-\frac{64}{27}T_F^2n_l^2\,.
\end{align}
The two-loop velocity-dependent cusp anomalous dimension is taken from refs.~\cite{Korchemsky:1987wg,Kidonakis:2009ev}
in the form given in ref.~\cite{Ferroglia:2009ii}.

\paragraph{Non-cusp terms of the hard anomalous dimension~\cite{Becher:2009qa,Becher:2009kw}.}
\begin{align}
  \gamma^q_1&=C_F^2\Big(-\frac32+2\pi^2-24\zeta_3\Big)+C_FC_A\Big(-\frac{961}{54}-\frac{11\pi^2}{6}+26\zeta_3\Big)\nonumber\\
  &\quad+C_FT_Fn_l\Big(\frac{130}{27}+\frac{2\pi^2}{3}\Big)\,,\\
  \gamma^g_1&=C_A^2\Big(-\frac{692}{27}+\frac{11\pi^2}{18}+2\zeta_3\Big)+C_AT_Fn_l\Big(\frac{256}{27}-\frac{2\pi^2}{9}\Big)
  +4C_FT_Fn_l\,,\\
  \gamma^Q_1&=C_FC_A\Big(\frac{2\pi^2}{3}-\frac{98}{9}-4\zeta_3\Big)+\frac{40}{9}C_FT_Fn_l\,.
\end{align}
The three-parton term of the two-loop anomalous dimension is given in refs.~\cite{Ferroglia:2009ep,Ferroglia:2009ii}.

\paragraph{Large-$x$ PDF anomalous dimensions~\cite{Becher:2007ty,Ahrens:2010zv}.}
\begin{align}
  \gamma^\phi_{q,1}&=C_F^2\Big(\frac32-2\pi^2+24\zeta_3\Big)+C_FC_A\Big(\frac{17}{6}+\frac{22\pi^2}{9}-12\zeta_3\Big)
  -C_FT_Fn_l\Big(\frac23+\frac{8\pi^2}{9}\Big)\,,\\
  \gamma^\phi_{g,1}&=C_A^2\Big(\frac{32}{3}+12\zeta_3\Big)-\frac{16}{3}C_AT_Fn_l-4C_FT_Fn_l\,.
\end{align}

\paragraph{Unmeasured jet function.}
For quark~\cite{Liu:2021xzi} and gluon~\cite{Cal:2024yjz} jets,
\begin{align}
  \gamma^J_{q,1}&=11.17\,C_F^2-181.30\,C_FC_A-7.916\,C_FT_Fn_l\,,\\
  \gamma^J_{g,1}&=52.898\,C_A^2-215.20\,C_AC_F-10.045\,C_F^2-5.1089\,C_AT_Fn_l-0.37279\,C_FT_Fn_l\,.
\end{align}
These coefficients enter only when $\mu_j\neq\mu_s$, see section~\ref{sec:jet}.

%% file: sections/appendix_mellin.tex
\section{Numerical inversion of the Mellin transform}
\label{app:mellin}

The closed form~\eqref{eq:kernelz} of the kernel in momentum space is available only when the dependence on $N$ is
that of $f_N(\eta)$, eq.~\eqref{eq:fN}. Multiplying the kernel by a function $D(N)$, as we do in
section~\ref{sec:damping}, destroys this property. In this appendix we describe how we then return to momentum space.

We split the matched cross section into the undamped result, which is computed with eq.~\eqref{eq:subtraction}, and
a remainder,
\begin{equation}
  R(z)=\frac{1}{2\pi i}\int_{\mathcal C}dN\,z^{-N}\,\bigl[D(N)-1\bigr]\,
  \bigl[K_{\rm res}(N)-K_{\rm thr}(N)\bigr]\,.
  \label{eq:Rz}
\end{equation}
Only the partonic kernel is inverted, so that no moments of the parton distributions are needed: $R(z)$ is convolved
with the luminosity in momentum space. Because $D(N)-1$ falls like $1/N$ or $1/N^2$ at large $N$ and
$K_{\rm res}-K_{\rm thr}$ vanishes at $\order{\as}$, the damping acts on the higher orders only and leaves the
fixed-order expansion to $\order{\as}$ unchanged.

The integrand is analytic in the right half plane, apart from the poles of $D(N)$: for $D(N)=N^2/(N^2+N_0^2)$ these
lie at $N=\pm iN_0$, and for $D(N)=N/(N+N_0)$ at $N=-N_0$. Since the integrand obeys $F(N^*)=F(N)^*$, it suffices to integrate along the upper half of the
contour and to take the imaginary part,
\begin{equation}
  R(z)=\frac{1}{\pi}\,{\rm Im}\int_{\mathcal C_+}dN\,z^{-N}F(N)\,.
\end{equation}
The path $\mathcal C_+$ first runs vertically, $N=c+is$ with $c=2$ and $0\le s\le S_1$, and then continues along the
ray $N=c+iS_1+te^{3\pi i/4}$ with $0\le t\le T$. The vertical segment passes above the poles at $\pm iN_0$ when $S_1>N_0$. On the ray, ${\rm Re}\,N$ decreases, so that $|z^{-N}|$ decays for $z<1$, and the contour may be
deformed in this way without crossing singularities of $f_N(\eta)$, which lie at ${\rm Re}\,N\le\eta$ on the real axis
for the values of $\eta$ we encounter. We use Gauss--Legendre rules with 200 nodes on the vertical segment and 600
nodes on the ray, with the mapping $t=Ts^3$ and $T=3000$.

We test this procedure on the function $f_N(\eta)$ itself, for which the exact result is eq.~\eqref{eq:kernelz}. For
$\eta=-0.4$, $-0.2$ and $0.15$ and $z=0.2$, 0.5, 0.8, 0.95 and 0.99, the relative deviation does not exceed
$2\times10^{-8}$. The test does not include the factor $D(N)$; the corresponding poles are handled by the choice of
$S_1$. The stability of the damped result is further tested by the comparison of the two forms of $D(N)$ in
section~\ref{sec:damping}.

%% file: sections/appendix_mg5.tex
\section{Set-up of the fixed-order and parton-shower calculations}
\label{app:mg5}

In this appendix we record the settings of our \textsc{MadGraph5\_aMC@NLO}~\cite{Alwall:2014hca} runs, so that they can be
reproduced. We use version 3.8.1 with the model \texttt{loop\_sm-no\_b\_mass} and the process \texttt{p p > t t\~{} j [QCD]}, where
the proton and the jet contain gluons and five massless quark flavours.

\paragraph{Fixed-order runs.}
The run settings are
\begin{itemize}
\item $\sqrt S=13.6$~TeV, or as stated in the text, with the \textsc{LHAPDF} set number 303600;
\item $m_t=172.5$~GeV, $\Gamma_t=0$ and fixed scales $\mu_R=\mu_F=M$;
\item a required accuracy of \texttt{req\_acc\_FO} $=0.005$ for the cross section in the window; at $M=3$~TeV we average two independent runs with \texttt{req\_acc\_FO} $=0.01$;
\item anti-$k_T$ jets, \texttt{jetalgo} $=-1$, $R=0.4$, without a rapidity requirement, and a jet threshold $p_T^j>0.15\,M$;
\item a user cut that requires at least one jet above threshold and an invariant mass of the top quarks and the leading jet inside the
  window $M\pm\Delta M$, with $\Delta M=10$~GeV, and $\Delta M=40$~GeV at $M=4$~TeV. The window is thus part of the generation, so
  that the integration effort is concentrated in it.
\end{itemize}
The histograms are filled with the nine combinations of $\mu_R$ and $\mu_F$ obtained by multiplying them by $1/2$, 1 and 2. We build the
seven-point uncertainty from these columns. For the PDF uncertainty, separate runs include the reweighting to the 100 replicas. These runs
retain the restriction on $\hat s$ described below. They are used only for the relative PDF uncertainty, on which it should have
little effect.

\paragraph{The lower limit on the partonic energy.}
\textsc{MadGraph5\_aMC@NLO} does not know that our user cut restricts $\Mttj$. It therefore samples the partonic energy from a low value
determined by the masses and the jet threshold. In an early stage of this work we raised the lower limit to
$\sqrt{\hat s}>0.75\,(M-\Delta M)$ at $M\ge1$~TeV in order to improve the efficiency. This limit applies to the
partonic energy of the Born configuration, which can be much smaller than that of a real-emission event with a hard emission, so that
the limit can remove physical contributions. The final results use a limit of $0.01\,(M-\Delta M)$, far below any value that contributes.
The size of the effect is given in table~\ref{tab:taucut}.

\paragraph{Parton-shower runs.}
For section~\ref{sec:shower} we generate samples with \textsc{MC@NLO} matching to \textsc{Pythia8}~\cite{Bierlich:2022pfr}, with 50\,000
events for each of five independent samples at each value of $M$. The samples cover the slice $0.75\,M<\Mttj<1.3\,M$, with a
parton-level jet threshold of $0.08\,M$, looser than the threshold $0.15\,M$ applied in the analysis. The top quarks are stable, and
hadronisation, multiple-parton interactions and QED radiation are switched off. We use the same lower limit on $\sqrt{\hat s}$ as in the fixed-order runs. The shower variations are as
follows: the renormalisation scale of the initial- and final-state shower is multiplied by 1/2 and 2, which corresponds to
\texttt{renormMultFac} $=0.25$ and 4 for the scale squared, and the global recoil of final-state emissions is switched off.